\documentclass[submitting]{nst}
\usepackage{subfigure,dcolumn}
\usepackage[T2A,T1]{fontenc}
\usepackage[russian,english]{babel}
\usepackage{appendix}
\usepackage{placeins}
\usepackage{color, colortbl}
\usepackage[autostyle]{csquotes}
\usepackage [utf8] {inputenc}
\usepackage{commath}
\usepackage{framed} % Framing content
\usepackage{nomencl} % Nomenclature package
\makenomenclature
\usepackage{tasks}
\usepackage{listings}

\usepackage{pbox,calc}

\begin{document}

%\title{Towards the inclusion of model uncertainties in nuclear data evaluations}

\title{Bayesian Model Averaging (BMA) for nuclear data evaluation}

\author{E. Alhassan}
\email[erwin.alhassan@sckcen.be ]{}%{(Now at SCK-CEN, Belgium)}
\affiliation{SCK-CEN (Belgian Nuclear Research Centre) Boeretang 200, 2400 Mol, Belgium}
\author{D. Rochman}
\email[dimitri-alexandre.rochman@psi.ch]{}
\affiliation{Laboratory for Reactor Physics and Thermal-Hydraulics, Paul Scherrer Institute, 5232 Villigen, Switzerland}
\author{G. Schnabel}
\email[G.Schnabel@iaea.org]{}
\affiliation{Nuclear Data Section, International Atomic Energy Commission (IAEA), Vienna, Austria}
\author{A.J. Koning}
\email[A.Koning@iaea.org]{}
\affiliation{Nuclear Data Section, International Atomic Energy Commission (IAEA), Vienna, Austria}
\affiliation{Division of Applied Nuclear Physics, Department of Physics and Astronomy, Uppsala University, Uppsala, Sweden}

\begin{abstract}
In order to ensure agreement between theoretical model calculations and experimental data, parameters to selected sets of nuclear reaction models coupled together in a nuclear physics code such as TALYS, are usually perturbed and fine-tuned in nuclear data evaluations. This approach assumes that the chosen set of models accurately represents the `true' distribution. Furthermore, the models are chosen globally, indicating their applicability across the entire energy range of interest. However, this approach overlooks uncertainties inherent in the models themselves. As a result, achieving satisfactory fits to experimental data within certain energy regions for specific channels becomes challenging, as the evaluation is constrained by the limitations or deficiencies of the selected models. In this work, we propose that instead of selecting globally a winning model set and proceeding with it as if it was the `true' model set, we take instead, a weighted average over multiple models available within a Bayesian Model Averaging (BMA) framework, each weighted by its posterior probability. The method involves the generation of a set of TALYS calculations by randomly varying multiple nuclear reaction models as well as their parameters to yield a vector of calculated observables. Next, the likelihood function was computed at each considered incident energy point for selected cross sections by comparing the vector of calculated observables with that of the selected differential experimental data. These were then combined with the prior distributions to obtain updated posterior distributions together with their means and corresponding standard deviations for the quantities of interests (QOI). The QOI considered in this work include the reaction, and residual production cross sections as well as the elastic angular distributions. As the cross sections and elastic angular distributions were updated locally on a per-energy-point basis, the approach typically results in discontinuities or "kinks" in the curves, and these were addressed using spline interpolation. The proposed BMA method has been applied to the evaluation of proton induced reactions on $^{58}$Ni within 1 - 100 MeV. The results demonstrate favorable comparisons with experimental data, as well as with the TENDL-2021 evaluation. 
\end{abstract}

\keywords{Bayesian Model Averaging (BMA); Nuclear data; Nuclear reaction models; model parameters; TALYS code system, covariances.}

\maketitle

\begin{table*}[!t]  
\begin{framed}
List of symbols and variables   \\
\begin{tasks}(2)
\task {$j:$ }{Model index represents the index corresponding to a model set among many model sets.} 
\task {$k:$ }{Parameter index denotes a specific parameter vector within a model set. Each $k$ parameter set can be mapped to $k$ cross sections at each incident energy, belonging to the $k$ random nuclear data file.} 
% \task {$k:$}{parameter index denotes a specific model vector within a model set.} 
\task {$i:$ }{Index denotes a single experimental data point.}
\task {$c:$ }{Index represents a particular nuclear reaction channel or cross section.}
\task {$\rm subscript$ $w:$ }{Denotes weighted.}
\task {$var:$ }{Denotes variance.}
\task {$\overrightarrow{\sigma^{cal}_{cik}}:$ }{Vector of TALYS calculated observables for channel, $c$, as a function of incident energy, $i$, and corresponding to the $k$ parameter set.}
\task {$\overrightarrow{\sigma^{exp}_{ci}}:$ }{Vector of experimental observables (cross sections and angular distributions) for the channel $c$, and incident energy,$i$.}
\task {$\overrightarrow{\theta_k}:$}{represents the $k^{th}$ parameter set associated with model, $\overrightarrow{\mathcal{M}_j}$.}
\task {$p(\overrightarrow{\mathcal{M}_j}):$ }{Prior probability for model $j$.}
\task {$p(\overrightarrow{\theta_k}|\overrightarrow{\mathcal{M}_j}):$ }{Prior distribution of the parameter, $\theta_k$, given model $\mathcal{M}_j$.}
\task {$p(\overrightarrow{\sigma^{exp}_{ci}}|\overrightarrow{\mathcal{M}_j},\overrightarrow{\theta_k}):$ }{Likelihood function which represents the probability of observing experimental data $\sigma^{exp}_{ci}$ given model, $\overrightarrow{\mathcal{M}_j}$, and parameter, $\overrightarrow{\theta_k}$.}
\task {$p(\overrightarrow{\mathcal{M}_j},\overrightarrow{\theta_k}):$ }{Joint probability of model, $\overrightarrow{\mathcal{M}_j}$, and parameter, $\overrightarrow{\theta_k}$, given experimental data, $\sigma^{exp}_{ci}$.}
\task {$p(\overrightarrow{\sigma^{exp}_{ci}}|\overrightarrow{\mathcal{M}_j}):$ }{Marginal likelihood or evidence given model, $\mathcal{M}_j$.}
\task {$\chi^2_{cik}:$ }{The reduced chi square computed at each considered incident energy, $i$, channel, $c$, and model parameter vector, $k$.}
\task {$\Delta \sigma^{exp}_{ci}:$ }{Denotes the experimental uncertainty at energy, $i$, for channel, $c$.}
\task {$\overline{\sigma^{\rm cal}_{cik}}:$ }{Calculated average cross section over the models and parameters at incident energy, $i$, channel, $c$, and model parameter vector, $k$.}
\task {$var(\sigma^{cal}_{ci}):$ }{Denotes the variance of the distribution of the calculated cross sections at energy, $i$, for channel, $c$.}
% \task {$E[\sigma^{cal}_{cik}]:$}{The expectation value of the calculated cross section, $\sigma^{cal}_{cik}$ at energy, $i$, channel $c$ for a total of $K$ samples.}
\task {$ w_{cik}:$ }{Denotes Bayesian Monte Carlo (BMC) weights for the channel, $c$, at incident energy, $i$, and random nuclear data file, $k$.}
\task {$\overline{\sigma^{\rm cal}_{cik}}_{w}:$ }{Weighted mean of the TALYS calculated cross sections for the channel, $c$, at incident energy, $i$, and random nuclear data file, $k$.}
\task {$cov_w:$ }{Weighted covariance.}
\task {$r:$ }{Correlation coefficient.}
\task {$\sigma_{\rm T_{a}}^{c}:$ }{Denotes TALYS ($T$) calculated cross sections at energy, $a$, for channel, $c$.}
\task {$\sigma_{\rm T_{b}}^{c}:$ }{Denotes TALYS ($T$) calculated cross sections at energy, $b$, for channel, $c$.}
\task {$var(\sigma^{cal}_{cik,comb}):$ }{Combined variance of the calculated cross section at energy, $i$, for channel, $c$, obtained from the variation of both models and their parameters.}
\task {$var(\sigma^{cal}_{cik,\mathcal{M}}):$ }{Variance of the calculated cross section obtained from the variation of many models ($\mathcal{M}$).}
\task {$var(\sigma^{cal}_{cik,\theta}):$ }{Variance of the calculated cross section obtained from the variation of only model parameters ($\theta$).}
\task {$U(\sigma^{cal}_{cik,\mathcal{M}}):$ }{The uncertainty due to models for channel, $c$, at incident energy, $i$}
\task {$p(\overrightarrow{\sigma^{cal}_{cik}}|\overrightarrow{\sigma^{exp}_{ci}}):$ }{Posterior distribution for the quantity of interest ($\overrightarrow{\sigma^{cal}_{cik}}$) given experimental data.}
\task {$p(\overrightarrow{\sigma^{cal}_{cik}}|\overrightarrow{\mathcal{M}_j},\overrightarrow{\theta_k},\overrightarrow{\sigma^{exp}_{ci}}):$ }{Joint posterior probability of our QOI given model $\overrightarrow{\mathcal{M}_j}$ and parameter $\overrightarrow{\theta_k}$.}
% \printnomenclature
\end{tasks}
\end{framed}

\end{table*}

\section{Introduction}
In a typical Monte Carlo method for nuclear data evaluation in the fast energy region as outlined in various references~\cite{bib:1aa,bib:2,bib:4,bib:5,bib:6,bib:8,bib:42Pu,bib:42Cu,bib:041}, parameters to pre-selected models are adjusted to fit selected experimental data. In Refs.~\cite{bib:42Pu,bib:42Cu} for example, a method based on the use of the minimum $\chi^2$ was used to determine the best nuclear data file from a large set of random files produced within the Total Monte Carlo method~\cite{bib:1}. In Ref.~\cite{bib:041}, a weighted $\chi^2$ which assigned large weights to reaction channels with a large number of experimental data, to experimental data with smaller uncertainties as well as to channels with large cross sections, was presented. 
%
% In Refs.~\cite{bib:1aa,bib:12}, a `best' model combination that yielded the smallest reduced $\chi^2$ value was selected from many candidate models, a process usually referred to as \emph{model selection}~\cite{bib:12wasserman}. The selected models were then used as if they were the `true' models for the parameter variation step. 
% 
Other works, such as Refs.~\cite{bib:1aa,bib:12}, followed a `model selection' process~\cite{bib:12wasserman,bib:13a}, wherein the `best' model combination, yielding the smallest reduced $\chi^2$ value, was chosen from a pool of candidate models. These selected models were then treated as if they represented the `true' models for subsequent parameter variation steps. In Refs.~\cite{bib:1aa,bib:12}, the reduced $\chi^2$ was obtained by comparing model calculations to three different types of experimental data which included the reaction cross sections, the residual production cross section and the elastic angular distributions, and particularly applied to the evaluation of proton-induced reactions. A more statistically rigorous Monte Carlo approach is to base the entire evaluation on Bayes' theory as presented in Refs.~\cite{bib:2,bib:4,bib:2a,bib:3,bib:5,bib:6}. This approach has the advantage that both posterior means and covariances can be obtained. In recent years, the Bayesian Monte Carlo (BMC) method has found application in the TALYS Evaluated Nuclear Data Library (TENDL) evaluations~\cite{bib:13}. In Ref.~\cite{bib:1EACd111}, the iterative Bayesian Monte Carlo ($iBMC$) method presented in Ref.~\cite{bib:1aa} was applied for the evaluation of p+$^{111}Cd$ between 1 and 100 MeV. Here, a `best' model set was selected by comparing calculation cross sections produced by varying numerous models and their parameters within TALYS~\cite{bib:1TALYS}, with experiments from the EXFOR database~\cite{bib:15} within a Bayesian framework and used as the starting point for new calculations in an iterative manner. 

The Monte Carlo methods presented above, however, relied exclusively on the variation of parameters to pre-selected models and are therefore limited by the constraints of these models. The underlying assumption in this approach has been that the chosen model set or combination accurately represents the `true' distribution. Additionally, the models were globally selected, implying that the chosen models are applicable across the entire energy range of interest. The conventional belief here is that the uncertainty in nuclear data arises solely from our imperfect knowledge of the parameters associated with these models~\cite{bib:27}. However, this approach tends to ignore uncertainties stemming from the models themselves. Consequently, it often leads to a difficulty in achieving satisfactory fits to experimental data within specific energy regions for certain channels as the evaluation is constrained by the shortcomings or deficiencies of the chosen models. 

\begin{figure}[htb] %tb]
  \centering
  \includegraphics[trim = 15mm 20mm 5mm 20mm, clip, width=0.4\textwidth]{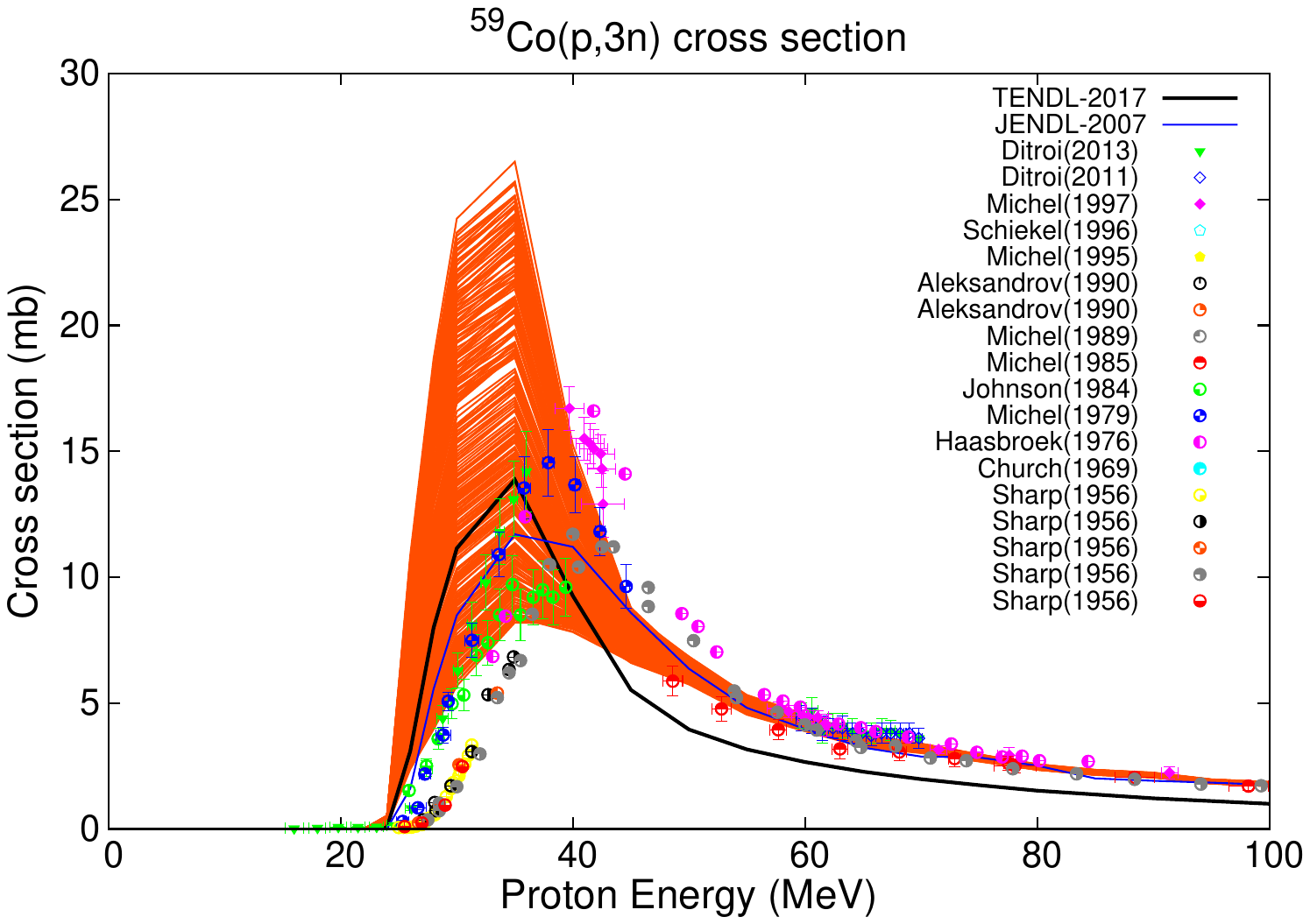}
   \caption{Random $^{59}Co$(p,3n) cross section curves computed using a single set of models but with perturbed model parameters, compared with differential experimental data sourced from EXFOR~\cite{bib:15}.}
   \label{vary_parameters}
 \end{figure}

A similar observation was made in Ref.~\cite{bib:2t}, where it was stated that \emph{`as long as a “near perfect model” is not available, a pure Monte Carlo solution based on model parameters alone cannot adequately combine theoretical results and microscopic experimental data'.} To attest to the validity of this observation, we present more than a 1000 random $^{59}Co$(p,3n) cross section curves produced by exclusively varying model parameters within a single model combination in the TALYS code~\cite{bib:1TALYS,bib:33} in Fig.~\ref{vary_parameters}. The cross section curves as presented in the figure were produced by executing the TALYS code with the following models~\cite{bib:1TALYS,bib:33}: 

\begin{enumerate}
    \item  \textbf{mass model 0}: Duflo-Zuker formula, 
     \item \textbf{level density 2}: Back-shifted Fermi gas model, and,
      \item \textbf{strength 1}: Kopecky-Uhl generalized Lorentzian
       \item Other default models
\end{enumerate}

The other default models include the following: for the pre-equilibrium model, \textbf{preeqmode 2}: Exciton model which employs numerical transition rates with an energy-dependent matrix element. Regarding the model utilized for width fluctuation corrections in compound nucleus calculations which is represented by the TALYS keyword, \textbf{widthmode}, the Moldauer model (\textbf{widthmode 1}) was used. It's important to mention that the default model for the level density used in the calculation is the TALYS keyword, \textbf{ldmodel 1}, which employs the Constant Temperature Model (CTM) for low energies and the Fermi gas model for high energies. For the mass model, the default is \textbf{massmodel 2}, which utilizes Goriely HFB (Hartree-Fock-Bogoliubov)-Skyrme tables. In terms of the gamma-ray strength functions, \textbf{strength 1}, the Kopecky-Uhl generalized Lorentzian is the default model for incident neutrons, while \textbf{strength 2}, the Brink-Axel Lorentzian, serves as the default for other incident particles such as protons~\cite{bib:33} in TALYS. From the figure, it can be seen that the random cross section curves overlap some but not all of the experimental data presented. As expected, there was difficulty in reproducing cross sections at the threshold energies. Furthermore, there is a noticeable narrowing of the cross-section spread between approximately 50 to 100 MeV. This makes it difficult to overlap all the experimental data presented in the figure for this energy range. We note however that no outliers were discarded in Fig.~\ref{vary_parameters} as the goal was to observe visually, the cross section spread due to the variation of only model parameters around a selected model set. The inability to overlap some of the experimental data even with parameter variation can be attributed to the underlying deficiencies in the models used. As mentioned in Ref.~\cite{bib:2t} and observed also in this work (see Fig.~\ref{vary_parameters}), by varying only model parameters it is sometimes impossible to reproduce the experimental data due to the deficiencies and rigidity of the selected models. 

% It is instructive to note that widening the parameter space did not result in a corresponding widening of the cross section spread between 50 to 100 MeV. Furthermore, the random cross section curves have a difficulty in reproducing experimental data from the threshold to about 40 MeV as can be observed in the figure.
%

\begin{figure}[htb] %tb]
  \centering
  \includegraphics[trim = 15mm 1mm 5mm 0mm, clip, width=0.4\textwidth]{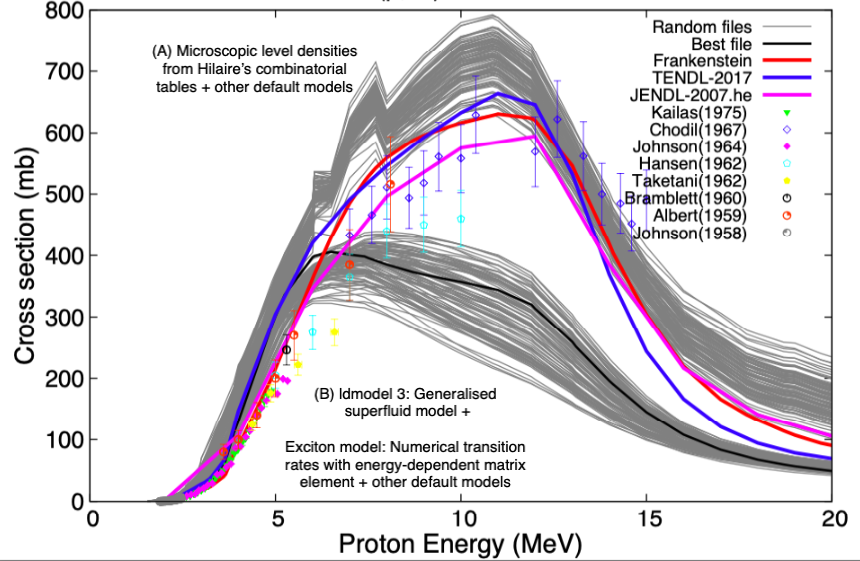}
   \caption{Random $^{59}Co$(p,n) cross section curves computed using two TALYS model sets compared with experimental data from EXFOR. For each model set, model parameters were perturbed to obtained the random cross section curves (in gray).}
   \label{vary_parameters_Co59pn}
 \end{figure}

Another example is given in Fig.~\ref{vary_parameters_Co59pn}, where models parameters are varied around two distinct model sets in the TALYS code. For model (A), the model used corresponds to $\textbf{6}$: microscopic level densities from Hilaire’s combinatorial tables, alongside other default TALYS models. Conversely, for model (B), the Generalised superfluid level density model ($\textbf{3}$) combined with the Exciton model (Numerical transition rates with energy-dependent matrix element) of the pre-equilibrium model, along with other default TALYS models, was used. It can be seen that model (A) generally follows the shape of the experimental data over the entire considered energy region while model (B) was observed to only reproduce experimental data from the threshold to about 8 MeV. It is important to state here that in the case of model (B), even with the variation of model parameters, it was still difficult to reproduce experimental data from about 8 to 20 MeV. This underscores the idea that different nuclear reaction models exhibit varying strengths in different energy regions. 

As shown in the figure, the `best' file is a globally optimized file achieved by comparing the reaction and residual production cross sections with differential experimental data for only model (B). Additionally, the `Frankenstein' file represents a locally optimized file obtained by comparing experimental data solely with cross-section curves for $^{59}$Co(p,n) in the case of model (A). The gray curves are random curves produced by perturbing model parameters around the chosen model sets.

\begin{figure*}[htb] %tb]
  \centering
  \includegraphics[trim = 15mm 12mm 5mm 20mm, clip, width=0.42\textwidth]{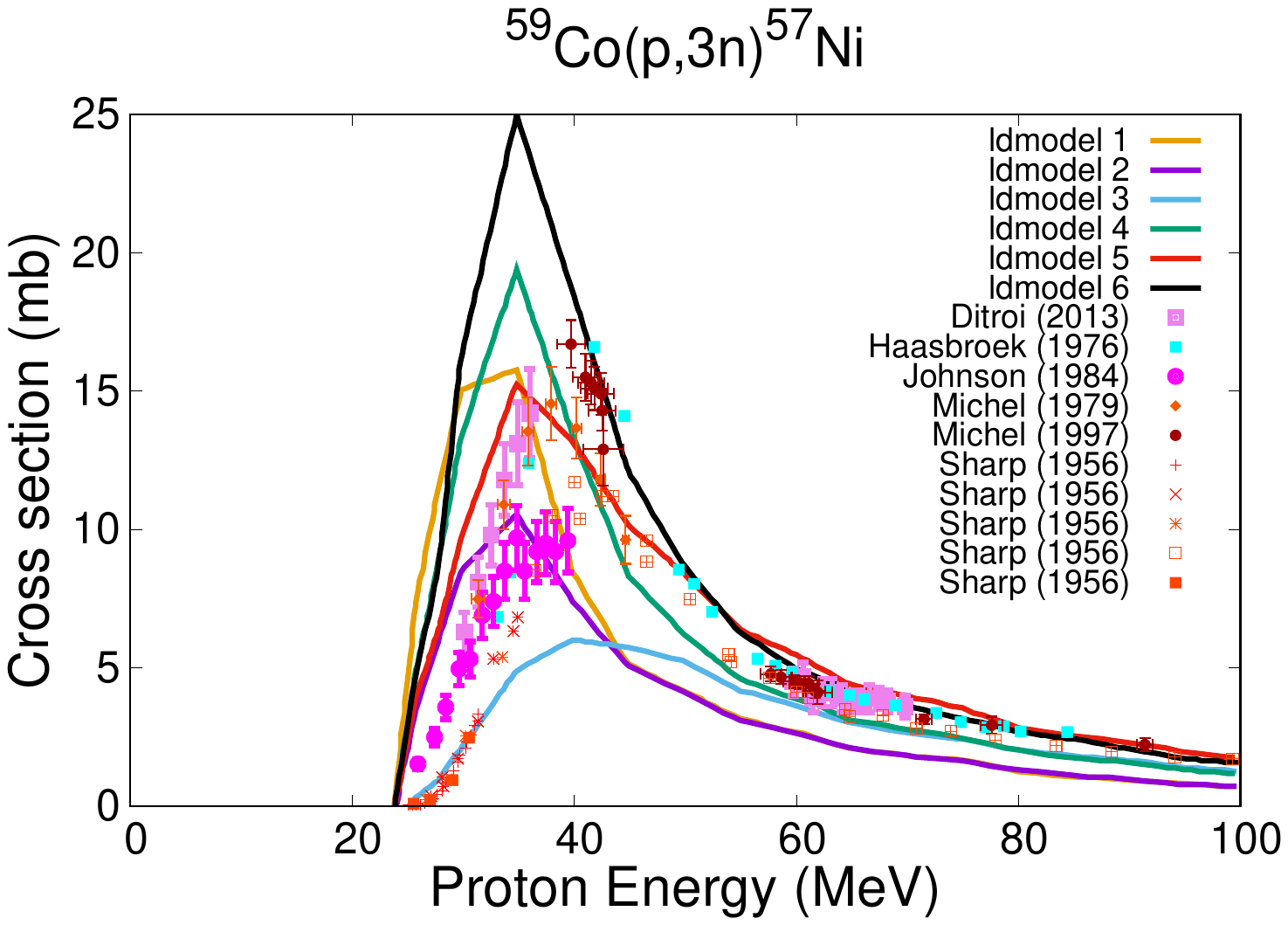}
  \includegraphics[trim = 15mm 12mm 5mm 20mm, clip, width=0.45\textwidth]{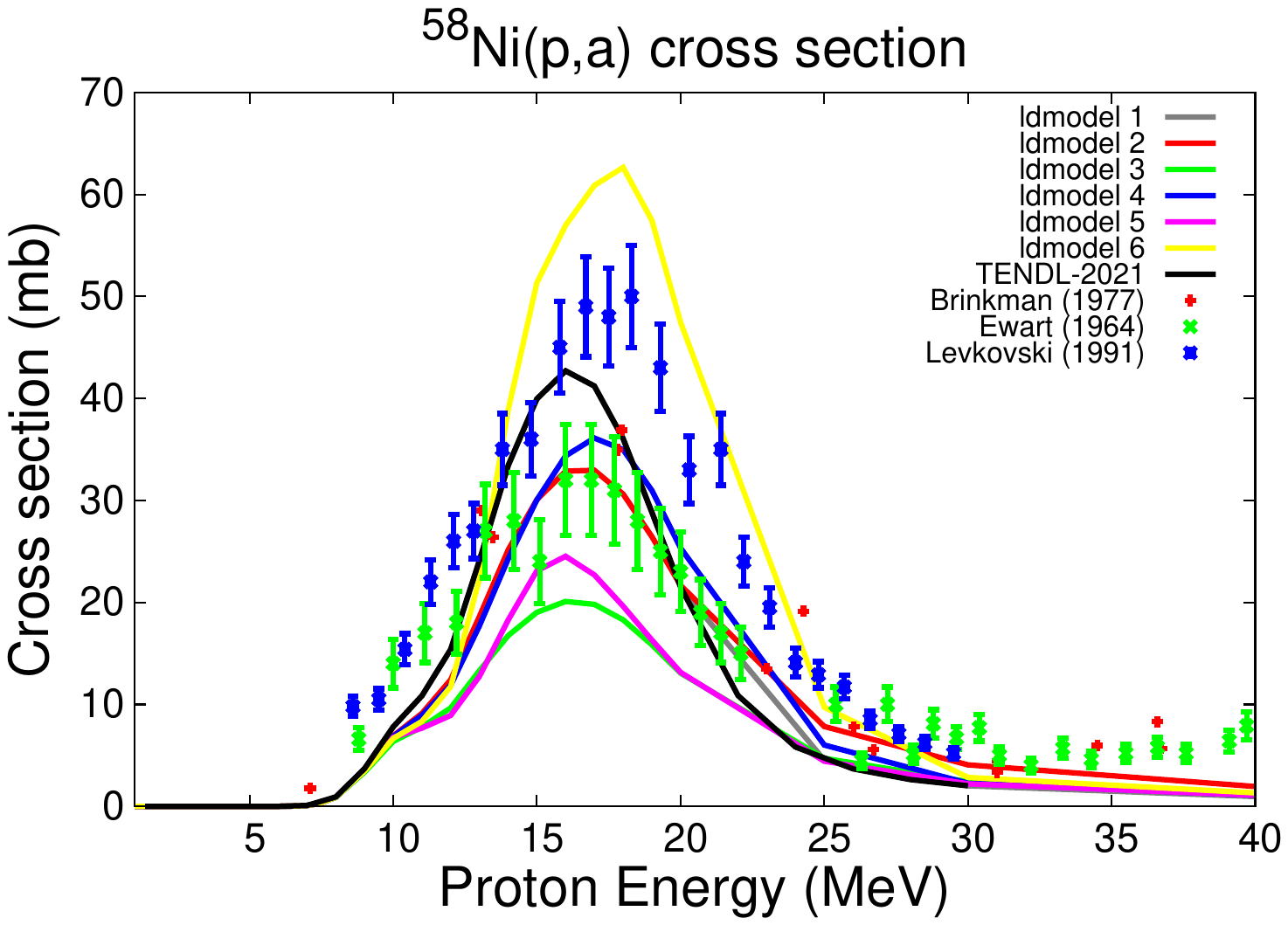}  
  \includegraphics[trim = 15mm 15mm 5mm 20mm, clip, width=0.45\textwidth]{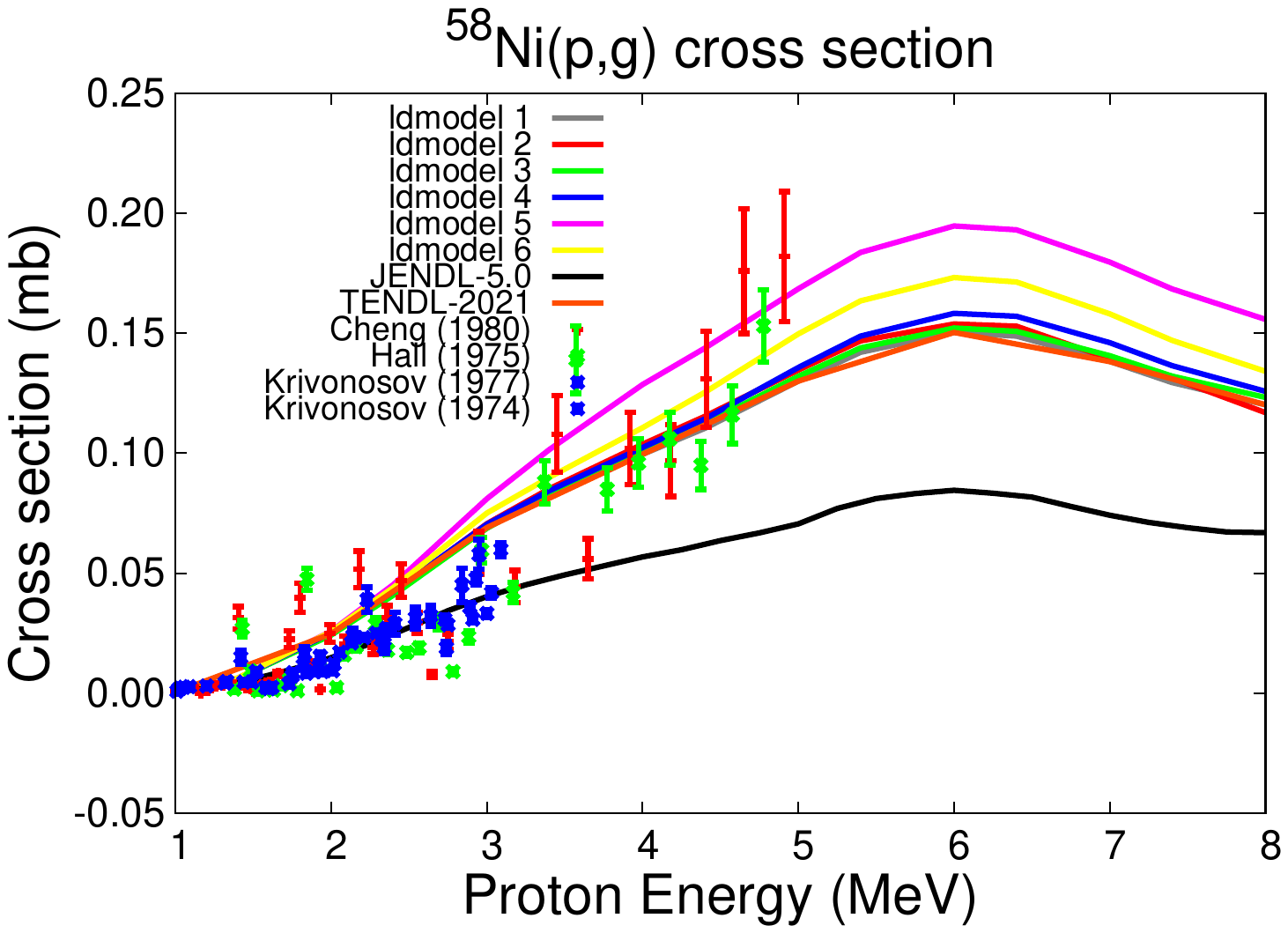}
  \includegraphics[trim = 15mm 15mm 5mm 20mm, clip, width=0.45\textwidth]{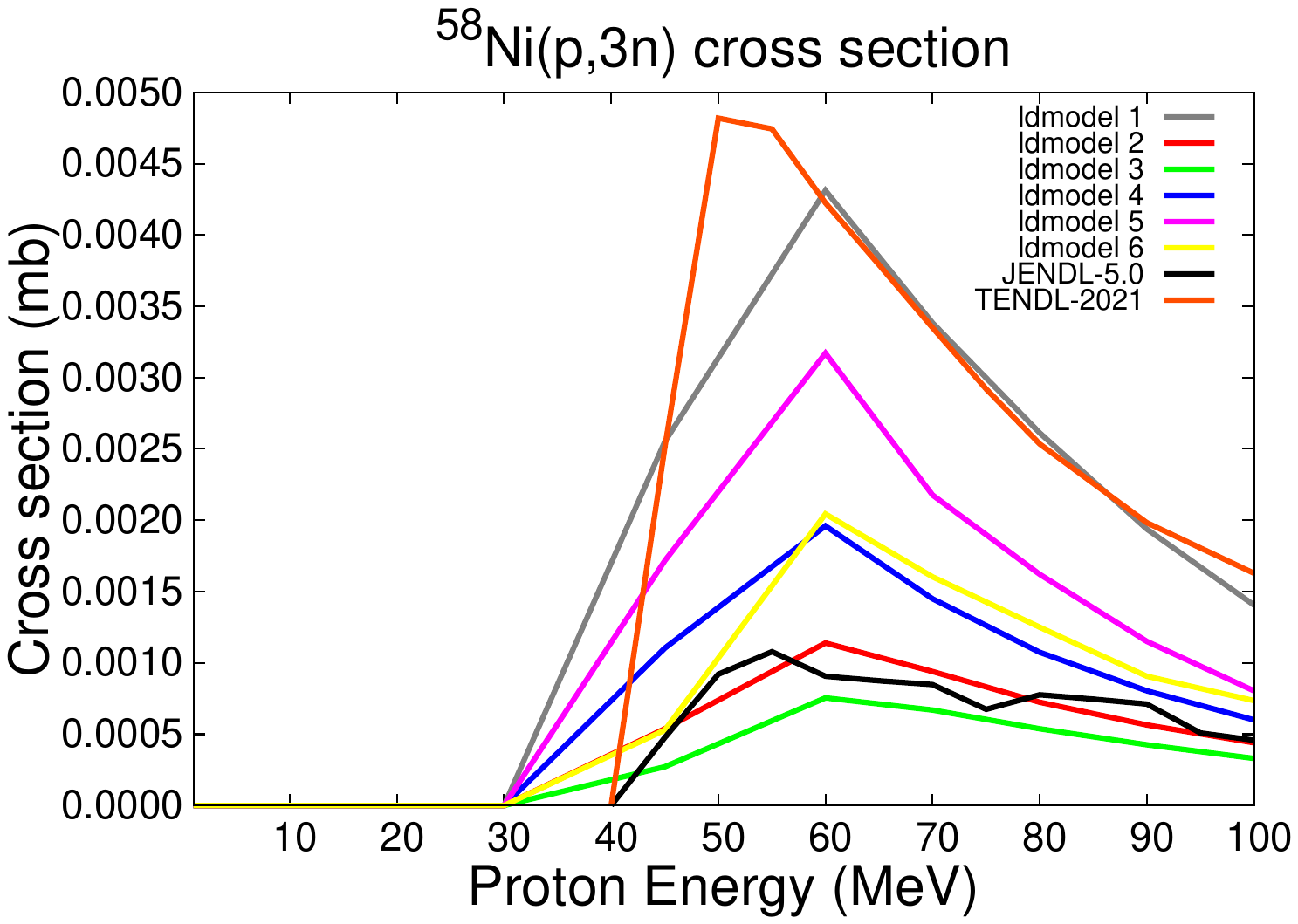}
   \caption{Cross sections curves computed with the six different level density models in TALYS, compared with differential experimental data from the EXFOR database~\cite{bib:15} (where available): (1) top left: $^{59}Co$(p,3n) cross section, (2) top right: $^{58}Ni$(p,$\alpha$) cross section, (3) bottom left: $^{58}Ni$(p,$\gamma$) cross section, and (4) bottom right: $^{58}Ni$(p,3n) cross section.}
   \label{ld_models}
 \end{figure*}

As demonstrated earlier in Refs.~\cite{bib:1ab,bib:13,bib:12}, the incorporation of both models and parameter uncertainties to obtain a widened prior space, results in a greater variability in the randomly generated cross-section curves. Hence, most of the available experimental data would fall within the spread of the combined model and parameter uncertainties. To illustrate this, in Fig.~\ref{ld_models} (top left), we revisit the $^{59}$Co(p,3n) cross-section curves, this time, varying each of the six different level density ($ld$) models within the TALYS code individually while holding all other models constant as their default TALYS values. The $ld$ models in TALYS are as follows~\cite{bib:1TALYS,bib:33}: 
\begin{itemize}
    \item $ld$ model 1 - Constant temperature + Fermi gas model; 
    \item $ld$ model 2 - Back-shifted Fermi gas model; 
    \item $ld$ model 3 - Generalised superfluid model; 
    \item $ld$ model 4 - Microscopic level densities (Skyrme force) from Goriely’s tables; 
    \item $ld$ model 5 - Microscopic level densities (Skyrme force) from Hilaire’s combinatorial tables, and
    \item $ld$ model 6 - Microscopic level densities (temperature dependent Hartree-Fock-Bogolyubov (HFB), Gogny force) from Hilaire’s combinatorial tables. 
\end{itemize}

\begin{figure}[htb] %tb]
  \centering
  \includegraphics[trim = 15mm 15mm 5mm 20mm, clip, width=0.45\textwidth]{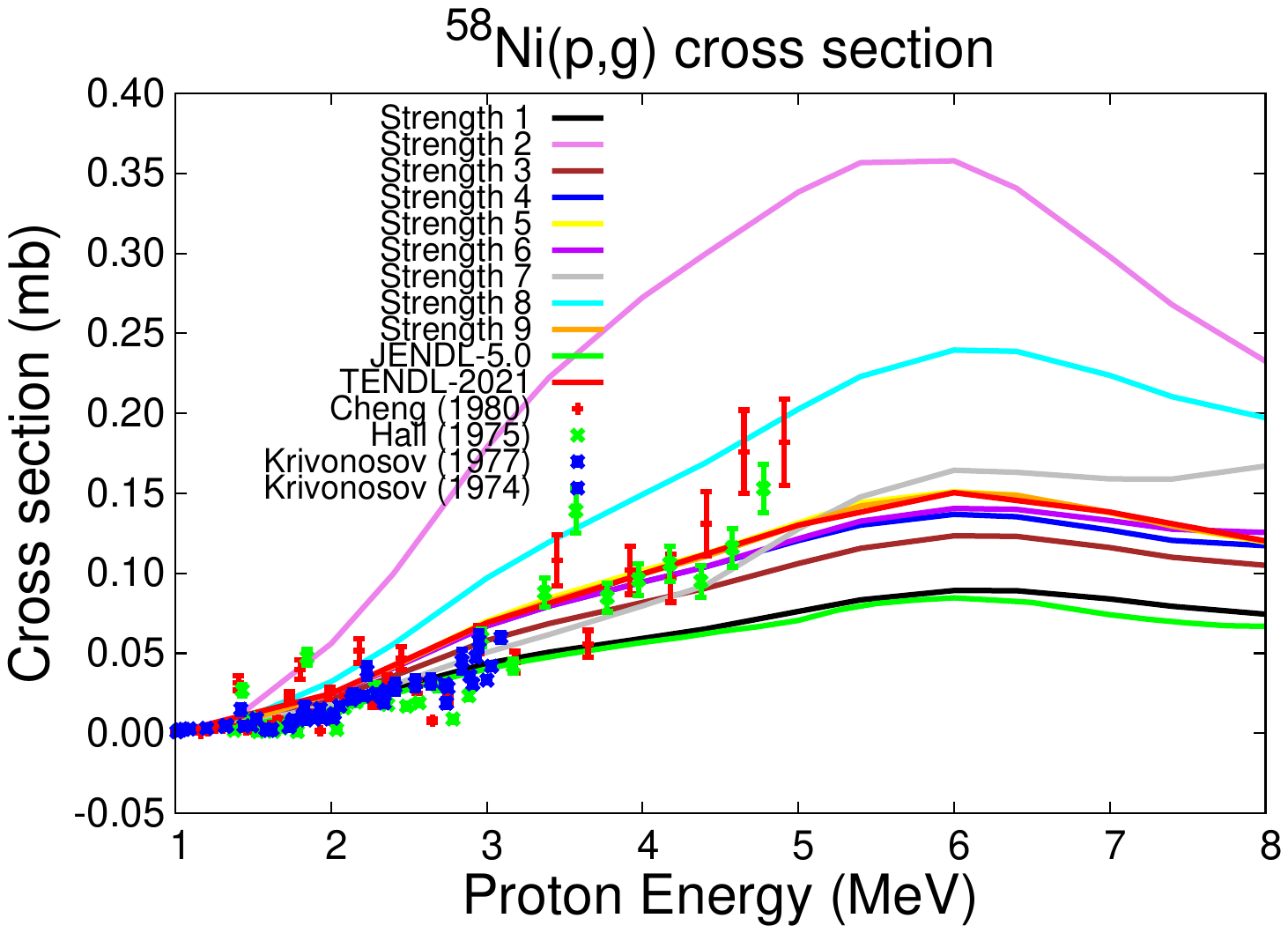}
   \caption{$^{58}Ni$(p,$\gamma$) cross section curves computed using the eight gamma-ray strength function models in TALYS, compared with differential experimental data and the TENDL-2021 and JENDL-5.0 evaluations.}
   \label{strength_models}
 \end{figure}

\begin{figure}[htb] %tb]
  \centering
  \includegraphics[trim = 15mm 15mm 5mm 20mm, clip, width=0.45\textwidth]{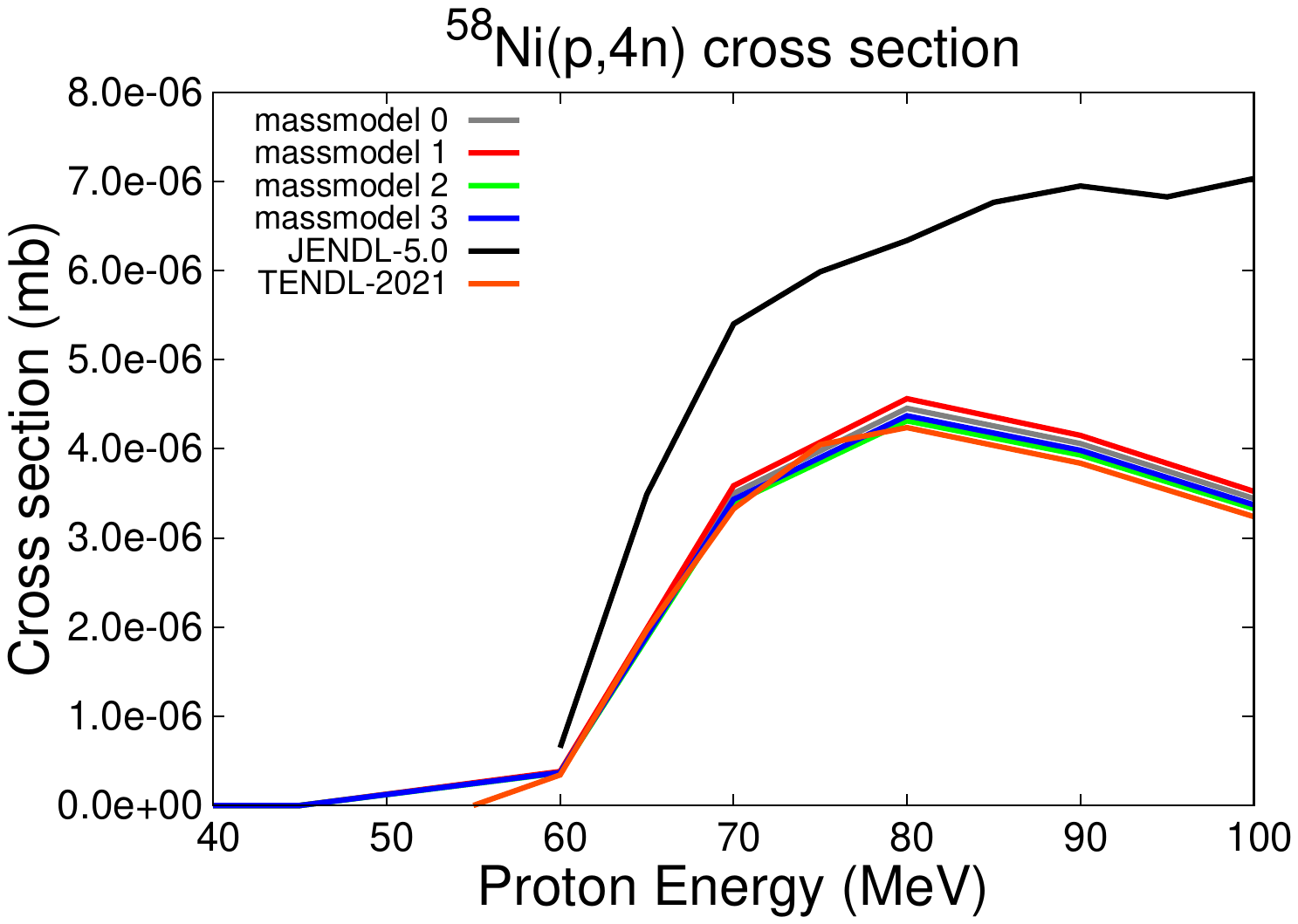}
   \caption{$^{58}Ni$(p,3n) cross section curves computed using the four mass models in TALYS, compared with differential experimental data and the evaluations from the TENDL-2021 and JENDL-5.0 libraries.}
   \label{mass_models}
 \end{figure}

It can be noticed from Fig.~\ref{ld_models} that each $ld$ model exhibits specific strengths with respect to reproducing the presented experimental data. For example, the cross section curves computed with $ld$ model 3 and 4, compared favourably with experimental data from about 60 to 100 MeV while $ld$ model 2 and 5 exhibit more favourably agreement with experimental data in the 20 to 40 MeV range. In general, it can be seen from the figure that most of the experimental data lie within the model spread or uncertainties as expected. Similarly, cross section curves produced with the six different level density models for the $^{58}Ni$(p,$\alpha$) (top right of figure), $^{58}Ni$(p,$\gamma$) cross sections (bottom left) and $^{58}Ni$(p,3n) (bottom right) cross section, are presented. By using different models, it can be observed that most of the experimental data fell within the model spread. In the case of the $^{58}Ni$(p,3n) cross section, no experimental data were not available in the EXFOR database and hence, the cross section curves produced with the $ld$ models are compared with the TENDL-2021 and JENDL-5.0 libraries. It is important to highlight that the threshold energy for the TENDL and JENDL evaluations are different from that of the $ld$ models as presented in the figure. In Fig.~\ref{strength_models}, the variations of the $^{58}Ni$(p,$\gamma$) cross section computed using the eight gamma-ray strength function models in TALYS are compared with experimental data as well as the TENDL-2021 and JENDL-5.0 evaluations. It is essential to point out that the gamma-ray strength function models are used in the description of the gamma emission channel. From the figure, \emph{strength 6} is observed to over predict the (p,$\gamma$) cross section from about 2 to 8 MeV. \emph{8} is observed to reproduce some experimental data from Cheng (1980) while \emph{strength 3}, \emph{6}, and \emph{7}, reproduced better, experimental data from Hall (1975). It can be observed that the curves from the JENDL-5.0 evaluation and that of \emph{strength 1} were similar. It is worth mentioning that there are cases were cross sections curves have low sensitivity to model variations. An example is presented in Fig.~\ref{mass_models} where the four different mass models in TALYS were varied one-at-a-time while keeping the other models as the default TALYS models to produce the (p,4n) cross section. The spread in the (p,4n) cross section curves due to the variation of the mass models is observed to be small.

 %Georg: The following sentence is not understandable. What does it mean? Does the approach introduce the prior correlations or the user? If *prior* correlation, then it is the user. Correlations are defined for two random variables that are real numbers. The selection of a model may be represented by an integer, but neither order nor differences are well applicable concepts here, so neither pearson nor spearman correlation applicable.
 %It is instructive to note that this approach could also introduce prior correlations between the different models, which according to Refs.~\cite{bib:12,bib:13} could have significant impact on the adjustment of nuclear data. [THIS HAS BEEN REMOVED - ERWIN]
 
 % Also, it was demonstrated in Refs.~\cite{bib:1aa,bib:12,bib:13}, that by varying many models together with their parameters, all or most of the experimental data could be overlapped.
 
%  \begin{figure}[htb] %tb]
%  \centering
%  \includegraphics[trim = 15mm 15mm 5mm 20mm, clip, width=0.45\textwidth]{MT107_ldmodel_p-Ni058.pdf}
   %  \includegraphics[trim = 15mm 10mm 5mm 10mm, clip, width=0.48\textwidth]{plotrandom_Co59_p3n_mass_all.pdf}
%   \caption{$^{58}Ni$(p,$\alpha$) cross section curves computed using the six different level density models implemented in TALYS, compared also with differential experimental and the TENDL-2021 evaluation.}
 %  \label{ld_models}
 % \end{figure} 

%   \begin{figure}[htb] %tb]
%  \centering
 % \includegraphics[trim = 15mm 15mm 5mm 20mm, clip, width=0.45\textwidth]{MT102_ldmodel2.pdf}
   %  \includegraphics[trim = 15mm 10mm 5mm 10mm, clip, width=0.48\textwidth]{plotrandom_Co59_p3n_mass_all.pdf}
 %  \caption{$^{58}Ni$(p,$\gamma$) cross section curves computed using the six different level density models implemented in TALYS, compared also with differential experimental and the TENDL-2021 evaluation.}
 %  \label{ld_models}
 % \end{figure} 

%  \begin{figure}[htb] %tb]
 % \centering
 % \includegraphics[trim = 15mm 15mm 5mm 20mm, clip, width=0.45\textwidth]{MT037_massmodel.pdf}
   %  \includegraphics[trim = 15mm 10mm 5mm 10mm, clip, width=0.48\textwidth]{plotrandom_Co59_p3n_mass_all.pdf}
 %  \caption{$^{58}Ni$(p,$\gamma$) cross section curves computed using the four mass models in TALYS, compared also with differential experimental and the TENDL-2021 evaluation.}
 %  \label{ld_models}
% \end{figure} 

 A potential remedy for addressing model deficiencies has been to use Gaussian processes for treating the defects as proposed and presented in various references (see~\cite{bib:9,bib:11,bib:17}) or related constructions, e.g., Refs.~\cite{bib:18,bib:19}. This approach however, treats the model defect using default TALYS models. In other studies presented in Refs.~\cite{bib:2} and \cite{bib:1aa,bib:1EACd111}, the models were selected from a pool of models globally for the entire considered energy range. However, due to limitations and inflexibility inherent in these selected models, achieving improvement in evaluations concerning certain channels and energy regions can still pose challenges. In this work however, in line with the search for a full Monte Carlo solution for combining theoretical and experimental data in nuclear data evaluations, we propose a departure from using a single fixed model set for the entire energy range, as done in the Bayesian Monte Carlo (BMC) approach~\cite{bib:2,bib:1EACd111} and the iterative Bayesian Monte Carlo (iBMC) outlined in Ref.~\cite{bib:1aa}. Rather, we propose to select models locally, at each incident energy or angle. This approach gives more flexibility to the adjustment process by assigning weights to each model based on their proximity to experimental data. Consequently, a weighted average was computed over all the considered models at each incident energy within a Bayesian Model Averaging (BMA) framework~\cite{bib:12wasserman,bib:13a}, using the likelihood function values as weights. It is important to note here that Bayesian Model Averaging has been applied to various fields (see for example, Refs.~\cite{bib:045a,bib:045b,bib:044MSMA}) as well as in nuclear physics~\cite{bib:044BMA}, among others. In Ref.~\cite{bib:1DimtriModel}, the covariance matrices generated from model variations, in contrast to the conventional parameter variations, were subsequently applied to quantities in astrophysics. In other studies such as in Ref.~\cite{bib:MLPhys}, machine learning techniques were applied in various aspects of nuclear physics such as, nuclear structure, nuclear reactions, and properties of nuclear matter at low and intermediate energies. In Ref.~\cite{bib:NeuralNW}, a prediction of nuclear charge density distribution with feedback neural networks was presented. In Ref.~\cite{bib:MBFox}, a standardized procedure for adjusting parameters in reaction modeling codes was proposed based on residual product excitation functions and utilizing the TALYS code, was proposed and presented.

 It is crucial to note that since the updates of the cross sections and angular distributions as proposed in this work were carried out locally on a per-energy-point basis, this approach typically results in discontinuities or "kinks" in the curves of the cross sections or angular distributions produced. By carrying the adjustments of the cross sections and elastic angular distributions at specific incident energies, we are better able to represent the behavior of the reactions at the considered energies. To address the kinks produced, a smoothing function using spline interpolation was applied. As proof of concept, the proposed BMA method was applied for nuclear data evaluation of p+$^{58}$Ni in the fast energy region between 1 and 100 MeV.

\section{Methods}
\label{Bayes_calibration}
As previously mentioned, Bayesian Model Averaging (BMA) is a statistical approach that accounts for model uncertainty by averaging over multiple models, each weighted by its posterior probability. In the context of nuclear reactions and nuclear data evaluation, the proposed models considered are nuclear physics models and their parameters. 
%
% within Bayesian Model Averaging (BMA), the conventional approach of selecting a definitive model set and treating it as the 'true' model set is replaced by averaging over all or a subset of the models, using the likelihood function values serving as weights. 

\begin{figure}[htb] %tb]
  \centering
  \includegraphics[trim = 50mm 68mm 5mm 40mm, clip, width=0.9\textwidth]{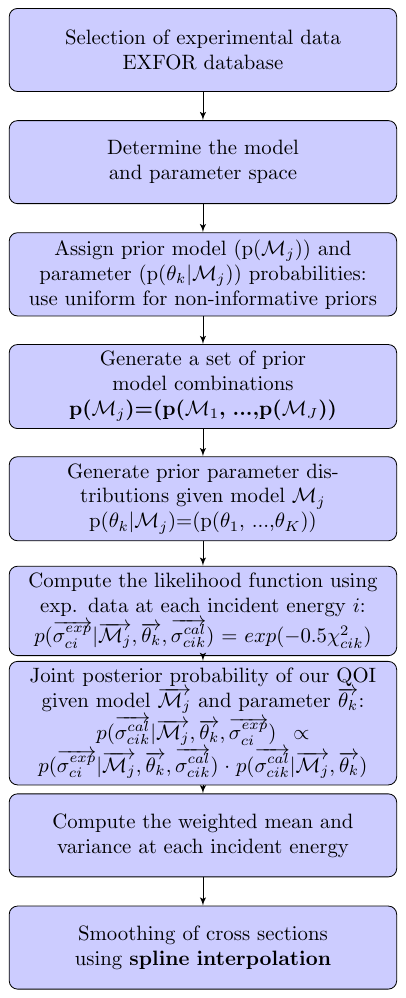}
   \caption{Flowchart outlining the proposed BMA method for nuclear data evaluation. In the chart, $\sigma^{exp}_{ci}$ and $\sigma^{cal}_{cik}$ respectively, are the experimental and calculated cross sections (which is our Quantity Of Interest (QOI)) for channel $c$ at incident energy $i$, which corresponds to the $k^{th}$ model parameter vector in the $k^{th}$ TALYS input file. $\theta_k$ denotes the $k^{th}$ model parameter vector and $\mathcal{M}_j$ is the model set or combination, $j$. The reduced chi square, $\chi^2_{cik}$, at each considered incident energy, $i$, channel $c$, and model parameter vector, $k$,}
   \label{flowchart_bma}
 \end{figure}

The flowchart in Fig.~\ref{flowchart_bma} outlines the BMA methodology proposed for nuclear data evaluation in this study. From the figure, we begin with a careful selection of experimental data from the EXFOR database~\cite{bib:15}. This is particularly important in BMA as outlier experiments have the potential to distort the shape of the updated cross section curves. An alternative approach to selecting experiments would involve assigning weights to each experimental data set based on a quality criteria. In this work however, we adopted a binary accept/reject approach for accepting and rejecting experimental data as outlined in Refs.~\cite{bib:1aa,bib:12}. For example, experiments lacking reported uncertainties were penalized with a binary value of zero, except in cases where these experiments were the sole experiments available for the considered channel. In such cases, a 10\% relative uncertainty was assigned to each data point of the experimental data set. Additionally, if an experimental data set is found to be inconsistent with other experimental data sets for a particular energy range, the inconsistent data set was excluded.
%%%%%

% The major objective of this work is to identify the optimal model combinations. 

%Georg: removed sentence as it is not necessary at this point: This was carried out taken into consideration, the computational resources available.
 
Next from the figure, we define the prior model and parameter spaces. As proof of concept, similar to Ref.~\cite{bib:1aa}, a total of 52 different nuclear reaction models in the TALYS code were considered in this work. A list of the selected nuclear reaction models are itemized in Table~\ref{models_varied}. It is important to clarify that, in the context used in this work, the term "models" encompasses sub-models and, at times, components of models or sub-models. From the table it can be seen that there are 4 Jeukenne-Lejeune-Mahaux (JLM) optical models, 6 level density models, 4 pre-equilibrium (PE) models, 4 mass models, and 8 strength functions, among others, available in the TALYS code. It is instructive to note here that each nuclear reaction calculation involves a combination of several of these models interconnected within a nuclear reaction code such as TALYS. Model calculations were performed using TALYS version 1.9~\cite{bib:33}.  

 \begin{table}[tb]
 \centering
  \caption{List of considered nuclear reaction models showing the number of the different models per model type. Note; PE denotes the pre-equilibrium model and JLM refers to the Jeukenne-Lejeune-Mahaux optical model~\cite{bib:16a}. These models are available in the TALYS code.}
  \label{models_varied}
  \begin{tabular}{lcl}  % Use 'c' to center the column, 'l' to have it left adjusted and 'r' for right adjusted
  \toprule
   TALYS keywords  &  \pbox{20cm}{Number of \\ models} & Model Name  \\
   \midrule
preeqmode    &  4 & Pre-equilibrium (PE) \\
ldmodel      &  6 & Level density models  \\
ctmglobal    & 1 & Constant Temperature \\
massmodel    & 4 & Mass model  \\
widthmode    & 4 & Width fluctuation \\
spincutmodel & 2 &  Spin cut-off parameter  \\
gshell       & 1 & Shell effects \\
statepot     & 1 & Excited state in Optical Model \\
spherical    & 1 & Spherical Optical Model \\
radialmodel  & 2 & Radial matter densities \\
shellmodel   & 2 & Liquid drop expression \\
kvibmodel    & 2 & Vibrational enhancement \\
preeqspin    & 3 & Spin distribution (PE) \\
preeqsurface & 1 & Surface corrections (PE) \\
preeqcomplex & 1 & Kalbach model (pickup) \\
twocomponent & 1 & Component exciton model \\
pairmodel    & 2 & Pairing correction (PE) \\
expmass      & 1 & Experimental masses \\
strength     & 8  & Gamma-strength function \\
strengthM1   & 2 & M1 gamma-ray strength function \\ 
jlmmode      & 4 & JLM optical model \\
\bottomrule
\end{tabular}
\end{table}

In the case of the parameters, Table~\ref{models_varied} provides the parameter widths (or uncertainties) which define the parameter space along with a comprehensive list of the parameters to the nuclear reaction models considered. The parameter widths given in the table were obtained from the TENDL library~\cite{bib:13}. It is important to emphasize that these parameter widths or uncertainties were obtained by comparing random cross-section curves produced through parameter variation with scattered experimental data. In Table~\ref{Table1modelp}, the model parameters are grouped under phenomenological and semi-microscopic optical models, level density and pre-equilibrium models, and gamma ray strength functions. The parameter widths or uncertainties as presented in the table are relative uncertainties (in \%) except in the case of the level density parameter $a$, and the $g_\pi$ and $g_\nu$ parameters, where the uncertainties are given in terms of the mass number $A$. Where $g_\pi$ and $g_\nu$ are the single-particle state densities used in the pre-equilibrium model. Similar tables have been provided in Refs.~\cite{bib:33} and \cite{bib:1aa}. A more complete list of all the model parameters can however be found in Ref.~\cite{bib:33,bib:1TALYS}.

Next, we assign prior probabilities to our models as well as their parameters. We assumed here that all models as well as the parameters have equal prior probabilities and hence the uniform distribution was assigned to both the models and the parameters. Employing uniform distributions for our priors ensures that the posterior distribution is predominantly shaped by the influence of experimental data used.

%Georg: the next sentence is not understandable. Maybe: ...and then its parameters varied simultaneously within their bounds... ? What does it mean: a model is varied within its bounds? 

\begin{figure*}[htb] %tb]
  \centering
  \includegraphics[trim = 15mm 12mm 5mm 12mm, clip, width=0.4\textwidth]{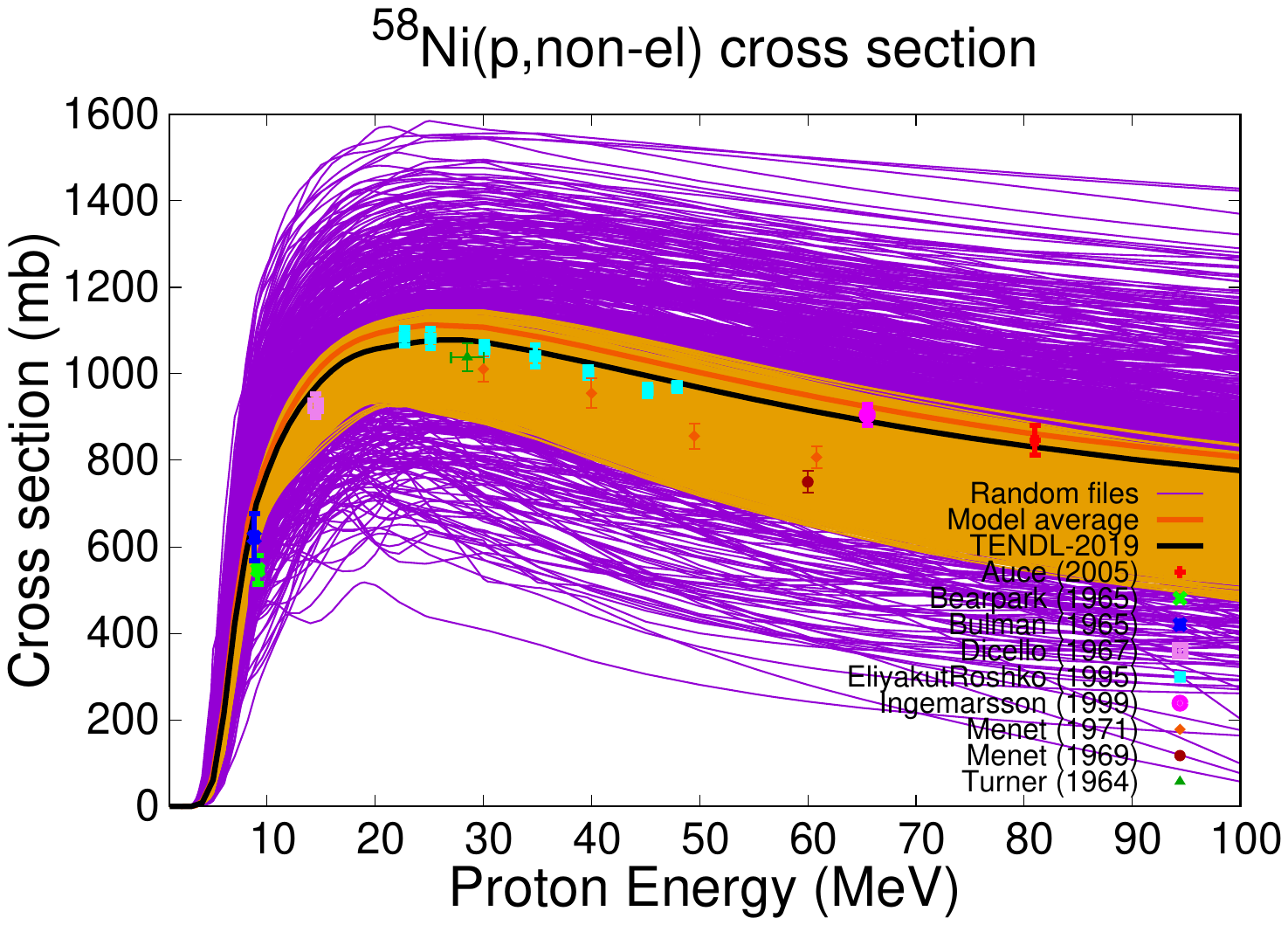}
  \includegraphics[trim = 15mm 12mm 5mm 12mm, clip, width=0.4\textwidth]{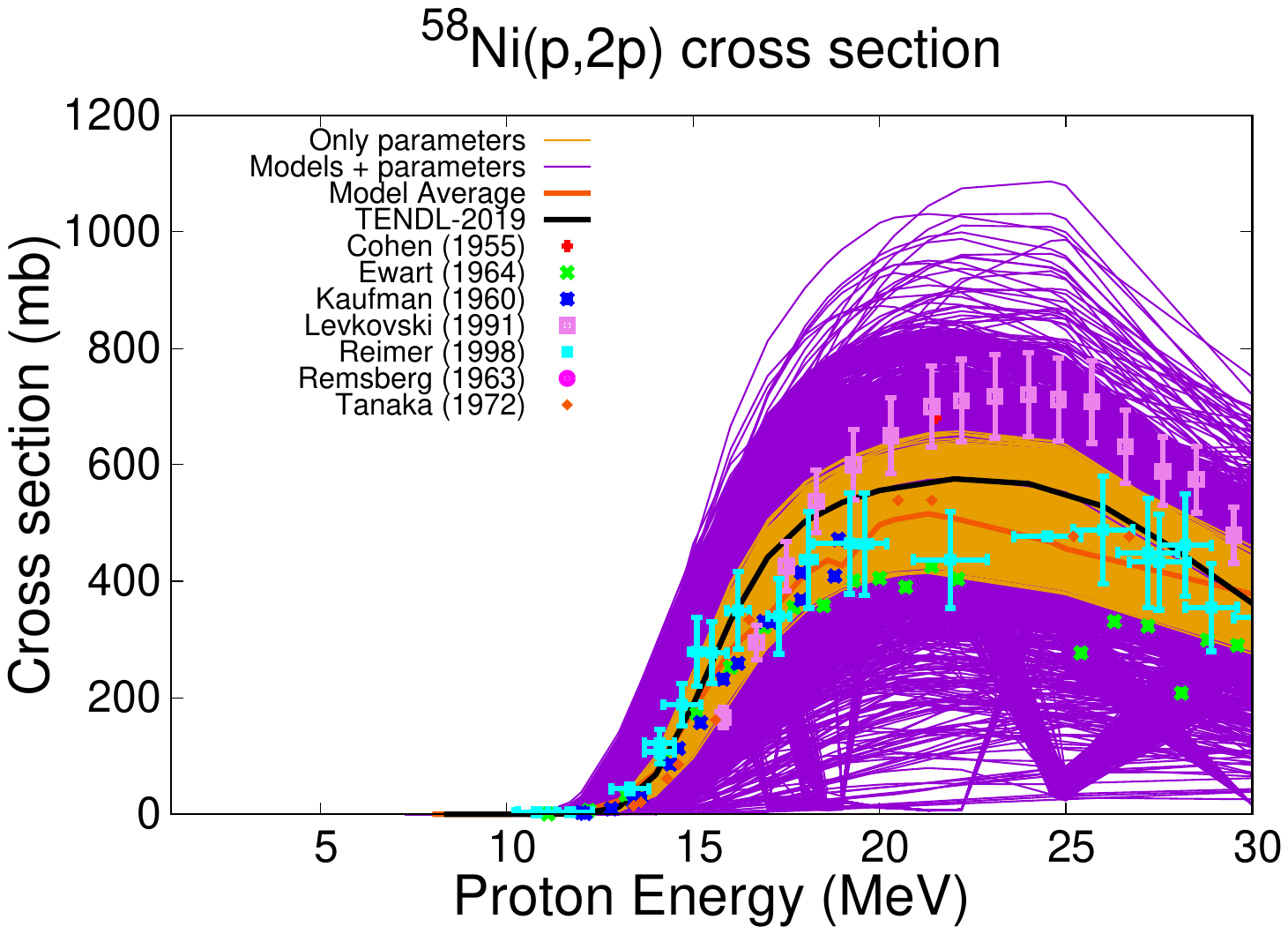}
  \includegraphics[trim = 15mm 12mm 5mm 12mm, clip, width=0.4\textwidth]{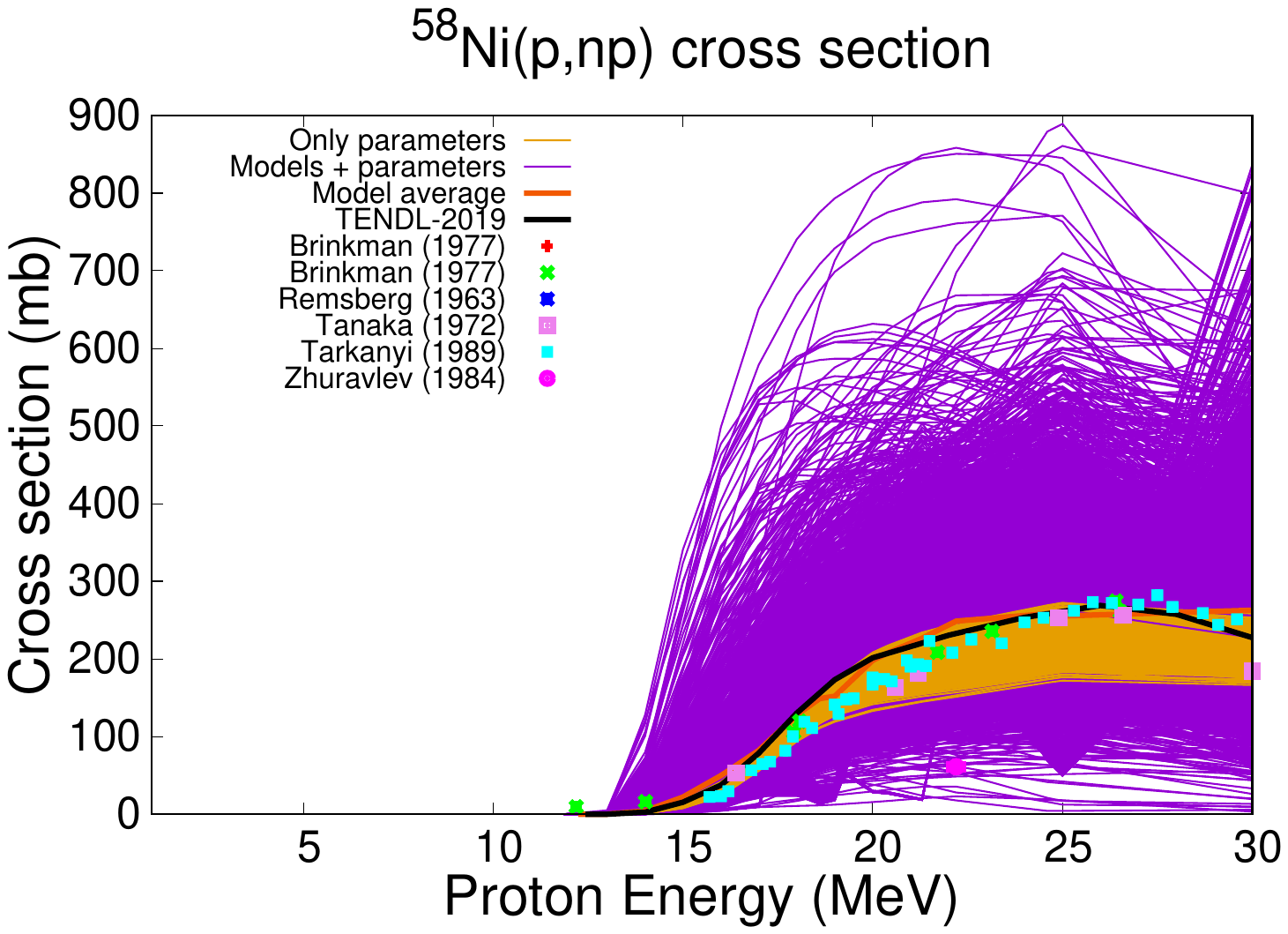}
    \includegraphics[trim = 15mm 12mm 5mm 12mm, clip, width=0.4\textwidth]{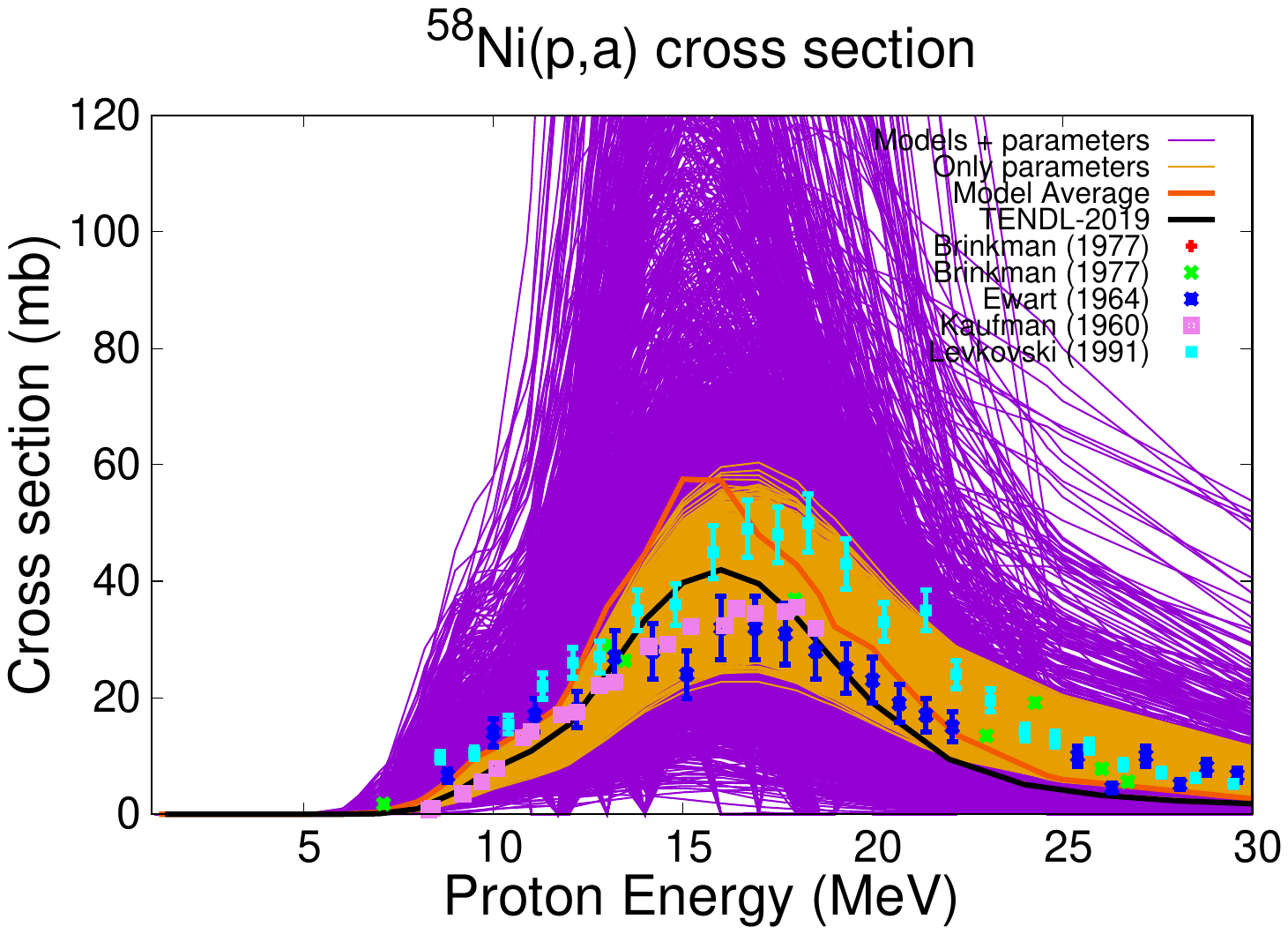}
   \caption{$^{58}Ni$(p,non-el), $^{58}Ni$(p,2p), $^{58}Ni$(p,np) and $^{58}Ni$(p,$\alpha$) cross sections showing the combined spread due to both models and parameters (in purple) as well as the spread due to only parameter variation (in orange). Note: all the 3030 random cross section curves produced by only parameter variation are plotted here. In the case of both model and parameter variation however, only 3000 files out of the 10000 files produced were plotted. The model average represents the average over all the cross section curves produced through the simultaneous variations of both models and their parameter.}
   \label{MT003_modelparam}
 \end{figure*}

Additionally, in order to estimate the uncertainty due to only parameter variation, a set of random ENDF nuclear data files were generated by varying only model parameters around default TALYS models. For this, a total of 3030 random nuclear data files were produced for p+$^{58}$Ni. In Fig.~\ref{MT003_modelparam}, random curves for the following cross sections: $^{58}Ni$(p,non-el), $^{58}Ni$(p,2p), $^{58}Ni$(p,np) and $^{58}Ni$(p,$\alpha$), are presented. These curves depict the variability due to the combined variation of both models and their parameters (shown in purple) as well as the variability arising solely from the variation of model parameters (shown in orange).

The figure reveals in the case of the $^{58}Ni$(p,2p) cross section that the exclusive variation of model parameters failed to capture several experimental data points from Levkovski (1991) and Ewart (1964). However, it can be seen that all the experimental data presented lie within the large prior spread obtained through the combined variation of both models and their parameters as expected. 

\begin{figure}[htb] %tb]
  \centering
    \includegraphics[trim = 15mm 20mm 15mm 20mm, clip, width=0.45\textwidth]{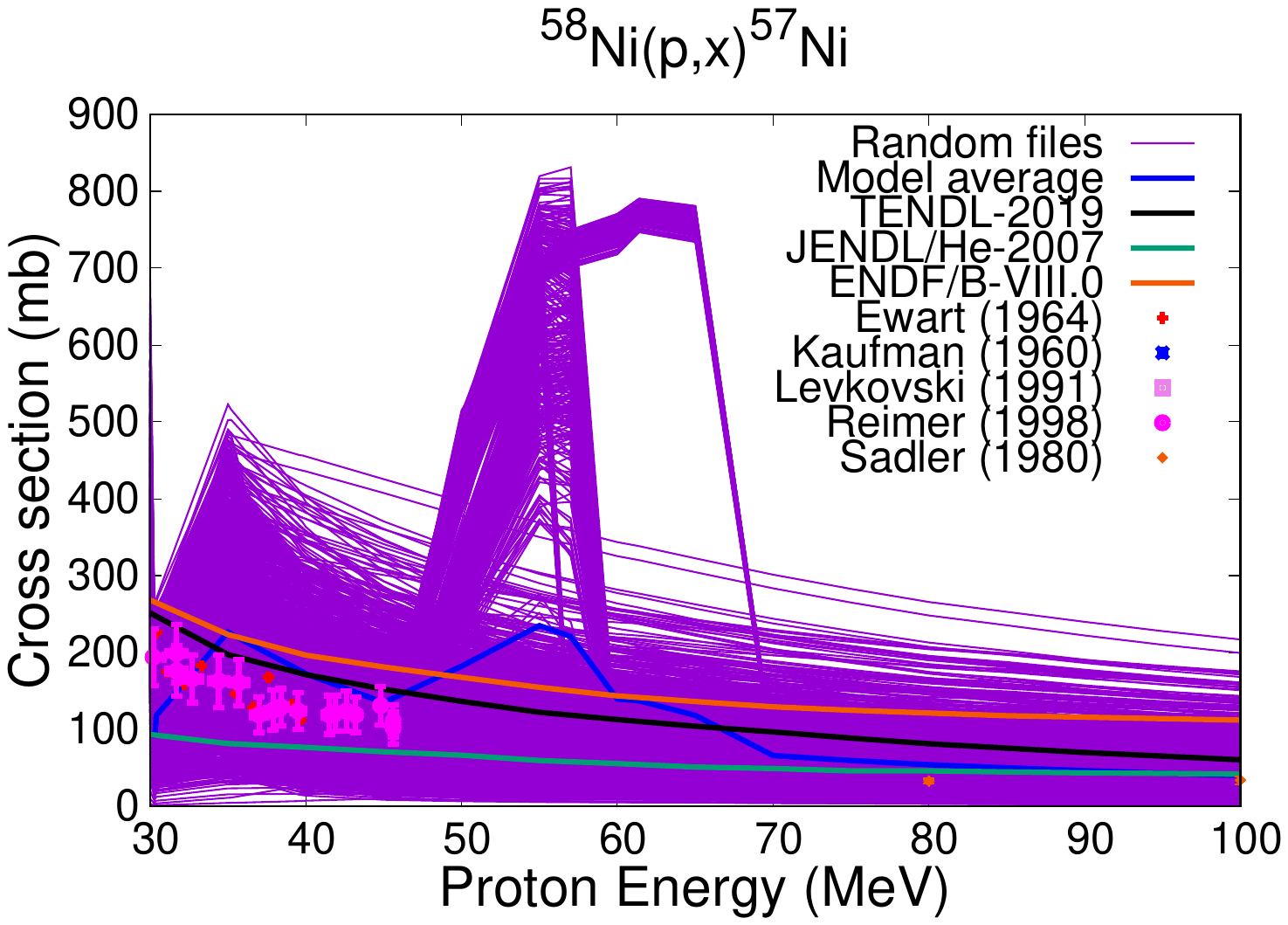}
   \caption{Prior $^{58}Ni(p,x)^{57}Ni$ cross section showing distortion in the curves produced by `unphysical' models. Note that both models and their parameters were varied altogether simultaneously.}
   \label{rp028057_avg}
\end{figure}

It is worth bearing in mind that a situation may arise where a considered data point or data set falls outside the spread of both the model and parameter uncertainties. This could be due to the fact that the model space (and/or parameter) was not fully explored. A possible solution to this problem would be to increase either the parameter widths or uncertainties (as presented in Table~\ref{Table1modelp}) in order to expand the parameter space. Alternatively, the model space can be enlarge further by introducing additional model (if available). However, if the considered experimental data point is deemed an outlier, it is generally advisable to exclude such data points. This precaution is particularly crucial as outlier experimental data can distort the overall shape of the final cross-section curves produced. Likewise, scenarios may arise where no experimental data are available for a given channel. This problem is discussed in more detail in section~\ref{BMAwoExpts}. In Fig.~\ref{rp028057_avg}, prior curves for the residual production cross section, $^{58}Ni(p,x)^{57}Ni$, is presented. Note that a particular combination of models yielded `unphysical' cross section shape which intend, distorted the model average between 45 and 70 MeV. How to treat `bad' models is presented in more detail in section~\ref{bad_models}. 

%Georg: To have consistency in terms of sum rules, the same model weights should be used for all reaction channels. As long as there are experimental in some of the channels, the weights given by the likelihood values are determined and should be used as they are also for the channels without data. I recommend to remove or rephrase the following sentence.

%%In such scenarios, a simple average should be taken over a carefully selected subset of the models. Note that model combinations that yield `unphysical' cross section shapes should be excluded from consideration as considering these cross sections could introduce distortion in the averaged cross section. These models, similar to the term used in experimental data selection, are referred to as outliers or `bad' models. For example, in Fig.~\ref{rp028057_avg}, prior $^{58}Ni(p,x)^{57}Ni$ cross section curves showing distortion in the curves produced by `unphysical' models, are presented. It can be observed in Fig.~\ref{rp028057_avg} that the average over the models compared fairly well with experimental data from Reimer (1998) and Ewart (1964) and in fact, outperforms the evaluation from the ENDF/B-VIII.0 evaluation. The distortion in the model average between about 48 to 70 MeV stems from the inclusion of all cross sections, including those produced by 'unphysical' models.

\begin{figure}[htb] %tb]
  \centering
    \includegraphics[trim = 15mm 20mm 5mm 20mm, clip, width=0.45\textwidth]{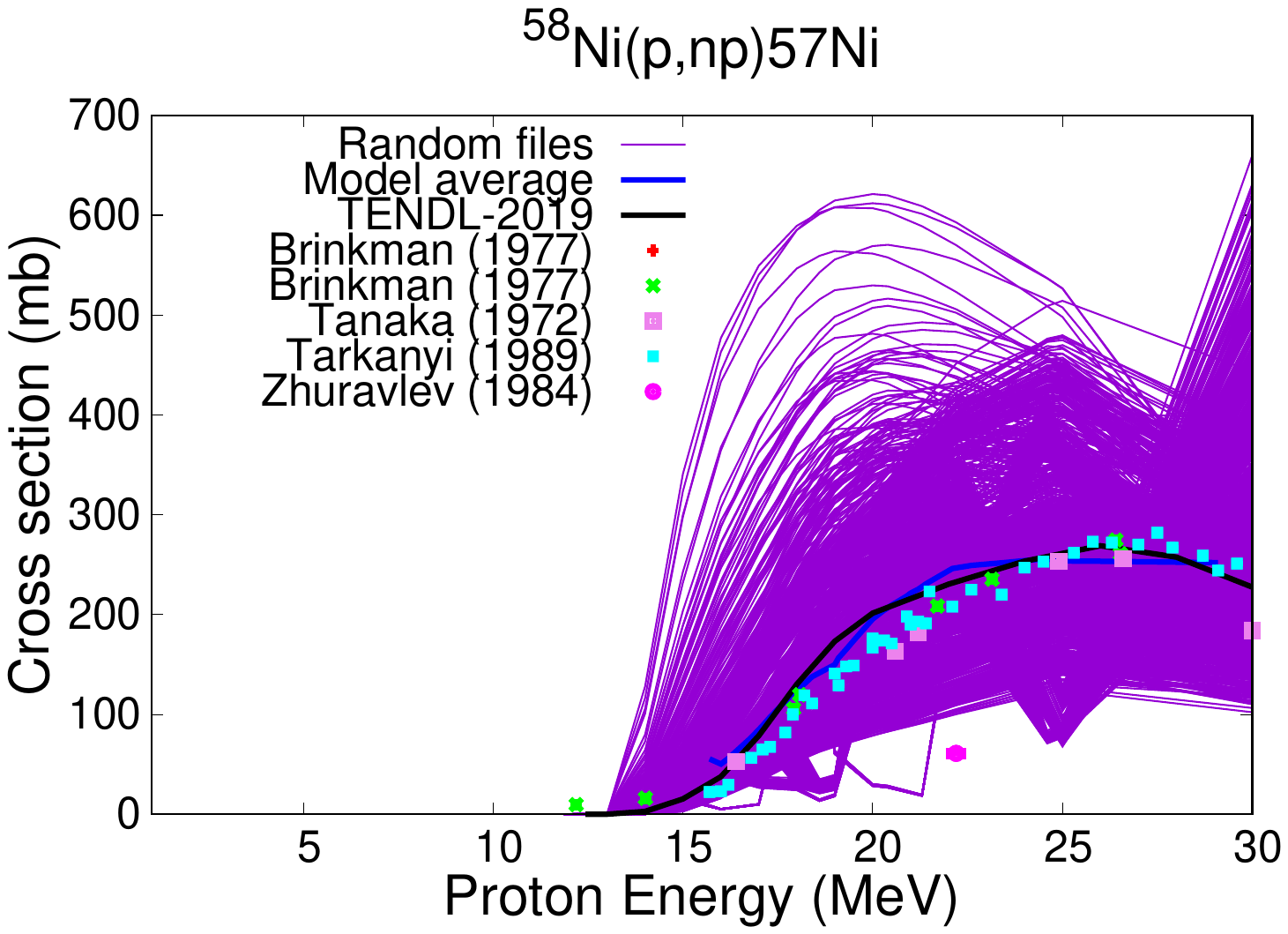}
   \caption{Prior $^{58}$Ni(p,np) cross section showing a combined spread due to both model and parameter variation and the corresponding model averaged value.}
   \label{mp_spread}
\end{figure}

In Fig.~\ref{mp_spread}, we present the prior random $^{59}Co$(p,3n) cross section curves showing the model average value obtained by simply taking the mean over all the models considered. It can be observed that the model average produced without the inclusion of experimental information, performs comparably well with the TENDL evaluation as well as with the experimental data. However, Zhuraviev (1984) dataset at about 22 MeV appears to be an outlier. Including this outlier in our Bayesian Model Averaging (BMA) calculations would significantly distort the cross-section curve at the energy point of the experimental data, and therefore was excluded from consideration.

Following the generation of the random cross sections, we compute the likelihood function by combining model predictions with experimental data at each data point, denoted as $i$, for the channel, $c$. The likelihood function values are then combined with the prior distributions to derive weighted averages and variances at each incident energy. Finally, a smoothing function using spline interpolation was utilized to smoothing-out any "kinks"or discontinuities in the cross section curves produced. The final product of the evaluation includes a central value accompanied by the corresponding prior and posterior variances and covariances.

 \begin{table}[htb]
       \centering
       \caption{List of selected model parameters with their corresponding uncertainties (or parameter widths). Except for the level density $a$ parameter, and the $g_\pi$ and $g_\nu$ parameters, where the parameter uncertainties are given in terms of the mass number $A$, the uncertainties of all the other parameters are given as a fraction (\%) of their absolute values. For a complete list of the parameters, see Ref.~\cite{bib:33}.}
       \label{Table1modelp}
       \begin{tabular}{cccc}  % Use 'c' to center the column, 'l' to have it left adjusted and 'r' for right adjusted 
       \toprule
       Parameter  &  Uncertainty [\%]  & Parameter  &  Uncertainty [\%]    \\
       \midrule
        & \multicolumn{3}{c}{OMP - phenomenological } \\
       \hline
       $r^p_V$  &  2.0 &   $a^p_V$ &  2.0  \\
       $v^p_1$ & 2.0 & $v^p_2$  &  3.0  \\        
       $v^p_3$ &  3.0  & $v^p_4$ & 5.0  \\
       $w^p_1$ & 10.0 & $w^p_2$  &  10.0  \\ 
       $w^p_3$ & 10.0 & $w^p_4$  &  10.0  \\ 
       $d^p_1$ &  10.0  & $d^p_2$ & 10.0  \\
       $d^p_3$  & 10.0 &  $r^p_D$ &  3.0 \\   
       $a^p_D$  & 2.0  &  $r^p_{SO}$ & 10.0 \\
       $a^p_{SO}$ & 10.0 & $v^p_{SO1}$ & 5.0 \\
       $v^p_{SO2}$ & 10.0 & $w^p_{SO1}$ &  20.0 \\
       $w^p_{SO2}$ & 20.0 & $r^p_c$ & 10.0 \\
       \hline
        \multicolumn{4}{c}{OMP - Semi-microscopic optical model (JLM) } \\
       \hline
       $\lambda_V$ & 5 & $\lambda_V1$ & 5 \\
       $\lambda_W$ & 5 & $\lambda_W1$ & 5 \\
      % $\lambda_{VSO}$ & 5 & $\lambda_{WSO}$ & 5 \\
       \hline
      & \multicolumn{3}{c}{level density parameters} \\
       \hline
       $a$ & 11.25-0.03125.A  &  $\sigma^2$ & 30.0 \\
       $E_0$ &  20.0 &  T &  10.0 \\
       $k_{rot}$  & 80.0 & $R_{\sigma}$ & 30.0 \\
       \hline
       & \multicolumn{3}{c}{Pre-equilibrium} \\
       \hline
       $R_{\gamma}$ &  50.0 & $M^2$  & 30.0 \\
       $g_\pi$ &  11.25-0.03125.A & $g_\nu$ & 11.25-0.03125.A \\
       $C_{break}$ &  80.0 & $C_{knock}$ &  80.0  \\
       $C_{strip}$ &  80.0 &   $E_{surf}$ & 20.0 \\
       $R_{\nu\nu}$ & 30.0  &  $R_{\pi\nu}$ & 30.0 \\
       $R_{\pi\pi}$ & 30.0  &  $R_{\nu\pi}$ & 30.0 \\
       \hline
      & \multicolumn{3}{c}{Gamma ray strength function} \\
       \hline
       $\Gamma_\gamma$ &  5.0 & $\sigma_{E\ell}$ & 20  \\
       $\Gamma_{E\ell}$ & 20 & $E_{E\ell}$ & 10 \\
       \bottomrule
       \end{tabular}
       \end{table}

% \subsection{Experimental data used}

\subsection{The case of a single model}
\label{single_M}
%\subsection{Bayesian Parameter estimation}
%georg: I think it makes sense to introduce the notation at the beginning of the method section, i.e., models, sub-models, parameter set, sets of parameter sets and what these quantities contain (can be after outlining the general idea or the references to past work).
% 
Let's assume that we have only one model set or combination ($\mathcal{\overrightarrow{M}}$) for describing a set of experimental data ($\overrightarrow{\sigma^{exp}_{ci}}$) for channel ($c$) and incident energy ($i$), as is the case of the Bayesian Monte Carlo (BMC) approach presented in Refs.~\cite{bib:1aa,bib:2,bib:5}. If we further assume that our single model is a vector encompassing several nuclear reaction models and sub-models ($\overrightarrow{\mathcal{M}}$) which is further characterized by a set of input model parameters denoted by $\overrightarrow{\theta_k}$ for the $k^{th}$ random parameter vector. In the BMC approach, the model vector, $\mathcal{\overrightarrow{M}}$, comprise of pre-selected models which come with each release of the TALYS code~\cite{bib:1TALYS}. These models are selected through methods such as chi square minimization and/or likelihood penalization, and sometimes, through visual comparison between model calculations and experimental data. Within a Bayesian framework, our posterior distribution which is defined as the updated probability distribution of our input parameters ($\overrightarrow{\theta_k}$) after the inclusion of experimental data ($\overrightarrow{\sigma^{exp}_{ci}}$) can be expressed as:

\begin{equation}
    p(\overrightarrow{\theta_k}|\overrightarrow{\sigma^{exp}_{ci}}) = \frac{p(\overrightarrow{ \sigma^{exp}_{ci}}|\overrightarrow{\theta_k})p(\overrightarrow{\theta_k})}{p(\overrightarrow{\sigma^{exp}_{ci}}|\overrightarrow{\mathcal{M}})}
    \label{Bayes_eq}
\end{equation}

where p($\overrightarrow{\theta_k}$) is the prior distribution of the parameters ($\theta_k$). The likelihood function, p($\sigma^{exp}_{ci}|\overrightarrow{\theta_k}$), is the sampling distribution of $\sigma^{exp}_{ci}$ given the $k^{th}$ random parameter vector, $\theta_k$. $\theta_k$ represents the parameters to the basic physics models in the TALYS code which together with codes are used to create random ENDF (Evaluated Nuclear Data File) files. ($\overrightarrow{\sigma^{exp}_{ci}}$) is a vector of differential experimental data obtained from the EXFOR database while $\overrightarrow{M}$ is a vector of nuclear reaction models, and $P(\overrightarrow{\sigma^{exp}_{ci}}|M)$ is the marginal likelihood. From Eq.~\ref{Bayes_eq}, p($\sigma^{exp}_{ci}|\overrightarrow{\mathcal{M}}$), is the marginal likelihood given our model vector, $\overrightarrow{M}$. Since the marginal likelihood is a normalisation constant, we can neglect this term. Once the parameters (including their widths or uncertainties) to these models are identified, the parameters were varied all-together within a Total Monte Carlo (TMC)~\cite{bib:1} framework to produce a large number of random cross sections for comparison with experimental data. The posterior distribution for our quantity of interest ($\overrightarrow{\sigma^{cal}_{cik}}$) can be given as:

\begin{equation}
    p(\overrightarrow{\sigma^{cal}_{cik}}|\overrightarrow{\sigma^{exp}_{ci}}) \propto p(\overrightarrow{\sigma^{exp}_{ci}}|\overrightarrow{\theta_k},\overrightarrow{\sigma^{cal}_{cik}})p(\overrightarrow{\sigma^{cal}_{cik}}|\overrightarrow{\theta_k}) 
\end{equation}

As previously discussed, a single model combination, as used here, ignores the modeling uncertainty associated with the selected models. However, given the current inadequacies of existing models, it becomes essential to consider model uncertainties in the nuclear data evaluation process as well. Therefore, we propose, leveraging all the knowledge in nuclear reactions accumulated from various models developed over the years through Bayesian Model Averaging (BMA), as presented in section~\ref{many_models}.

\subsection{The case of many models}
\label{many_models}
In situations involving two or more computational models where no single model set can be designated as the `best' model set as is often the case with nuclear reaction models, this work recommends a departure from the conventional approach of obtaining the posterior distribution for parameters, and hence covariances, based on a single model vector. Instead, we propose computing a weighted combination of the posterior distributions derived from different models as well as their parameters.

Let's suppose we have $J$ computing models, $\mathcal{M}_1,...,\mathcal{M}_J$. $\mathcal{M}$ in this case denotes a vector of nuclear reaction models such as the Exciton model of the pre-equilibrium model, or the constant temperature in combination with the Fermi gas model of the level density models. For our BMA approach, we begin with assigning prior probabilities to each model ($p(\mathcal{M}_j$)). In this work, the models were all drawn from uniform distributions. By using a uniform prior distribution, we assign the same $\emph{ex ante}$ probability to each model as follows:

\begin{equation}
    p(\mathcal{M}_j) = \frac{1}{J}
\end{equation}

In a simple form, the  Bayesian Model-Averaged can be given as: 

\begin{equation}
    \text{Prediction} = \sum_{\text{all models}} \text{p(model$|$data) x Model Prediction}
\end{equation}

where \text{p(model$|$data)} denotes the likelihood function given experimental data while the \emph{Model Prediction} denotes the prior distribution or model prediction. The average of the posterior distribution under each considered model, $\mathcal{M}$, for the quantity of interest, $\overrightarrow{\sigma^{cal}_{cik}}$ in our case, can then be expressed as:

\small
\begin{equation}  
 p(\overrightarrow{\sigma^{cal}_{cik}}|\overrightarrow{\sigma^{exp}_{ci}}) = \sum_{j=1}^{J}p(\overrightarrow{\sigma^{cal}_{cik}}|\overrightarrow{\mathcal{M}_j},\overrightarrow{\theta_k},\overrightarrow{\sigma^{exp}_{ci}})p(\overrightarrow{\mathcal{M}_j}|\overrightarrow{\sigma^{exp}_{ci}})
\end{equation}
\normalsize

with $p(\overrightarrow{\mathcal{M}_j}|\overrightarrow{\sigma^{exp}_{ci}})$, the posterior probability distribution for model $\overrightarrow{\mathcal{M}_j}$ given the data, $\overrightarrow{\sigma^{exp}_{ci}}$, expressed within Bayes' theorem as:

\begin{equation}
p(\overrightarrow{\mathcal{M}_j}|\overrightarrow{\sigma^{exp}_{ci}}) = \frac{p(\overrightarrow{\sigma^{exp}_{ci}}|\overrightarrow{\mathcal{M}_j})p(\overrightarrow{\mathcal{M}_j})}{p(\overrightarrow{\sigma^{exp}_{ci}})}
\end{equation}

where $p(\overrightarrow{\sigma^{exp}_{ci}})$ is the marginal likelihood of the data and $p(\overrightarrow{\mathcal{M}_j})$ is the prior probability of model, $\overrightarrow{\mathcal{M}_j}$. 

$p(\overrightarrow{\theta_k}|\overrightarrow{\mathcal{M}_j},\overrightarrow{\sigma^{exp}_{ci}})$, the posterior distribution of parameter $\theta_k$ given model $\overrightarrow{\mathcal{M}_j}$ and data $\overrightarrow{\sigma^{exp}_{ci}}$, is expressed as: 
\begin{equation}    p(\overrightarrow{\theta_k}|\overrightarrow{\mathcal{M}_j},\overrightarrow{\sigma^{exp}_{ci}}) = \frac{p(\overrightarrow{\sigma^{exp}_{ci}}|\overrightarrow{\mathcal{M}_j},\overrightarrow{\theta_k})p(\overrightarrow{\theta_k}|\overrightarrow{\mathcal{M}_j})}{p(\overrightarrow{\sigma^{exp}_{ci}}|\overrightarrow{\mathcal{M}_j})}
\end{equation}
Since the prior formulation induces a joint probability distribution ($p(\overrightarrow{\mathcal{M}_j},\overrightarrow{\theta_k})$), the joint posterior distribution of model, $\overrightarrow{\mathcal{M}_j}$, and parameter, $\overrightarrow{\theta_k}$, $p(\overrightarrow{\mathcal{M}_j},\overrightarrow{\theta_k}|\overrightarrow{\sigma^{exp}_{ci}})$, can be given as:
\small
\begin{equation}
p(\overrightarrow{\mathcal{M}_j},\overrightarrow{\theta_k}|\overrightarrow{\sigma^{exp}_{ci}}) = \frac{p(\overrightarrow{\sigma^{exp}_{ci}}|\overrightarrow{\mathcal{M}_j},\overrightarrow{\theta_k}) \cdot p(\overrightarrow{\theta_k}|\overrightarrow{\mathcal{M}_j}) \cdot p(\overrightarrow{\mathcal{M}_j})}{\sum_{j}p(\overrightarrow{\sigma^{exp}_{ci}}|\overrightarrow{\theta_k},)\overrightarrow{\mathcal{M}_j}) \cdot p(\overrightarrow{\theta_k}|\overrightarrow{\mathcal{M}_j}) \cdot p(\overrightarrow{\mathcal{M}_j}) } 
\label{margLjoint}
\end{equation}
\normalsize

where the denominator of Eq.~\ref{margLjoint} is the marginal (integrated) likelihood which is the probability of the observed data given the model and parameter sets, integrated over all possible models and parameter values. Since the marginal likelihood is a normalization constant, it is usually neglected. From Eq.~\ref{margLjoint}, $p(\overrightarrow{\mathcal{M}_j})$ is the prior probability of the model, $\mathcal{M}_j$, and $p(\overrightarrow{\theta_k}|\overrightarrow{\mathcal{M}_j})$ is the prior distribution of the parameters, $\overrightarrow{\theta_k}$, given the model, $\overrightarrow{\mathcal{M}_j}$. Since the models and parameters were varied altogether simultaneously, the joint posterior distribution of the models and parameters in Eq.~\ref{margLjoint} for our quantity of interest, $\overrightarrow{\sigma^{cal}_{cik}}$,  can be re-written as: 
\small
\begin{equation}
p(\overrightarrow{\sigma^{cal}_{cik}}|\overrightarrow{\mathcal{M}_j},\overrightarrow{\theta_k},\overrightarrow{\sigma^{exp}_{ci}}) \propto p(\overrightarrow{\sigma^{exp}_{ci}}|\overrightarrow{\mathcal{M}_j},\overrightarrow{\theta_k},\overrightarrow{\sigma^{cal}_{cik}}) \cdot p(\overrightarrow{\sigma^{cal}_{cik}}|\overrightarrow{\mathcal{M}_j},\overrightarrow{\theta_k})
\label{margLjoint2}
\end{equation}
\normalsize

The likelihood function, which represents the probability of our observed data given the parameter values, $\overrightarrow{\theta_k}$, and model, $\overrightarrow{\mathcal{M}_j}$, is expressed as: 

\begin{equation}
p(\overrightarrow{\sigma^{exp}_{ci}}|\overrightarrow{\mathcal{M}_j},\overrightarrow{\theta_k},\overrightarrow{\sigma^{cal}_{cik}}) \propto exp \left[ -\frac{1}{2} \overrightarrow{\chi^2_{cik}}\right]
\label{likelihoodfun}
\end{equation}

with the reduced chi square, denoted as $\chi^2_{i}$, at each considered incident energy $i$, channel $c$, and model parameter vector, $k$, expressed as:
\begin{equation}
    \overrightarrow{\chi^2_{cik}} = \left(\frac{\overrightarrow{\sigma^{cal}_{cik}} - \overrightarrow{\sigma^{exp}_{ci}}}{\Delta \overrightarrow{\sigma^{exp}_{ci}}}\right)^2
    \label{chi_square}
\end{equation}

The terms $\sigma^{exp}_{ci}$ and $\Delta \sigma^{exp}_{ci}$ represents respectively, the experimental cross sections and  its corresponding uncertainty, and $\overrightarrow{\sigma^{cal}_{cik}}$ is the calculated cross sections for channel $c$ and incident energy $i$. $\theta_k$ denotes the $k^{th}$ model parameter vector and $\mathcal{M}_j$ is the model $j$. It is instructive to highlight that this work does not take into account uncertainties from nuisance parameters in the calculation as carried out in, for example, Ref.~\cite{bib:12EA2016}. Consequently, this could lead to a relatively narrow posterior uncertainty band.

To standardize the weights, we normalize the likelihood function presented in Eq.~\ref{likelihoodfun} to have a maximum of 1 as follows: 

\begin{equation}
    p(\vec{\sigma^{exp}_{ci}}|\overrightarrow{\mathcal{M_j}},\overrightarrow{\theta_k},\overrightarrow{\sigma^{cal}_{cik}})_R = \frac{p(\vec{\sigma^{exp}_{ci}}|\overrightarrow{\mathcal{M}_j},\overrightarrow{\theta_k},\overrightarrow{\sigma^{cal}_{cik}})}{p(\overrightarrow{\sigma^{exp}_{ci}}|\overrightarrow{\mathcal{M}_j},\overrightarrow{\theta_k},\overrightarrow{\sigma^{cal}_{cik}})_{max}}
\end{equation}

where $p(\overrightarrow{\sigma^{exp}_{ci}}|\overrightarrow{M}_j,\overrightarrow{\theta_k},\overrightarrow{\sigma^{cal}_{cik}})_{max}$ is the maximum likelihood and the $p(\overrightarrow{\sigma^{exp}_{ci}}|\overrightarrow{M}_j,\overrightarrow{\theta_k},\overrightarrow{\sigma^{cal}_{cik}})_R$ is the relative likelihood. 

If we let the relative likelihood function equal to the file weight ($w_{cik}$), also called Bayesian Monte Carlo (BMC) weights, the weights for each considered channel ($c$), incident energy ($i$) and for the file $k$ can be given as:

\begin{equation}
    w_{cik} = \frac{e^{-\frac{1}{2}\chi^2_{cik}}}{e^{-\frac{1}{2}\chi^2_{cik}(min)}}
    \label{fileweight}
\end{equation}

where $\chi^2_{cik}(min)$ denoted the minimum $chi^2$, and $w_{cik}$ is the file weight at energy $i$, for channel $c$ and in the random file $k$. Then, we can compute a weighted cross section at energy $i$ for channel $c$ averaged over all the models (i.e. over all the $K$ random cross sections and angular distributions produced) as:
\begin{equation}
    \overline{\sigma^{\rm cal}_{cik}}_{w} = \frac{\sum_{i=1}^{i}\overrightarrow{w_{cik}}.\overrightarrow{\sigma_{\rm ci}^{cal}}}{\sum_{k=1}^{K}\overrightarrow{w_{cik}}}
    \label{xs_avg}
\end{equation}
where $\overline{\sigma^{\rm cal}_{cik}}_{w}$ is the weighted mean of the TALYS calculated cross sections ($\overrightarrow{\sigma^{\rm cal}_{cik}}$) taken from $K$ files, $w$ denotes weighted and $w_{cik}$ represents the weights at energy $i$ for channel $c$ in random file $k$. 

It is important to note here that instead of using the reduced $\chi^2$ for the computation of the weights as presented in Eg.~\ref{fileweight}, other weight specifications such as the Bayesian information criterion (BIC)~\cite{bib:044BIC} or the Akaike Information Criterion (AIC)~\cite{bib:044AIC} which are sometimes used within the BMA approach could have been used. These weights were however not utilized in this work. 

The corresponding weighted variance at incident energy, $i$, and channel, $c$ ($var_{w}(\overline{\sigma_{\rm cik}^{cal}}_{w})$) over the models and parameters can be expressed as:
\begin{equation}
    var_{w}(\overline{\sigma_{\rm cik}^{cal}}_{w}) = \frac{\sum_{k=1}^{K}\overrightarrow{w_{cik}}.\overrightarrow{\sigma_{\rm cik}^{cal}}^2}{\sum_{k=1}^{K}\vec{w_{cik}}} - \overline{\sigma_{\rm cik}^{cal}}^2
    \label{xs_var}
\end{equation}
% 
% $\vv{\sigma_{\rm T_a}^{c}}$ $\overline{\sigma_{\rm T_a}^{c}}$
% 
Similarly, a weighted or the posterior covariance matrix between cross sections at energy $a$ ($\overrightarrow{\sigma_{\rm T_a}^{c}}$) and $b$ ($\overrightarrow{\sigma_{\rm T_b}^{c}}$) can be given as:
\begin{equation}
    cov_{w}(\sigma_{\rm T_{a}}^{c},\sigma_{\rm T_{b}}^{c})=\frac{\sum_{k=1}^{K}\overrightarrow{w_{cik}}(\overrightarrow{\sigma_{\rm T_{a(k)}}^{c}}-\overline{\sigma_{\rm T_{a}}^{c}})(\overrightarrow{\sigma_{\rm T_{b(k)}}^{c}}-\overline{\sigma_{\rm T_{b}}^{c}})}{\sum_{k=1}^{K}\vec{w_{cik}}}
    \label{cova}
\end{equation}
Finally, a sample weighted correlation coefficient ($r_{w}$) can be obtained using the following expression~\cite{bib:1aa}: 
\begin{equation}
\small
    r_{w}=\frac{\sum_{k=1}^{K}\overrightarrow{w_{cik}}(\overrightarrow{\sigma_{\rm T_{a(k)}}^{c}}-\overline{\sigma_{\rm T_{a}}^{c}})(\overrightarrow{\sigma_{\rm T_{b(k)}}^{c}}-\overline{\sigma_{\rm T_{b}}^{c}})}{\sqrt{\sum_{k=1}^{K}\vec{w_{cik}} (\overrightarrow{\sigma_{\rm T_{a(k)}}^{c}}-\overline{\sigma_{\rm T_{a}}^{c}})^2} \sqrt{\sum_{k=1}^{K}\vec{w_{cik}}(\overrightarrow{\sigma_{\rm T_{b(k)}}^{c}}-\overline{\sigma_{\rm T_{b}}^{c}}})^2}
    \label{cova}
\end{equation}
It is important to highlight here that the correlations were obtained based on the cross sections and not on the model parameters, as presented in for example, Refs.~\cite{bib:1aa,bib:2}. Hence, the final evaluation includes a central file with its corresponding covariance information for the nuclide and incident particle of interest. 

% From these random cross sections for the considered channel, the total variance at each considered energy can be extracted. 

\subsection{BMA in the absence of experimental data}
\label{BMAwoExpts}
\begin{figure}[htb] %tb]
  \centering
    \includegraphics[trim = 15mm 13mm 5mm 20mm, clip, width=0.45\textwidth]{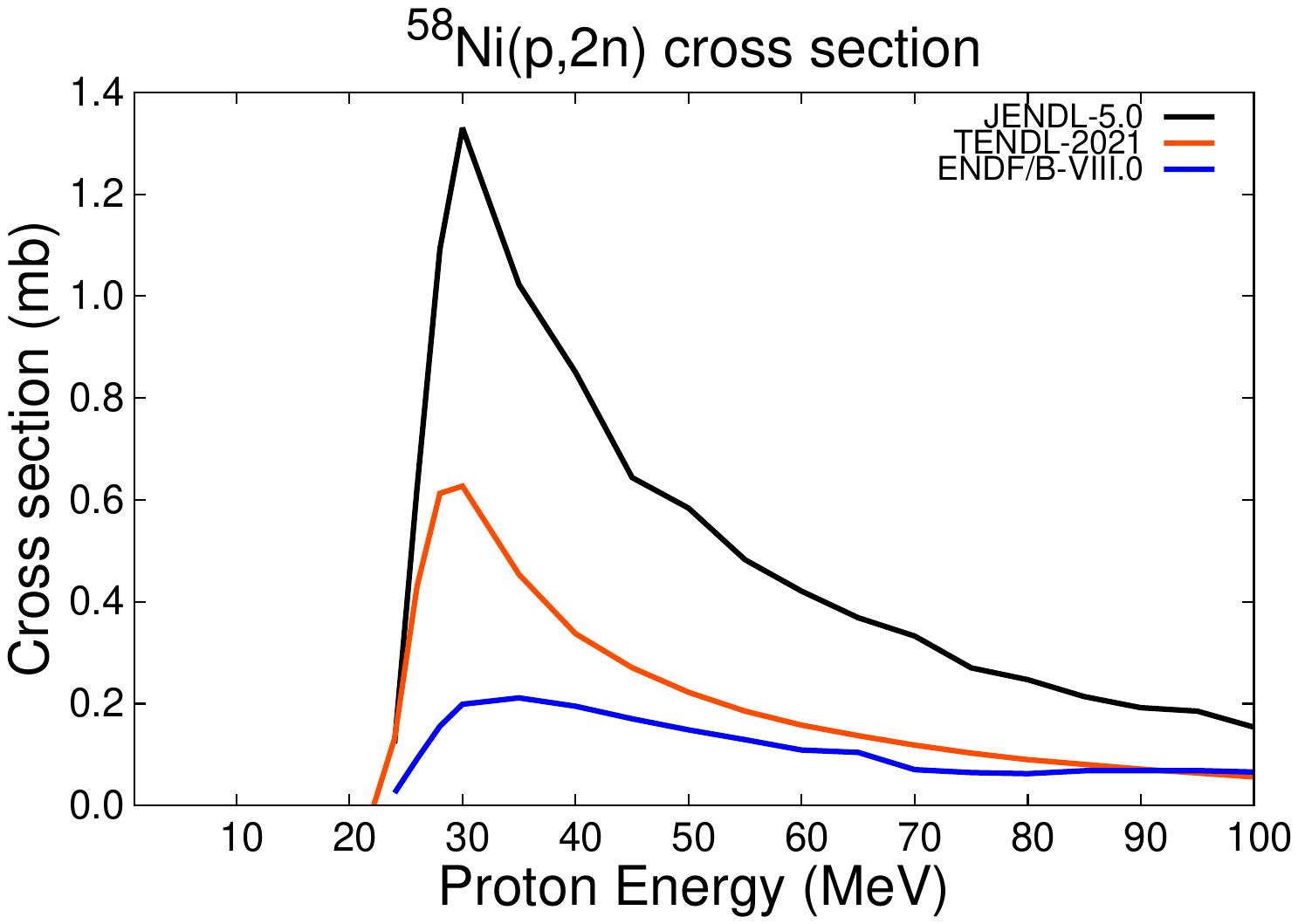}
   \caption{$^{58}$Ni(p,2n) cross section showing spread between evaluations from three nuclear data libraries: ENDF/B-VIII.0, JENDL-5.0 and TENDL-2021 libraries.}
   \label{spreadOfLibs}
\end{figure}

% There are cases during the evaluation process where there are no experimental data available. 
In the absence of experimental data, the Bayesian Model Averaging (BMA) solution for a joint model and parameter distribution involves combining only the prior information from multiple models and their parameters without relying on observed data. As an example, the $^{58}$Ni(p,2n) cross section showing evaluations from three different libraries (ENDF/B-VIII.0, JENDL-5.0 and TENDL-2021) is presented in Fig.~\ref{spreadOfLibs}. The ENDF/B-VIII.0 p+$^{58}Ni$ evaluation was produced using the GNASH code system~\cite{bib:Nashcode} which utilizes the Hauser-Feshbach statistical model, pre-equilibrium and direct-reaction theories. For the evaluation, the particle transmission coefficients used for the Hauser-Feshbach calculations as well as for the elastic proton angular distributions, were obtained from the spherical optical model calculations. The Gamma-ray transmission coefficients were calculated using the Kopecky-Uhl model. Similar to TENDL library, the ECIS95 code was used for the optical model calculation. For the JENDL-5.0 evaluation, the CCONE code system~\cite{bib:CCONE} which integrates various nuclear reaction models needed to describe nucleon, light charged nuclei up to alpha-particle and photon induced reactions, was used. For the JENDL-5.0 p+$^{58}$Ni evaluation, the two-component exciton model with global parametrization of Koning-Duijvestijn, was used. For the level density, the constant temperature and Fermi-gas model with shell energy corrections was used. In the case of the gamma-ray strength functions, the enhanced generalized Lorentzian form was used for E1 transition. For M1 and E2 transitions the standard Lorentzian form was adopted. For the calculation of angular distribution for emitted particles, Kalbach Systematics, was used. For  the TENDL-2021 evaluation, the TALYS code system was used with TALYS default models and parameters were used. 

From the figure, a wide spread between the evaluations presented can be observed. Ideally in such situations, `trusted' microscopic models (if available) can be utilized.   

\begin{figure}[htb] %tb]
  \centering
    \includegraphics[trim = 10mm 10mm 5mm 10mm, clip, width=0.45\textwidth]{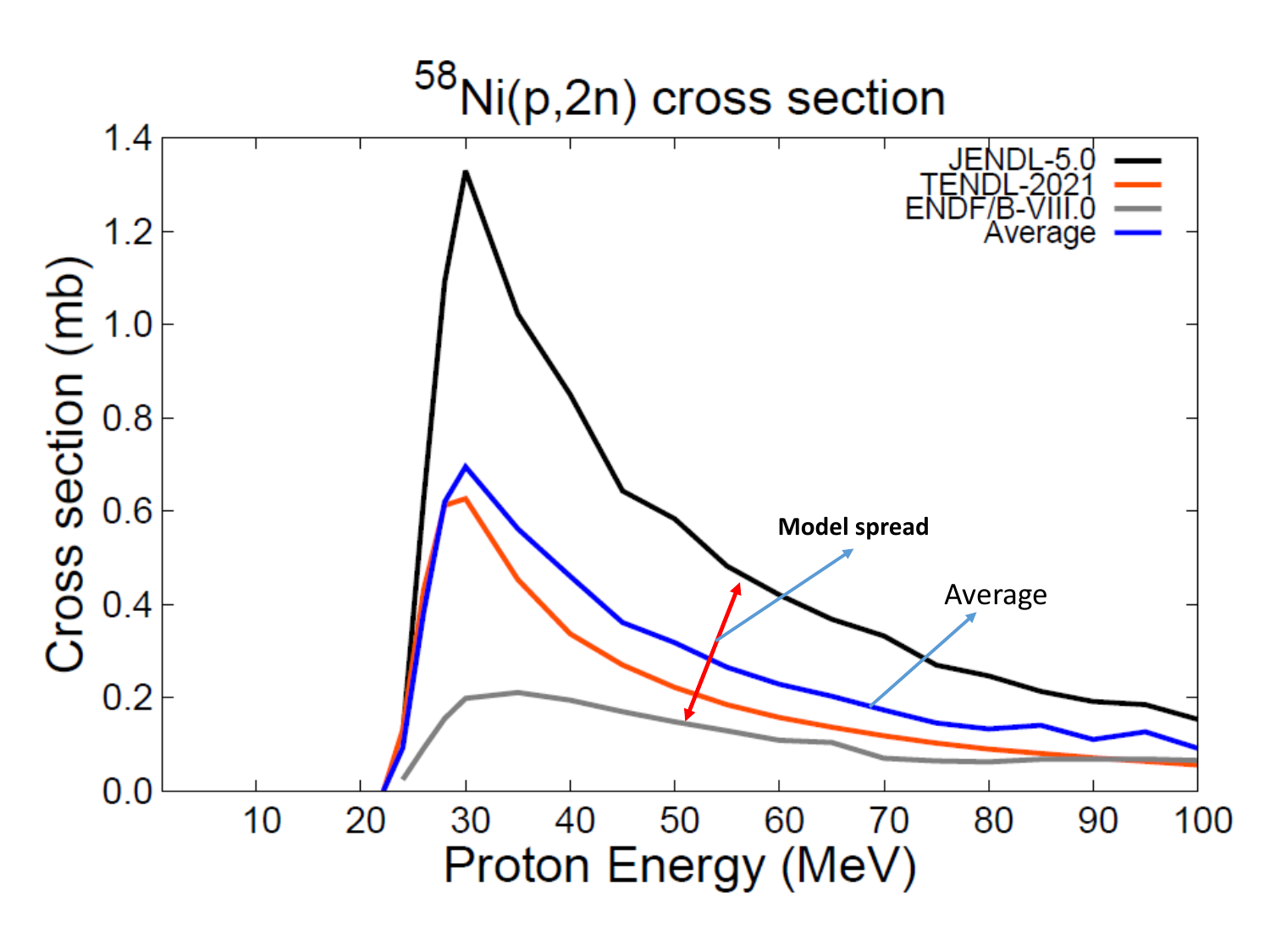}
   \caption{$^{58}$Ni(p,2n) cross section showing spread between three evaluations from the ENDF/B-VIII.0, JENDL-5.0 and TENDL-2021 libraries as well as the BMA average over the libraries.}
   \label{spreadOfLibs2}
\end{figure}

The Bayesian Model Averaging solution, assuming that each of the evaluation from the different library was a different model, and assuming that all the models were assigned equal weights, would be to take an average over the three evaluations and proceed with it as though it was our best estimate (see Fig.~\ref{spreadOfLibs2}). 

\begin{figure}[htb] %tb]
  \centering
    \includegraphics[trim = 10mm 10mm 5mm 10mm, clip, width=0.45\textwidth]{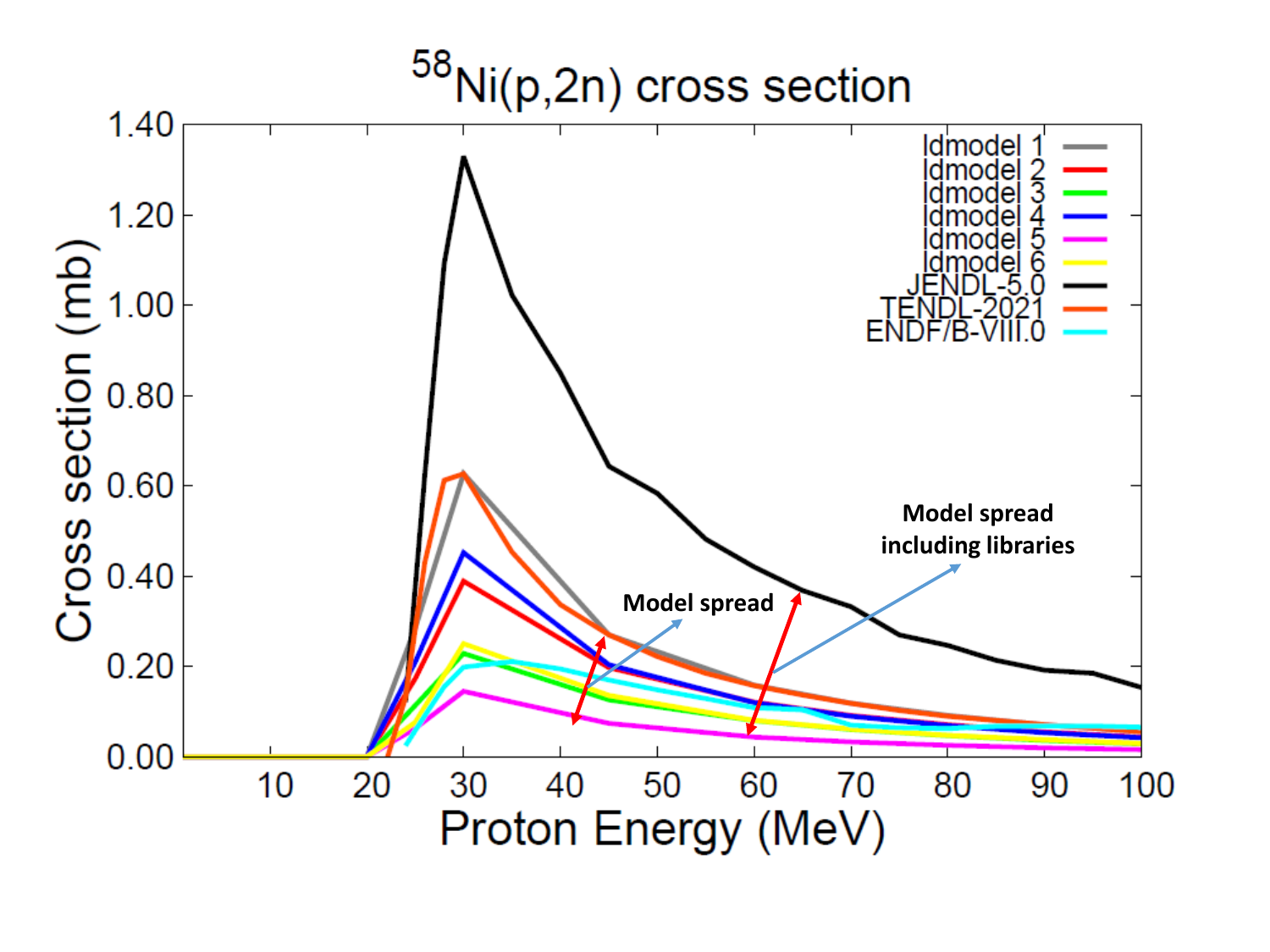}
   \caption{$^{58}$Ni(p,2n) cross section showing spread between the six level density models in TALYS compared with the evaluations from the ENDF/B-VIII.0, JENDL-5.0 and TENDL-2021 libraries.}
   \label{spreadOfLibs3}
\end{figure} 

The posterior distribution of our quantity of interest, $\overrightarrow{\sigma^{cal}_{cik}}$, given our model vector, $\overrightarrow{\mathcal{M}_j}$ and parameters, $\overrightarrow{\theta_k}$, in the absence of experimental data, can be given as:

\begin{equation}
    p(\overrightarrow{\sigma^{cal}_{cik}}|\overrightarrow{\mathcal{M}_j},\overrightarrow{\theta_k}) \propto p(\overrightarrow{\mathcal{M}_j}|\overrightarrow{\sigma^{cal}_{cik}},\overrightarrow{\theta_k}) \cdot p(\overrightarrow{\sigma^{cal}_{cik}}|\overrightarrow{\theta_k},\overrightarrow{\mathcal{M}_j})
\end{equation}

where $p(\overrightarrow{\sigma^{cal}_{cik}}|\overrightarrow{\mathcal{M}_j},\overrightarrow{\theta_k})$ is the posterior distribution of our quantity of interest, $\overrightarrow{\sigma^{cal}_{cik}}$, given our model, $\overrightarrow{\mathcal{M}_j}$, and parameters, $\overrightarrow{\theta_k}$. The likelihood function in this case, $p(\overrightarrow{\mathcal{M}_j}|\overrightarrow{\sigma^{cal}_{cik}},\overrightarrow{\theta_k})$, is the probability of the model, $\overrightarrow{\mathcal{M}_j}$, given our quantity of interest, $\overrightarrow{\sigma^{cal}_{cik}}$, and the parameters, $\overrightarrow{\theta_k})$. $p(\overrightarrow{\sigma^{cal}_{cik}}|\overrightarrow{\theta_k},\overrightarrow{\mathcal{M}_j})$, is the prior distribution which represents our prior knowledge about the distribution of our quantity of interest ($\overrightarrow{\sigma^{cal}_{cik}}$) given the model and parameters. 

In the absence of experimental data, the posterior distribution is significantly shaped by the prior distribution and any assumptions made about the considered models. Hence, the likelihood function may be constructed based on our prior beliefs regarding the models in the absence of experimental data. For nuclear reactions, as in our case, it is recognized that microscopic models exhibit greater predictive accuracy than their phenomenological counterparts. Consequently, in the absence of experimental data, substantial weights can be assigned to the microscopic models, reflecting their strong predictive capabilities, while relatively lower weights are assigned to the phenomenological models because of their known limited predictive power. In this work however, all models were assumed to have equal weights for channels where experimental data are unavailable. 

% The Bayesian Model Average of the joint distributions over models and parameters ($p(\overrightarrow{\mathcal{M}_j},\overrightarrow{\theta_k})$) in the absence of experimental data can be expressed as:

% \begin{equation}
%    p(\overrightarrow{\mathcal{M}_j},\overrightarrow{\theta_k}) = \sum_j % p(\overrightarrow{\mathcal{M}_j},\overrightarrow{\theta_k},\overrightarrow{\sigma^{cal}_{cik}})
% \end{equation}

% respect to this work, the 
In the case of this work, the calculated cross sections were obtained by simultaneously varying both the models and their parameters. Hence, the average cross section over the models and parameters at incident energy, $i$, and channel, $c$, in the absence of experimental data ($\overline{\sigma^{\rm cal}_{cik}}$), can be given as:

\begin{equation}
    \overline{\sigma^{\rm cal}_{cik}} = \frac{1}{K}\sum_{k=1}^{K} \overrightarrow{\sigma^{\rm cal}_{cik}}
\end{equation}

It should be highlighted here that by taking a simple average across models and their parameters, we assigned equal weights to the models (as well as their parameters). This approach is, however, highly sensitive to the presence of `bad' models, which have the potential of distorting the average cross-section curves and angular distributions. Consequently, identifying and discarding these `bad' models is important in model averaging for cases where no experimental data is available (see Figs.~\ref{MT111_badmodels} and \ref{rp028057_avg}). Also, as mentioned, arbitrary weights, reflecting their strong predictive capabilities, can be assigned to each model. In this way, model combinations considered as `bad' models would be assigned low weights and hence contribute little to the BMA estimate in the absence of experimental data. 

The unbiased estimate of the variance which is a measure of how spread out the distribution of the cross section at each energy, $i$, for channel, $c$ ($var(\sigma^{cal}_{ci})$), can be expressed as: 

\begin{equation}
    var(\sigma^{cal}_{cik}) = \frac{\sum_{k=1}^{K}(\sigma^{cal}_{cik} - \overline{\sigma^{cal}_{cik}})^2}{K-1}
% E(\sigma^{cal}_{cik} - E[\sigma^{cal}_{cik}])^2]
\label{unbiasedV}
\end{equation}

where $K$ represents the total number of random cross section (or samples) curves produced. Note that each random cross section curve was calculated with a set of TALYS models and unique parameters. It is also instructive to note that by dividing Eq.~\ref{unbiasedV} by $K-1$ instead of $K$, we obtain an unbiased estimate of the variance. From Eq.~\ref{unbiasedV}, 
the unbiased estimate of the standard deviation at each energy point $i$, which denotes the uncertainty, can be extracted.  

\begin{figure}[htb] %tb]
  \centering
    \includegraphics[trim = 15mm 13mm 5mm 20mm, clip, width=0.4\textwidth]{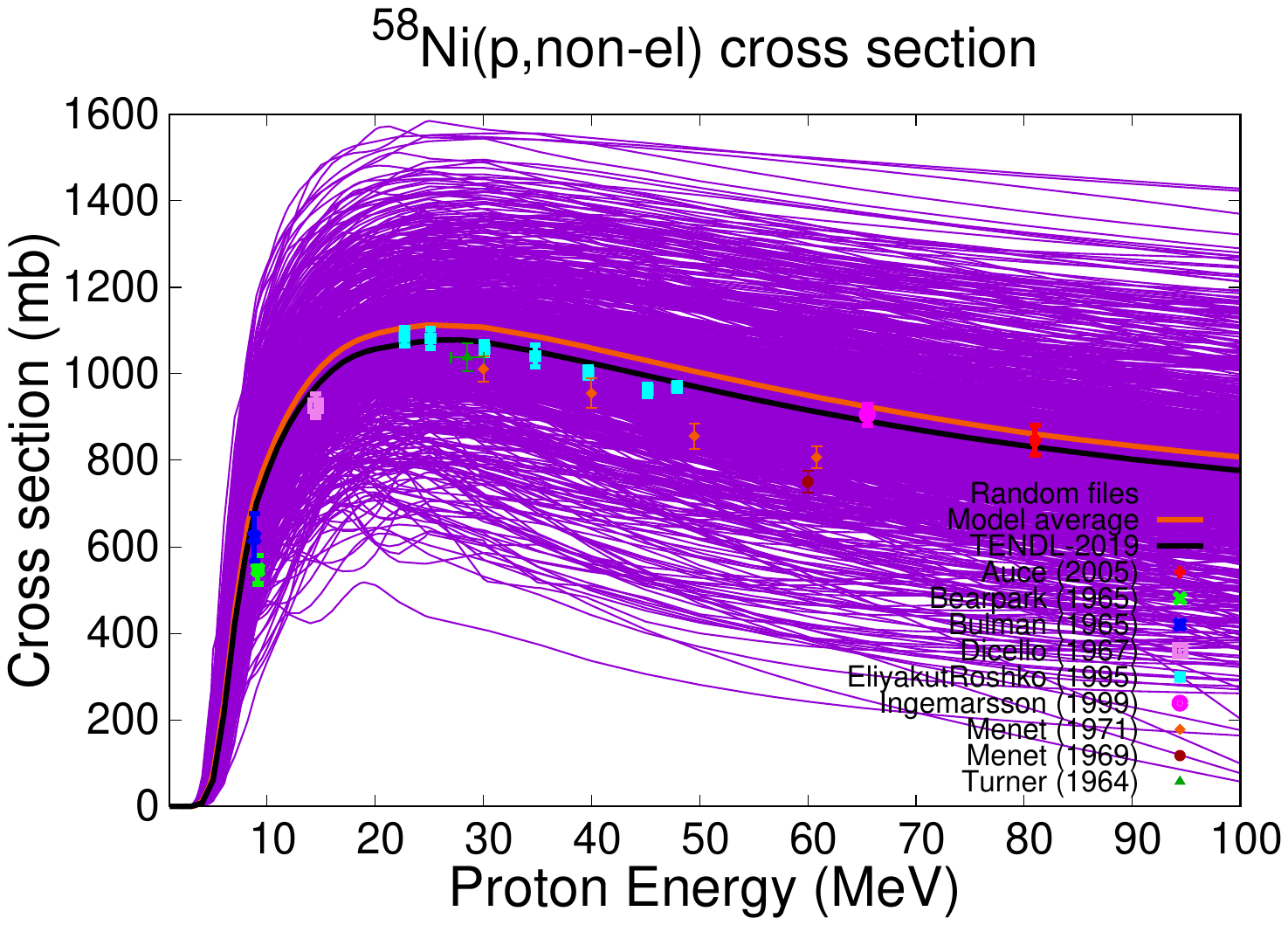}
   \caption{$^{58}$Ni(p,non-el) cross section showing the Bayesian Model Average over the prior spread of cross section curves in the absence of experimental data.}
   \label{MT003_avg}
\end{figure}

\begin{figure}[htb] %tb]
  \centering
    \includegraphics[trim = 15mm 13mm 5mm 20mm, clip, width=0.4\textwidth]{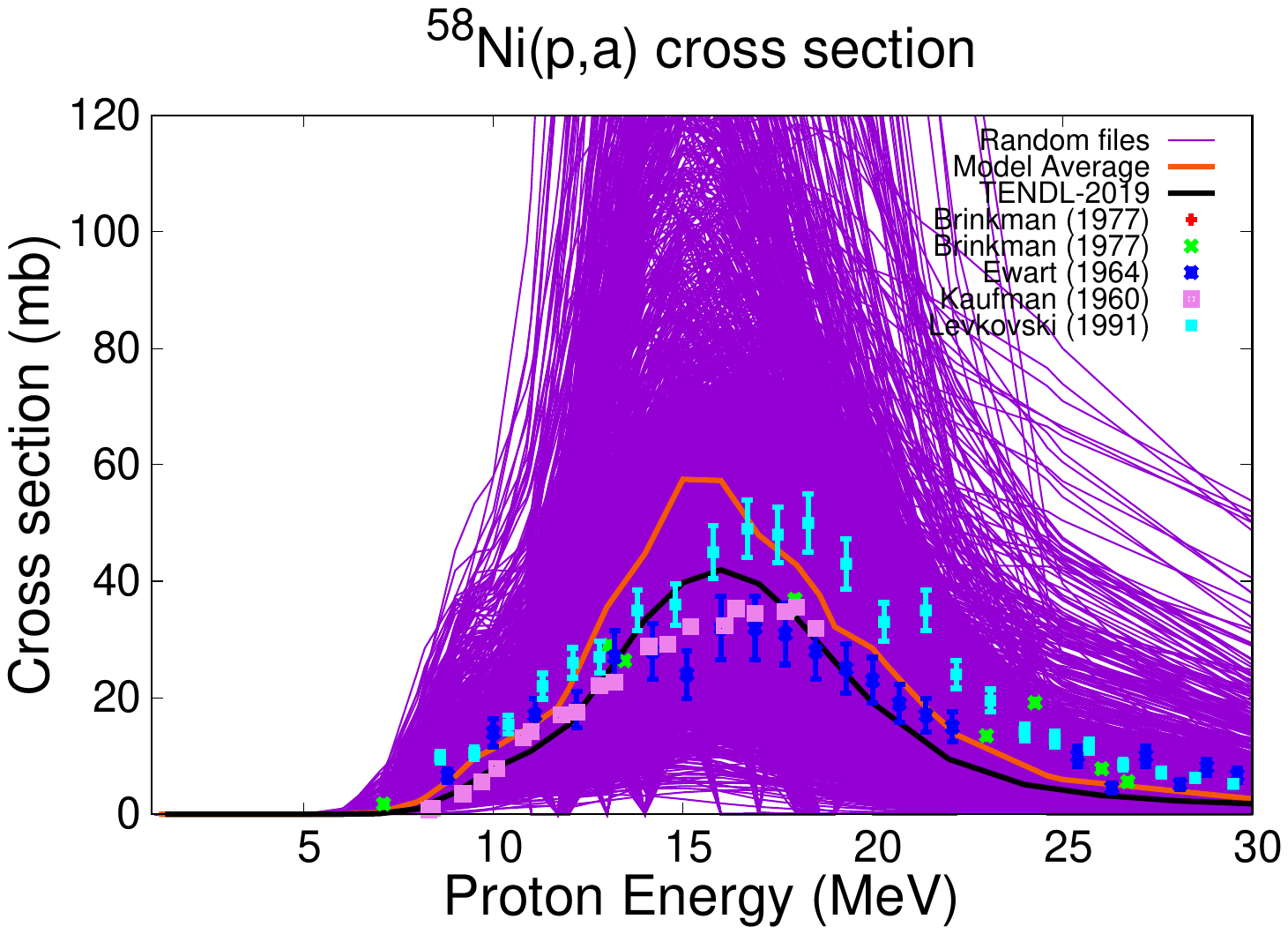}
   \caption{$^{58}$Ni(p,$\alpha$) cross section showing the Bayesian Model Average over the prior spread of cross section curves in the absence of experimental data.}
   \label{MT007_avg}
\end{figure}

To verify the predictive power of this approach, the model average over the prior distribution as discussed is compared with experimental data for the $^{58}$Ni(p,non-el) cross section in Fig.~\ref{MT003_avg}. The non-elastic cross section is made up of all interactions between a particle (a proton in this case) and a target nucleus, excluding the elastic scattering cross section. This include processes like inelastic scattering, capture reactions, and others. It can be observed from the figure that the BMA average (in the absence of experimental data) slightly over predicts most of the experimental data available but compares quite favourably with the TENDL evaluation over the entire energy region. This gives an indication that the Bayesian Model Average Prediction (BMAP) can give relatively good evaluations in the absence of experimental data. Further, it can be observed that the experimental data falls within the model-parameter uncertainty or spread. In Fig.~\ref{MT007_avg}, the $^{58}$Ni(p,$\alpha$) cross section showing the Bayesian Model Average over the prior spread in the absence of experimental data of cross section curves, is presented. It can be observed from the figure that even though the model average value over estimated the experimental data between about 12 and 16 MeV, it compared relatively well with experimental data for the rest of the energy range considered. 

% \overrightarrow{\sigma^{exp}_{ci}}

\subsection{Extracting model and parameter uncertainties}
A step by step algorithm for the extraction of model uncertainties at each incident energy is outlined in Table~\ref{modelUncert_algorithm}. As mentioned earlier, we start by selecting the distribution from which the models and their parameters were sampled. Next, we vary the models and their parameters simultaneously using the TALYS code system to generate a large set of random cross section curves as a function of incident energy as well as angular distributions. From the combined spread due to the variation of the models and their parameters together, a distribution in the cross section of interest can be obtained at each incident energy $i$ in the case of the reaction and residual production cross section, or in angle, in the case of the angular distributions, can be extracted. The total spread from this distribution can be attributed to the simultaneous variations of both the models and their parameters.

\begin{table}[thb]
 \centering
  \caption{Step by step algorithm for computation of model and parameter uncertainties.}
  \label{modelUncert_algorithm}
  \begin{tabular}{l}  % Use 'c' to center the column, 'l' to have it left adjusted and 'r' for right adjusted
  \toprule
   algorithm   \\
   \midrule
  1: Choose distribution from which models and their parameters \\ 
  \hspace{4mm} would be sampled \\
  2: Vary many model and their parameters simultaneously to  \\ 
  \hspace{4mm} produce a large number of random cross sections \\
  3: Determine the total variance at incident energy $i$ ($var(\sigma^{cal}_{ci,comb})$)  \\
  \hspace{4mm} for each considered channel ($c$)  \\
 % 4: Compute a global weight for each file $k$, L_{MLE} = exp(-0.5\chi_{k}^2)\\
 % 5: Select the wining model (WM) combination by selecting  \\
  %   \hspace{7mm} \(\displaystyle  WM \rightarrow L_{MLE} = \operatorname*{arg\,max}_m [L(\vec{\sigma_E}|\vec{M_j},\vec{p_k})]\) \\
  6: Vary only model parameters around a single model set \\
  \hspace{4mm} combination using uniform distributions \\
  7: Compute the variance at energy $i$ due to only model parameters \\ 
   \hspace{4mm} for each channel $c$ \\ 
  8: Extract the uncertainty due to only models ($U(\sigma^{cal}_{ci,\mathcal{M}})$): \\ 
  \hspace{4mm} $U(\sigma^{cal}_{ci,\mathcal{M}}) = \sqrt{var(\sigma^{cal}_{ci,comb}) - var(\sigma^{cal}_{ci,\theta})}$ \\
\bottomrule
\end{tabular}
\end{table}

Next, we determine the variance due to only parameter variation at a considered energy $i$. To achieve this, more than 3000 random cross section curves were produced with varying only model parameters, around a single (fixed) set of models. The following model vector, alongside other models not explicitly listed here, was used~\cite{bib:33}: 
\textbf{ld model 2:} Back-shifted Fermi gas model; \textbf{strength function model 6:} Goriely Time (T)-dependent Hartree-Fock-Bogolyubov (HFB); \textbf{pre-equilibrium model 3:} Exciton model using numerical transition rates with optical model for collision probability; \textbf{preeqspin 2:} spin distribution from total level densities is adopted; \textbf{pair model 2:} Compound nucleus pairing correction; \textbf{widthmode 2:} Hofmann-Richert-Tepel-Weidenm\"{u}ller (HRTW) model. 

% Model for liquid drop expression for nuclear mass, to be used to calculate the shell correction;

\begin{figure}[htb] %tb]
  \centering
   \includegraphics[trim = 25mm 85mm 25mm 85mm, clip, width=0.50\textwidth]{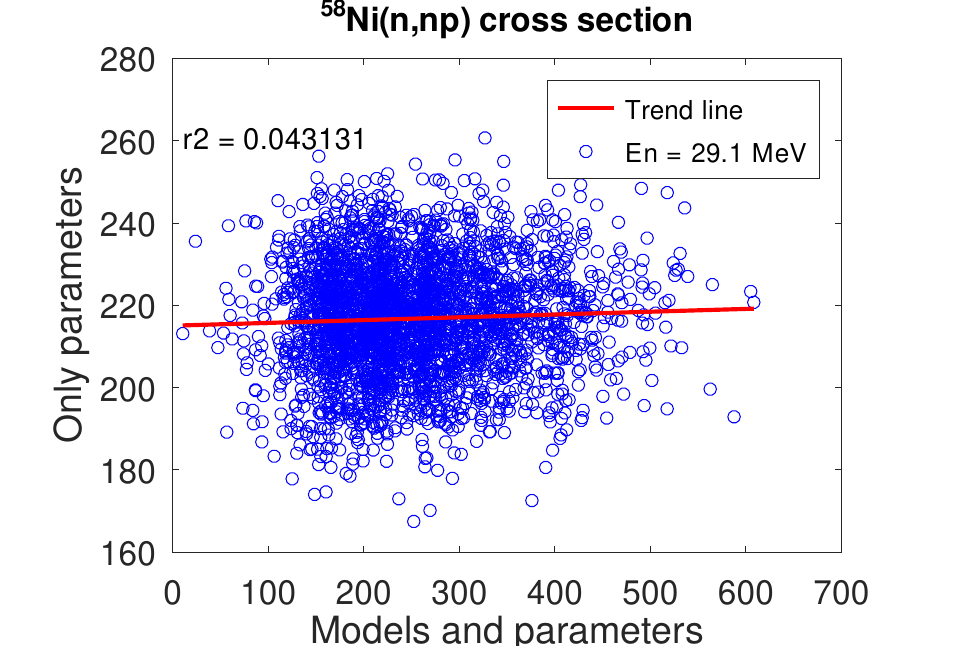}
   \caption{Scatter plot illustrating no significant correlation between random cross sections generated solely through model parameter variations and those produced with the simultaneous variation of both models and their parameters for the $^{58}$Ni(n,np) cross section at 29.1 MeV. A total of 3030 random cross sections were utilized in both cases.}
   \label{scatter_MT028_10}
 \end{figure}

From the distribution of cross sections or angulation distributions at each incident energy of interest, the variance can be calculated. If we assume that there are no strong correlations between the models and the parameters, the total or combined variance of the calculated cross section at energy $i$ for channel $c$ ($var(\sigma^{cal}_{ci,comb}$) can be expressed as a quadratic sum of the model (var($\sigma^{cal}_{ci,\mathcal{M}})$) and the parameter variance ($var(\sigma^{cal}_{ci,\theta})$):

\begin{equation}
 var(\sigma^{cal}_{ci,comb}) = var(\sigma^{cal}_{ci,\mathcal{M}}) + var(\sigma^{cal}_{ci,\theta})
% V^{ci}_{tot} = V_{mod}^{ci} + V^{ci}_{par} 
 \label{tot_var}
\end{equation}

where $\sigma^{cal}_{ci,\mathcal{M}}$ and $\sigma^{cal}_{ci,\theta}$ represents the cross sections of channel $c$ at energy $i$, obtained from the variation of models and parameters, respectively. The correlation between the models and the parameters was investigated and no significant correlation was observed. For example, in Fig.~\ref{scatter_MT028_10}, we present a scatter plot illustrating no significant correlation between random cross sections generated solely through model parameter variations and those produced with the simultaneous variation of both models and their parameters for the $^{58}$Ni(n,np) cross section at 29.1 MeV. Note that, in the case involving only model parameter variations, the models were kept constant, while their parameters were simultaneously varied. In the other case case, both the models and their parameters were simultaneously varied altogether.

% In the case of only model parameter variations, the models were fixed while their parameters were varied simultaneously. In the other case, both the models and their parameters were varied altogether. % We note however that there exist dependency relationships between the various models and their parameters. 

Since we can compute the uncertainty due to only parameter variations, the model uncertainties can easily be extracted from Eq.~\ref{tot_var} as follows: 

\begin{equation}
% U^{ci}_{mod} = \sqrt{V_{tot}^{ci} - V^{ci}_{par}} 
 U(\sigma^{cal}_{ci,\mathcal{M}}) = \sqrt{var(\sigma^{cal}_{ci,comb}) - var(\sigma^{cal}_{ci,\theta})}
 \label{modelUncert}
\end{equation}

where $U(\sigma^{cal}_{ci,\mathcal{M}})$ is the uncertainty due to models for channel $c$ at incident energy $i$.

\section{Application of BMA methodology}
The key steps of the BMA methodology outlined in this work can be summarized as follows: 1) begin with a very large non-informative prior as much as possible which ensures that a significant number, if not all, of the selected experimental data falls within the prior spread of random cross section curves or angular distributions, 2) allow the shape and spread of the posterior distribution to be solely shaped by the experimental data; and 3) locally select models at each incident energy point rather than globally, that is, over the entire energy range under consideration.

\subsection{Prior distribution of models and parameters}
\begin{figure}[htb] %tb]
  \centering
  \includegraphics[trim = 15mm 85mm 5mm 85mm, clip, width=0.45\textwidth]{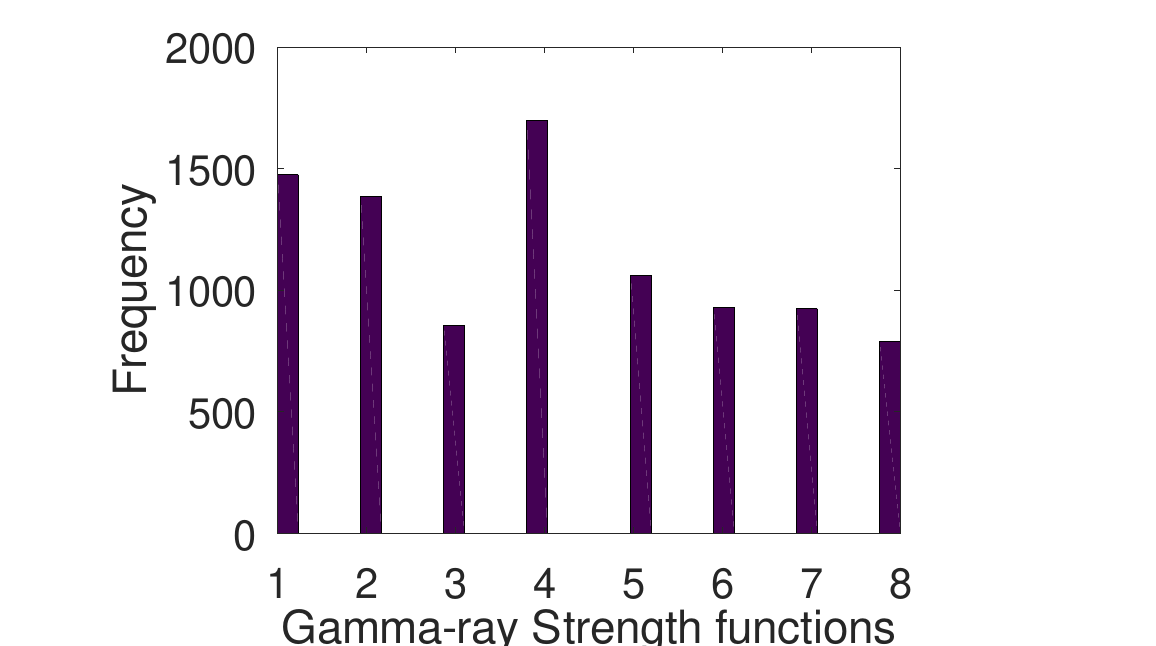}
   \caption{Prior distribution for the gamma-ray strength function utilized in this work, and based on more than 9000 random samples.}
   \label{strengthModels}
 \end{figure}

\begin{figure}[htb] %tb]
  \centering
  \includegraphics[trim = 15mm 85mm 5mm 85mm, clip, width=0.45\textwidth]{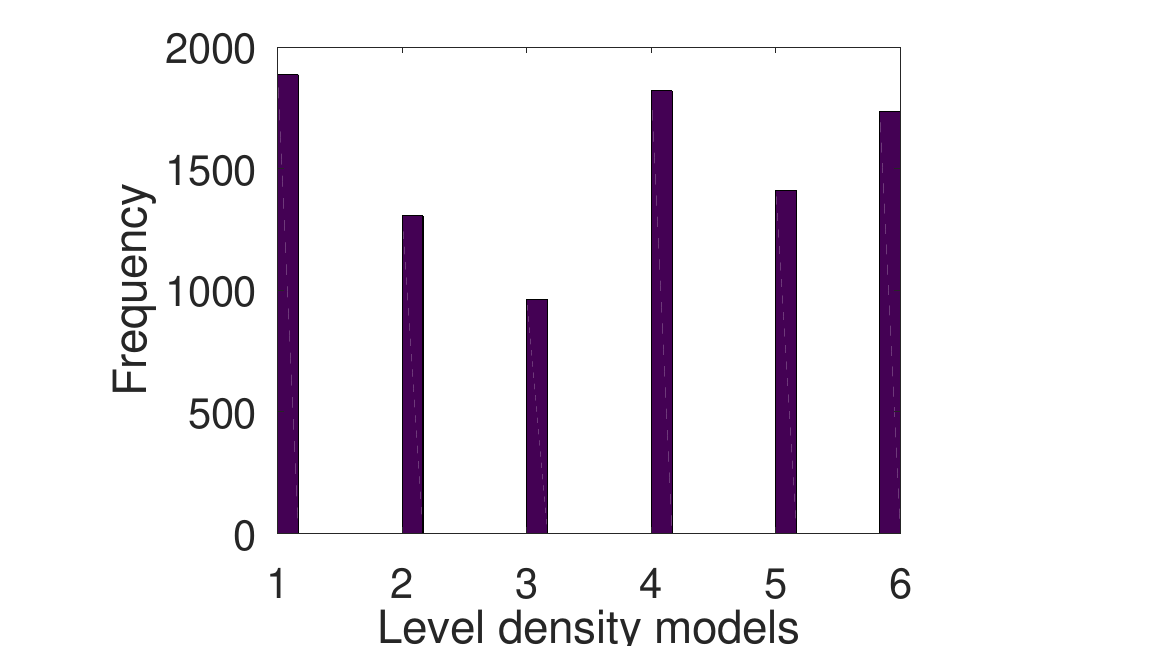}
   \caption{Prior distribution for the level density utilized in this work, and based on more than 9000 random samples.}
   \label{ldmodelHist}
 \end{figure}

In this work, we adopted a uniform distribution for the prior models and parameters distributions. By using the uniform distribution to each model type, we assign a constant probability to each model within the lower and upper bounds of the model type under consideration. For example, in the case of the level density model type, the six available $ld$ models were each assigned unique identifiers ($ld1, .., ld6$). These models were drawn randomly many times within the assigned lower and upper bounds of each model type. In Figs.~\ref{strengthModels} and \ref{ldmodelHist}, the distribution of the gamma-ray strength functions and the level density models for more than 9000 random samples are presented respectively. To assess whether our model distributions, as shown in Figs.~\ref{strengthModels} and \ref{ldmodelHist}, conform to uniform distributions, we computed the p-value for each distribution. The obtained p-values for each distribution were well below 0.05, leading us to reject the null hypothesis (H0 is: The model prior distributions are not uniform) within 95\% confidence interval. Applying the same methodology to the different model types resulted in different model vectors. A total of 100 different model combinations, each used as input to the TALYS code, were produced. An example of a list of models contained in random file number, 2025, is provided in Table~\ref{modelVector}.

\begin{table}[thb]
 \centering
  \caption{An exmple of a list of some models in random file number 2025.}
  \label{modelVector}
  \begin{tabular}{l}  % Use 'c' to center the column, 'l' to have it left adjusted and 'r' for right adjusted
  \toprule
   Model combination in random file: 2025   \\
   \midrule
  1: ldmodel 1: Constant Temperature + Fermi gas model (CTM) \\ 
  2: ctmglobal y: Flag to enforce global formulae for the Constant \\ 
  \hspace{4mm} Temperature Model (CTM)  \\
  3: strength 8: Gogny D1M HFB+QRPA   \\
  4: widthmode 0: no width fluctuation, i.e. pure Hauser-Feshbach  \\
  5: preeqmode 3: Exciton model: Numerical transition rates with \\ 
   \hspace{4mm} optical model for collision probability \\ 
  6: preeqspin 3: the pre-equilibrium spin distribution is based  \\ 
   \hspace{4mm} on particle-hole state densities \\ 
  7: kvibmodel 1: Model for the vibrational enhancement of  \\ 
  \hspace{4mm} the level density \\
  8: spincutmodel 2: Model for spin cut-off parameter for   \\ 
  \hspace{4mm} the ground state level densities  \\
  9: strengthm1 2: Normalize the M1 gamma-ray strength function   \\ 
  \hspace{4mm} with that of E1 as fE1/(0.0588A0.878)  \\
  10: preeqcomplex y: Flag to use the Kalbach model for pickup,  \\ 
  \hspace{4mm} stripping and knockout reactions, in addition to the\\
  \hspace{4mm}  exciton model, in the pre-equilibrium region.  \\
\bottomrule
\end{tabular}
\end{table}

\begin{figure}[htb] %tb]
  \centering
  \includegraphics[trim = 15mm 85mm 5mm 85mm, clip, width=0.45\textwidth]{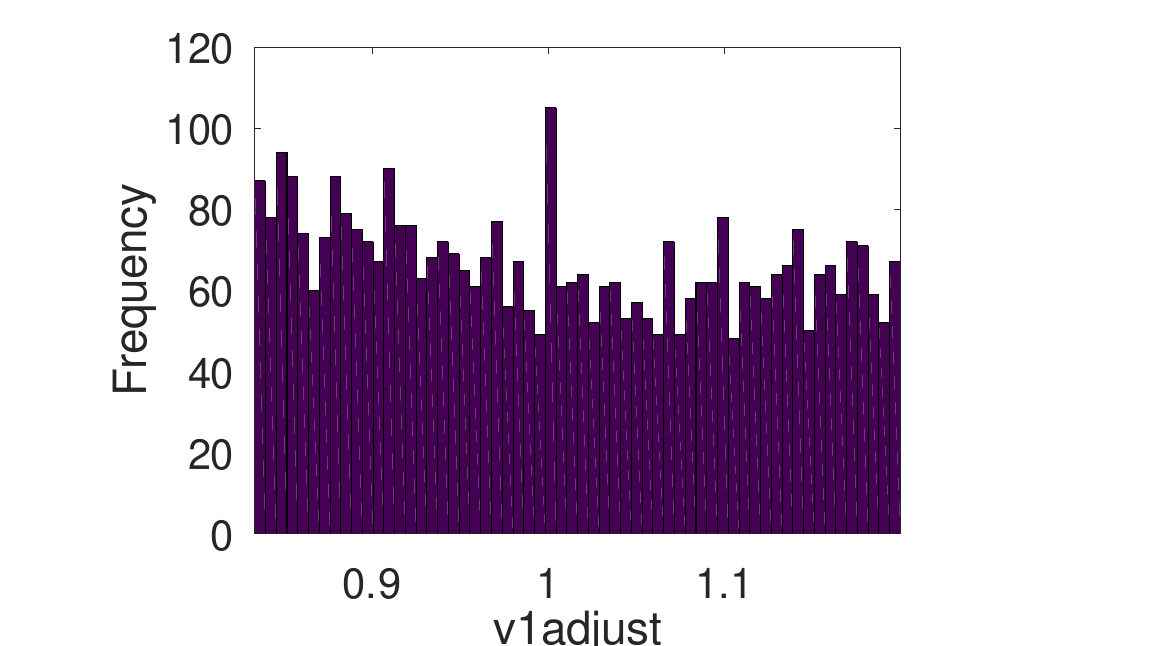}
   \caption{Prior distribution for the $v_1$, an adjustable parameters used in the computation of the depth of the real central potential of the optical model. The values presented are TALY's multipliers which were multiplied by the parameter mean values given in Table~\ref{Table1modelp}.}
   \label{v1adjustOMP}
 \end{figure}

\begin{figure}[htb] %tb]
  \centering
  \includegraphics[trim = 15mm 85mm 5mm 85mm, clip, width=0.45\textwidth]{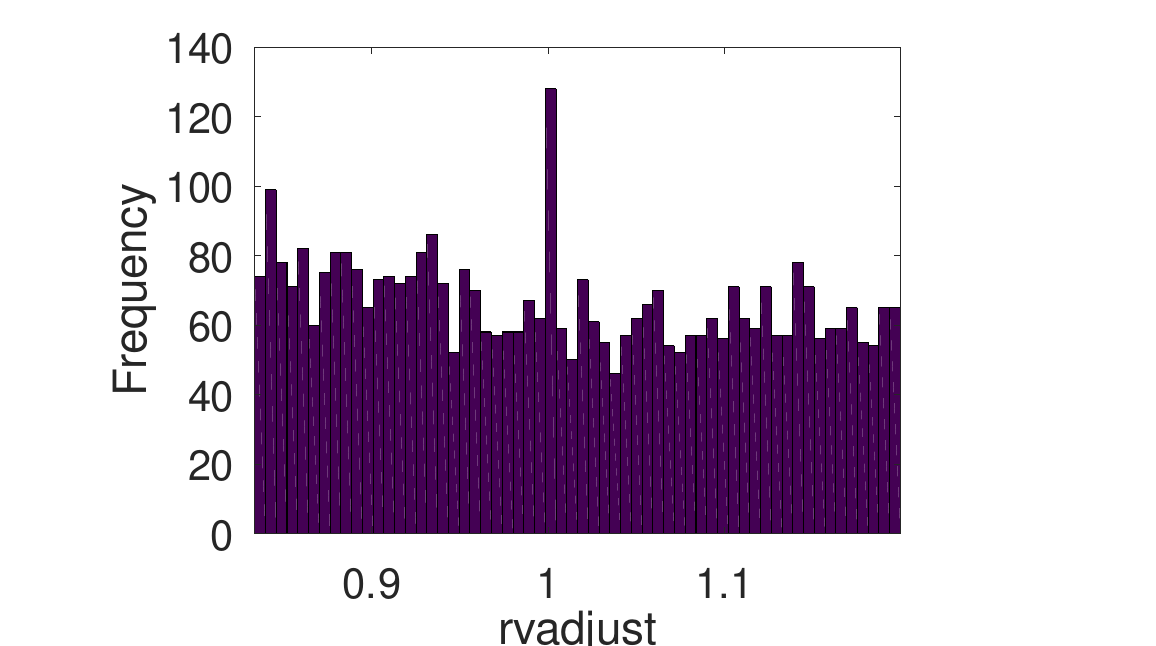}
   \caption{Prior distribution for the radius parameter ($r^p_V$) of the real central potential of the optical model. The values presented are TALY's multipliers which were multiplied by the parameter mean values given in Table~\ref{Table1modelp}.}
   \label{rvadjustOMP}
 \end{figure}

\begin{figure}[htb] %tb]
  \centering
  \includegraphics[trim = 15mm 85mm 5mm 85mm, clip, width=0.45\textwidth]{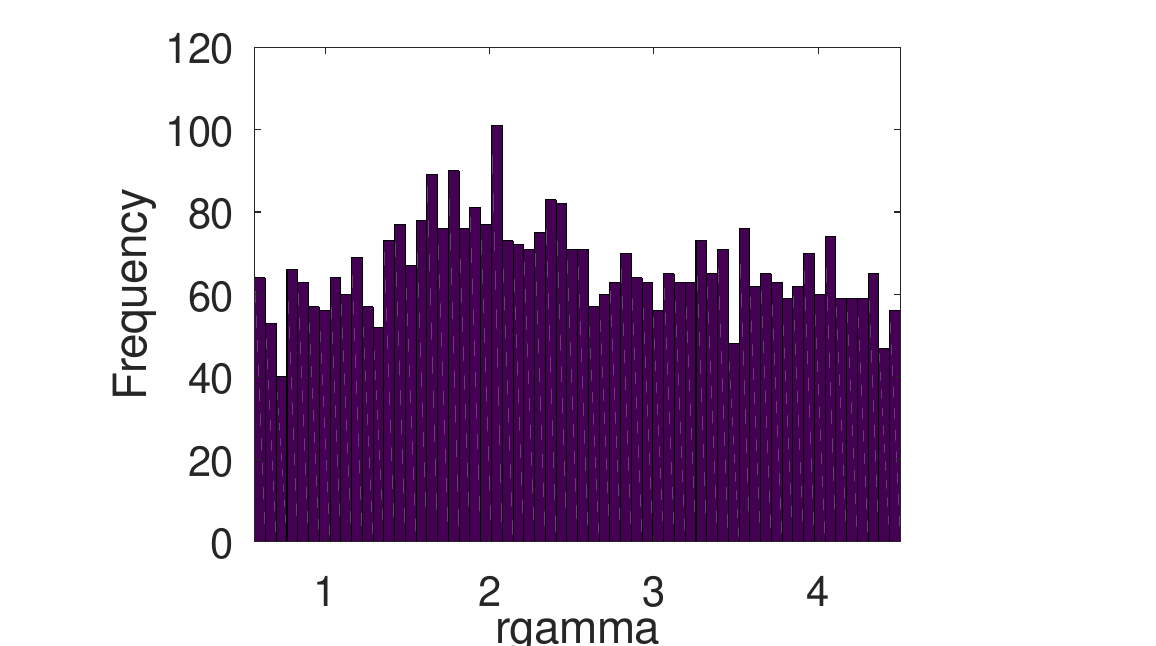}
   \caption{Prior distribution for $\gamma$-decay parameter ($R_{\gamma}$) for the pre-equilibrium model. The values presented are TALY's multipliers which were multiplied by the parameter mean values given in Table~\ref{Table1modelp}.}
   \label{Rgama}
 \end{figure}

Thereafter, TALYS parameters to each model combination, as listed for example in Table~\ref{modelVector}, were varied within their widths or uncertainties (see Table~\ref{Table1modelp}) using the TALYS code system. This process generated approximately 100 random nuclear data files per model combination, resulting in more than 9000 random ENDF-formatted nuclear data files for p+$^{58}$Ni. It is worth noting that the incident energies considered for p+$^{58}$Ni ranged from 1 to 100 MeV. In Fig.~\ref{rvadjustOMP}, the prior distribution of the geometrical parameters ($r^p_V$) which denotes the radius of the real central potential of the optical model is presented. Prior distribution for the $\gamma$-decay parameter ($R_{\gamma}$) for the pre-equilibrium model is presented in Fig.~\ref{Rgama}. Additionally, in fig.~\ref{v1adjustOMP}, we present the prior distribution for the $v_1$, an adjustable parameters used in the computation of the depth of the real central potential of the optical model. The values presented are TALY's multipliers for each parameter. The values presented are TALY's multipliers which were multiplied by the parameter mean values given in Table~\ref{Table1modelp}. As depicted in the figures, the parameters were drawn from uniform distributions with the minimum and maximum values. It is important to note, however, that due of the large number of parameters considered, not all the parameter prior distributions converged to the uniform distribution for the total number of samples considered in this work. 

 \subsection{Experimental data}
 The experimental data used in this work were selected from threshold up to 100 MeV. The Bayesian Model Averaging (BMA) method proposed in this work was applied to the following:

 \begin{itemize}
     \item Reaction cross sections: $(p,non-el)$, $(p,n)$, $(p,np)$, $(p,p)$, $(p,\alpha)$, $(p,2p)$.
     \item Residual production cross sections: $^{58}Ni(p,x)^{55}Co$, $^{58}Ni(p,x)^{56}Co$, $^{58}Ni(p,x)^{56}Ni$, $^{58}Ni(p,x)^{57}Ni$.
     \item Elastic angular distributions at the following incident energies: 9.51, 16.00, 20.00, 21.30, 35.20, 39.60, 40.00, and 61.40 MeV between 1 and 180 degrees. 
 \end{itemize}
 
The experimental data for each of the categories listed were obtained from the EXFOR database.

\subsection{Case of one and two experimental point}
\begin{figure*}[htb] %tb]
  \centering
   \includegraphics[trim = 15mm 65mm 5mm 65mm, clip, width=0.45\textwidth]{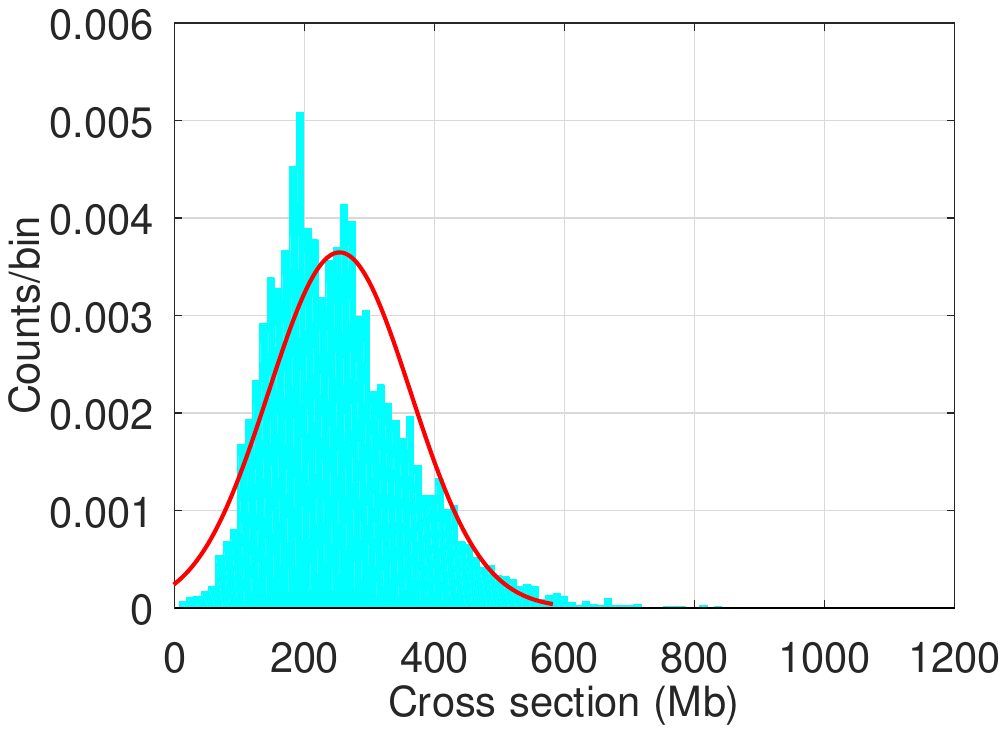}
  \includegraphics[trim = 15mm 75mm 5mm 65mm, clip, width=0.45\textwidth]{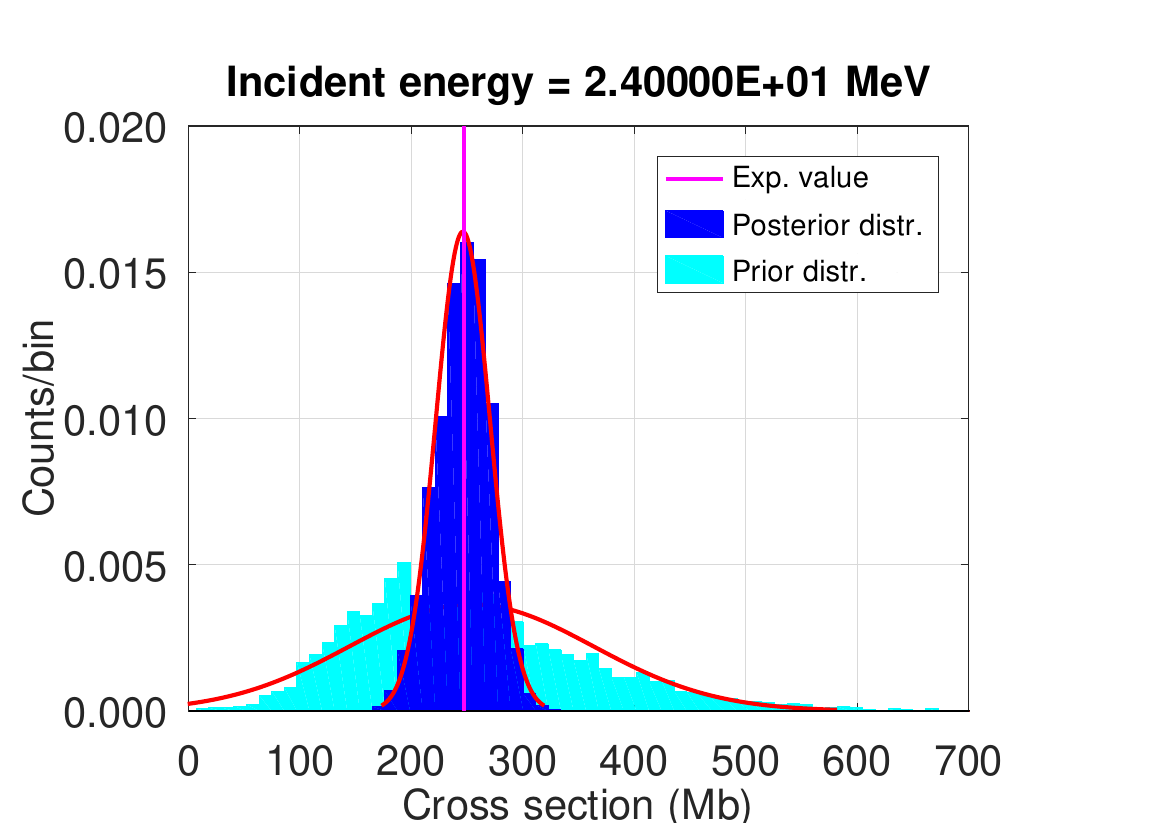}
    \includegraphics[trim = 25mm 75mm 5mm 75mm, clip, width=0.45\textwidth]{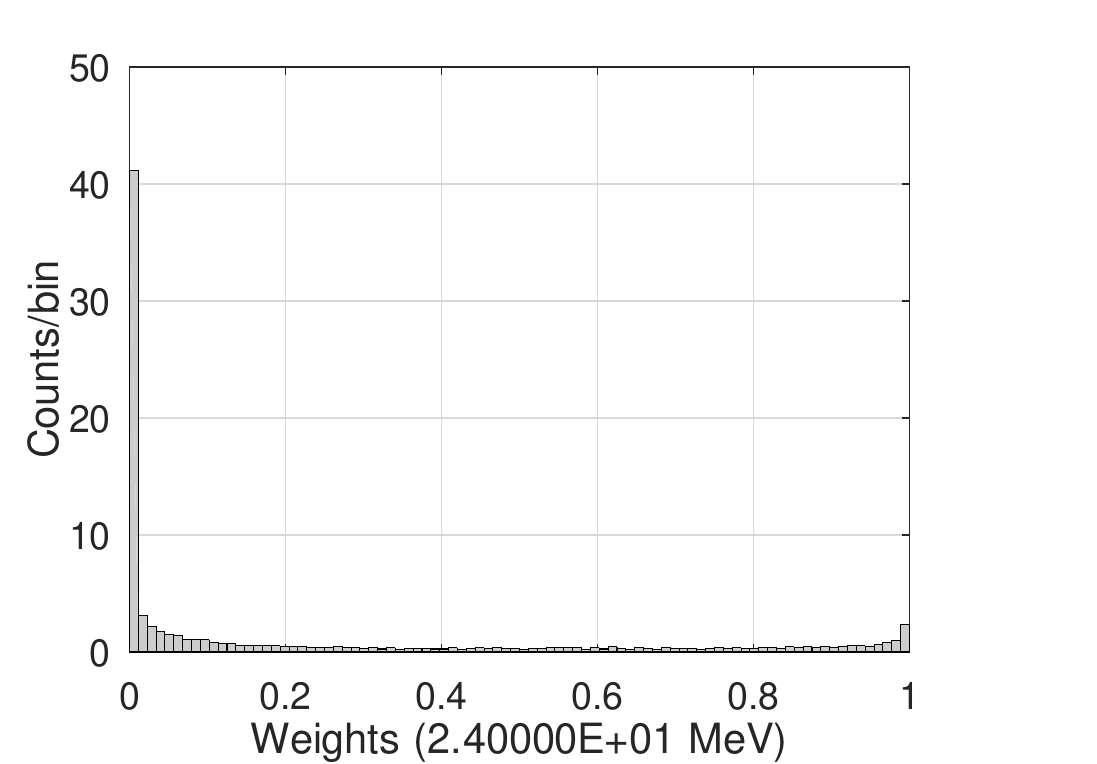}
  \includegraphics[trim = 30mm 82mm 5mm 82mm, clip, width=0.45\textwidth]{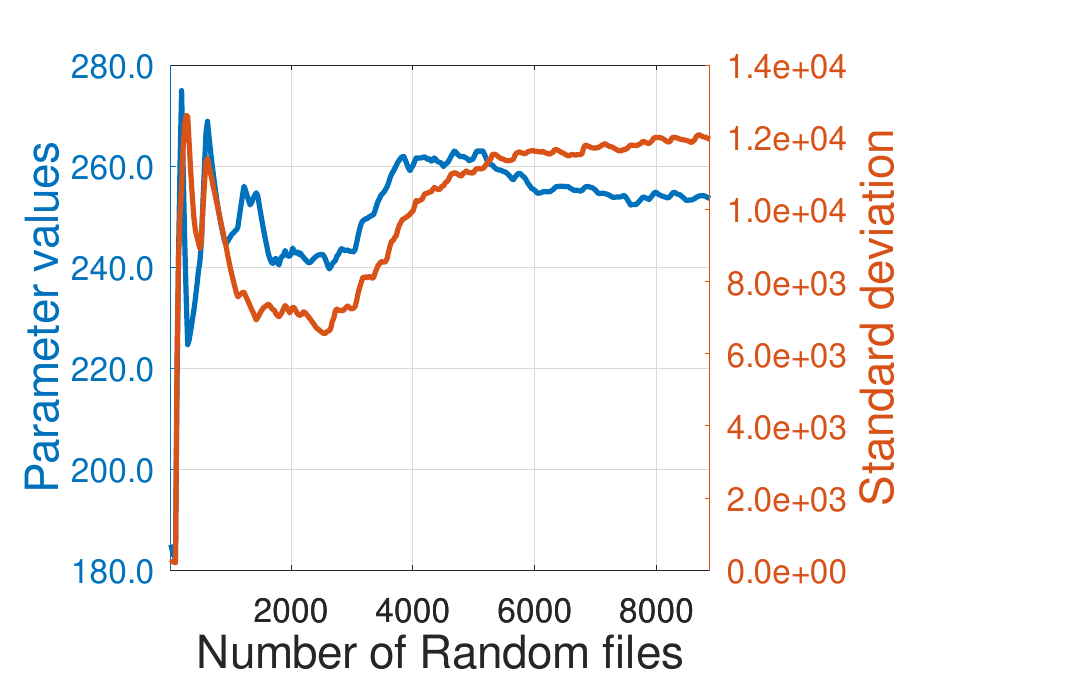}
   \caption{Prior (upper left), and prior and posterior (bottom) distributions, and distribution of file weights (lower left) for each random $^{58}Ni$(p,np) cross section value computed at 24 MeV. In the bottom right panel, a plot illustrating the convergence of the mean and $1\sigma$ standard deviation of the prior distribution is presented. Note: both model and their parameters were varied.}
   \label{MT003_prior_post_dist}
 \end{figure*}

To illustrate the practical implementation of the method, we consider a single experimental data point for the $^{58}$Ni(p,np) cross section at the incident energy, $i$ = 24 MeV. The prior distribution at this energy is made up of more than 9000 random cross section values (in mb) generated through the variation of numerous TALYS models and their parameters. Subsequently, we compared the calculated cross sections with the experimental data at this incident energy by computing a reduced $\chi^2$. The resulting file weights for each random file at the given energy are then combined with the prior distribution to obtain the posterior distribution, from which the updated mean and 1$\sigma$ standard deviations were extracted. In Fig.~\ref{MT003_prior_post_dist}, we present an illustrative example showcasing the prior (upper left), and prior and posterior (bottom) distributions, and distribution of file weights (lower left) for each random $^{58}Ni$(p,np) cross section value computed at 24 MeV. In the bottom right panel of the figure, a plot illustrating the convergence of the mean and $1\sigma$ standard deviation of the prior distribution is presented. We note here that, in the case presented in Fig.~\ref{MT003_prior_post_dist}, both models and their parameters were varied. The prior distribution reflects the cross section extracted from the cross section curves before the inclusion of experimental information while the posterior distribution represents the cross section distribution at 24 MeV after taken experimental data into consideration.

\begin{figure}[htb] %tb]
  \centering
   \includegraphics[trim = 5mm 20mm 5mm 5mm, clip, width=0.45\textwidth]{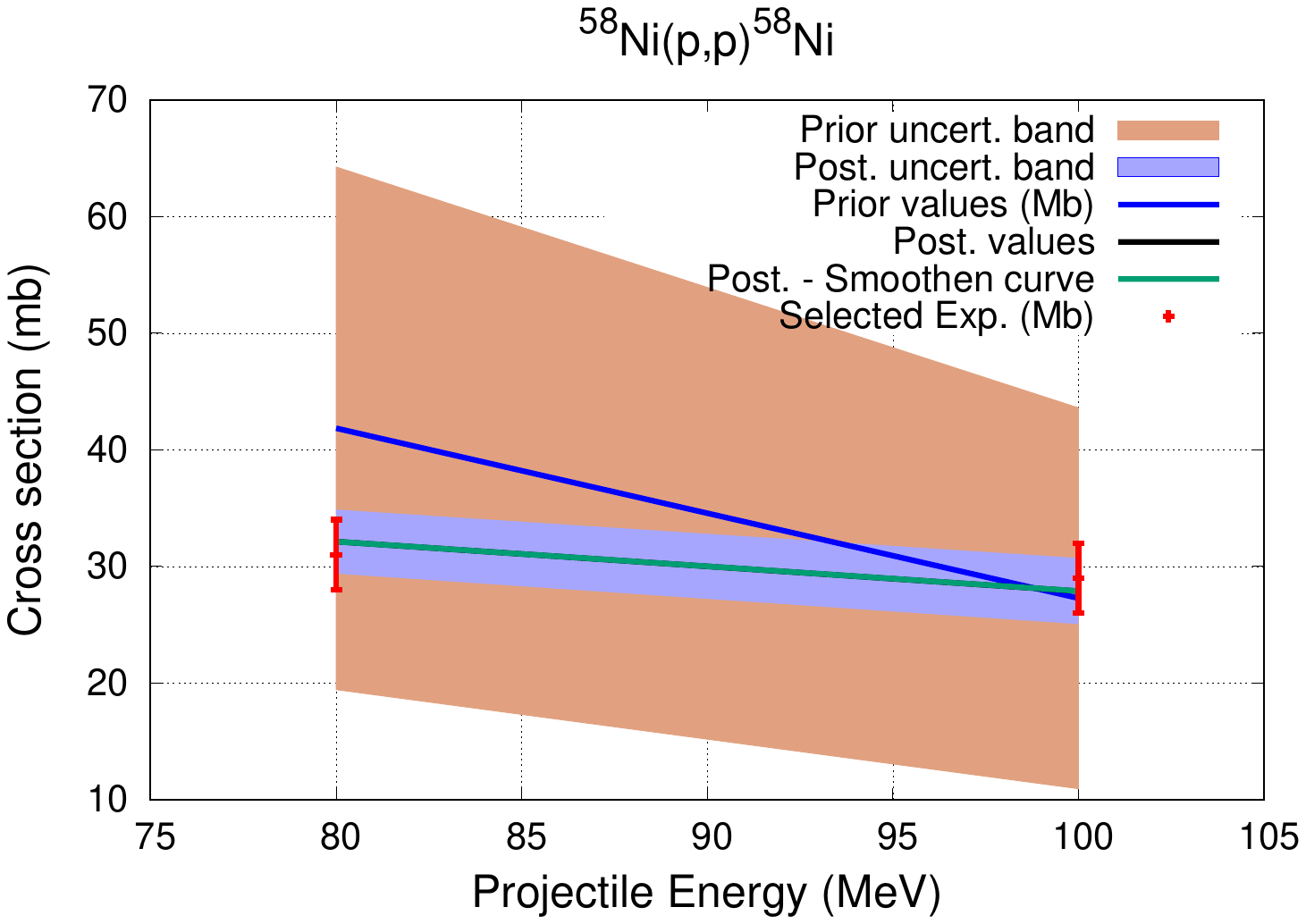}
   \caption{Application of the Bayesian Model Averaging (BMA) methodology with two experimental data points for the $^{58}$Ni(p,p)$^{58}$ cross section at 80 and 100 MeV.}
   \label{application_BMA_2pts}
 \end{figure}

From Fig.~\ref{MT003_prior_post_dist}, a relatively large prior can be observed as expected. This large prior then narrows around the mean of the experimental data for the posterior distribution. The large prior distribution observed from the figure can be attributed to the relatively large non-informative prior used. It can also be observed that the prior distribution is slightly skewed to the right of the distribution were larger cross section values were obtained. The skewed prior distribution was then shaped into a normal distribution after combining experimental data through the likelihood function (see upper right panel of Fig.~\ref{MT003_prior_post_dist}). Since the prior involve variations of many models and parameters, achieving convergence of the random cross sections at each incident energy can be difficult. From the convergence plot (bottom right of Fig.~\ref{MT003_prior_post_dist}), it can be observed that both the mean and the 1$\sigma$ standard deviation converges after about 7000 random samples. The final mean value of 252.58 Mb is observed to be close to the experimental cross section value of 247 Mb for the $^{58}Ni$(p,np) cross section at 24 MeV as expected. It is important to note that this value falls within the experimental uncertainty of $\pm$24.7. It can also be observed that there is a significant reduction in the 1$\sigma$ uncertainty of the prior uncertainty of 109 to a posterior uncertainty of 25. From the weight distribution presented in the figure, it can be observed that a considerable number of files were assigned with low and insignificant weights between 0 and 0.2. This is expected as the use of a large informative prior resulted in many cross section curves  to be positioned far from the experimental data and hence, resulted in large chi square values. In Fig.~\ref{application_BMA_2pts}, the BMA methodology is applied to two experimental data points of the $^{58}$Ni(p,p)$^{58}$ cross section at 80 and 100 MeV. From the figure, the prior values reflect the simple average over the models while the posterior values denote the BMA values. The prior uncertainty band is a combination of both model and parameter uncertainties.

% In order to predict in energy regions where no experimental data are available, a simple average over the model can 

\subsection{Treating `bad' models in the absence of experimental data}
\label{bad_models}
As previously discussed, the presence of `bad' models for the computation of the prior mean and in the case where no experimental data is available, can lead to significant distortions in the shape of the cross section or angular distribution curves as well as in the corresponding updated uncertainty bands. It is instructive to note however that the BMA approach (in the presence of experimental data) inherently handles the issue of `bad' models by assigning them lower weights compared to experimental data, thereby, minimizing their impact on the posterior distribution.

\begin{figure}[htb] %tb]
  \centering
  \includegraphics[trim = 15mm 22mm 5mm 12mm, clip, width=0.48\textwidth]{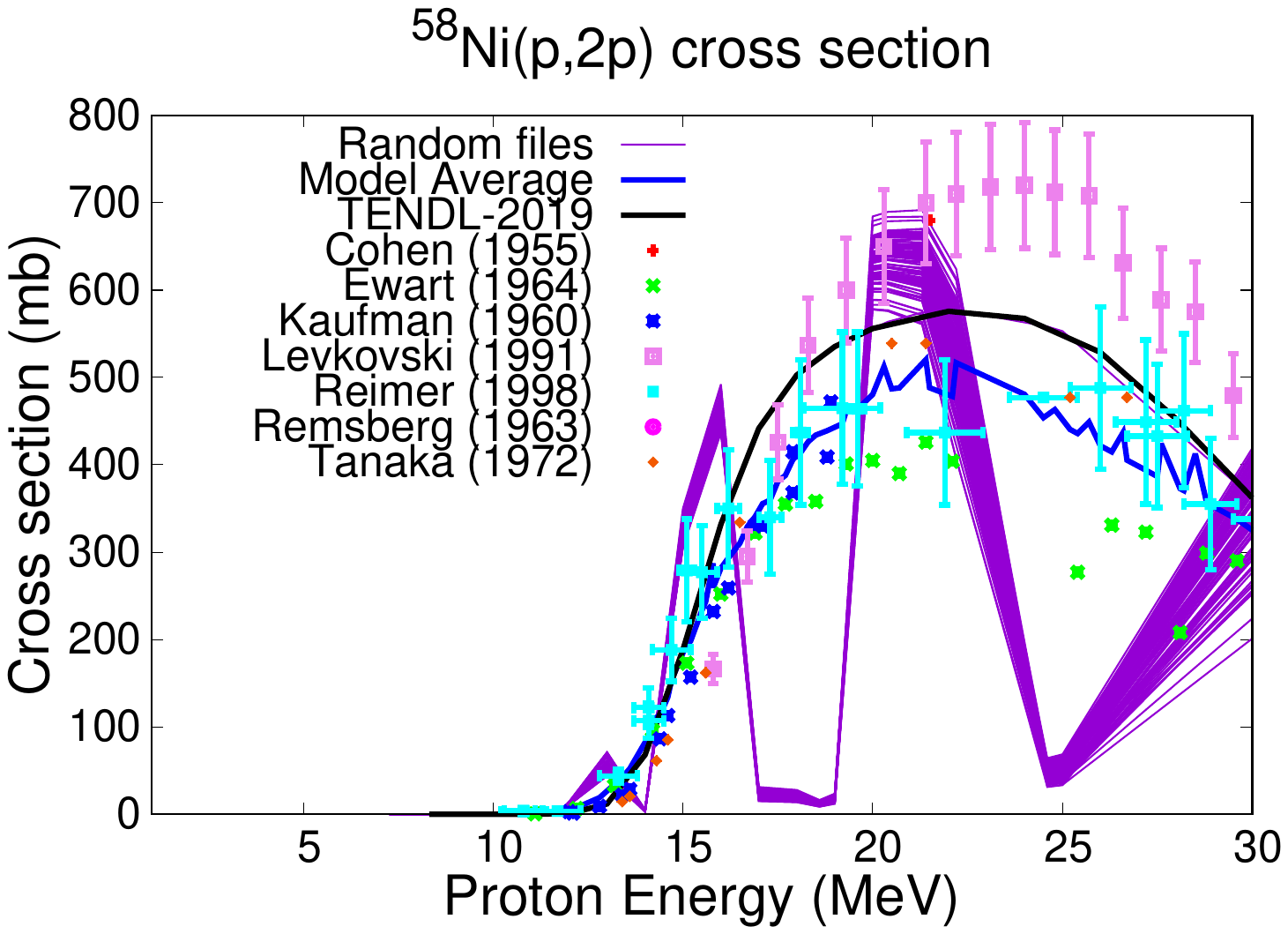}
   \caption{$^{58}Ni$(p,2p) cross section showing an example of highly distorted non-smooth curves produced with a model set labelled as a `bad' model combination.}
   \label{MT111_badmodels}
 \end{figure}

As a rule of the thumb, it is crucial to identify and exclude these `bad' models in BMA in the absence of experimental data. In Fig.~\ref{MT111_badmodels} for example, we present the $^{58}Ni$(p,2p) cross section, highlighting non-smooth curves in the energy range between 17 to 25 MeV. These curves look `unphysical' and hence, are assumed to have been generated with a `bad' model combination. The spread in the cross section curves is due to the variation of parameters around the `bad' model combination. The curves were compared against experimental data from EXFOR and the TENDL evaluation. The prior mean in the figure was calculated by averaging over all models at each energy point (see $^{58}Ni$(p,2p) cross section in Fig.\ref{MT003_modelparam}). It can be observed that the non-smooth cross sections deviated from the observed trend of the experimental data as well as the TENDL evaluation. Despite the non-smoothness in the prior mean curve, it was observed that it compared favorable agreement with some experimental data, particularly at threshold energies and at higher energies, especially with data from Reimer (1998) and Kaufman (1960). The non-smoothness in the prior mean curves is attributed to the inclusion of both `good' and `bad' models in the computation of the averaged cross section values. A potential solution to `bad' models involves implementing the Occam's razor as suggested in Ref.~\cite{bib:046a}. This approach eliminates models that globally perform poorly in their prediction of experimental data for the considered channels. 
The non-smooth cross section curves as depicted in Fig.~\ref{MT111_badmodels}, were produced with the following model combination among other models in the TALYS code~\cite{bib:33}: 
\begin{enumerate}
    \item \textbf{level density model 6:} Microscopic level densities (temperature dependent Hartree-Fock-Bogolyubov (HFB), Gogny force) from Hilaire’s combinatorial tables; 
    \item  \textbf{gamma strength function model 5:} Goriely’s hybrid model; 
    \item \textbf{width fluctuation correction model 2:}  Hofmann-Richert-Tepel-Weidenm\"{u}ller (HRTW); 
    \item \textbf{pre-equilibrium model 4:} Multi-step direct/compound model; 
    \item \textbf{Jeukenne-Lejeune-Mahaux (JLM) model 0:} standard Jeukenne-Lejeune-Mahaux (JLM) imaginary potential; 
    \item  \textbf{mass model 3:} HFB-Gogny D1M table;
    \item Other default models.
\end{enumerate}

This specific model combination was therefore treated as a `bad' model and consequently, excluded from subsequent analyses. It is noteworthy to point out that in the case of the optical model, the Jeukenne-Lejeune-Mahaux (JLM) semi-microscopic model was utilized instead of the default local and global parameterisations of Koning and Delaroche which are typically used as the default optical model parameterisations in TALYS for non-actinides (as applicable in this case). Additionally, it's essential to recognize here that certain model combinations may not exert a significant impact or sensitivity to cross sections or angular distributions of interest. For instance, it has been observed in Ref.~\cite{bib:1aa} that the use of the HRTW model instead of the Moldauer model has no noticeable impact on proton-induced reaction cross sections for p+$^{59}Co$ between 1 and 100 MeV. 

For consistency in the evaluated files, it is important that each complete evaluation obeys the cross section sum rules. However, in the case of BMA in the absence of experimental data as proposed in this work, we may encounter situations where the sum rules are not obeyed. In this work, as suggested in Ref.~\cite{bib:2t}, our approach is to reassign the excess cross sections to "less important channels". By "less important channels", we refer to those channels with relatively small cross sections, and whose products hold minimal significance for the scientific community.

\section{Results}
Table~\ref{modelparam_uncert_MT028} presents $1\sigma$ model and parameter uncertainties for selected incident energies ranging from 15 to 30 MeV in the case of the $^{58}$Ni(p,np) cross section. From the table, relatively smaller parameter uncertainties can be observed from threshold to about 18 MeV. This observation can be made also from Fig.~\ref{MT003_modelparam} where a narrow spread was observed in the low energy region in the case of only parameter variation. In contrast, the model uncertainties exhibit relatively larger uncertainties across the entire considered energy region, ranging from 46.44 to 108.12 as can be seen in Table~\ref{modelparam_uncert_MT028}. 
% 
% This is expected since it is well known that TALYS has difficulty in reproducing experimental data at the threshold energies. Also, the parameter uncertainties are observed to be generally lower than the model uncertainties as expected. 

\begin{table}[h!]
 \centering
  \caption{1$\sigma$ model and parameter uncertainties for selected incident energies for the $^{58}$Ni(p,np) cross section. The model uncertainties were extracted using Eq.~\ref{modelUncert}.}
  \label{modelparam_uncert_MT028}
  \begin{tabular}{cccc}  % Use 'c' to center the column, 'l' to have it left adjusted and 'r' for right adjusted
  \toprule
   \pbox{20cm}{ Incident energy \\ (MeV)}  &  \pbox{20cm}{Total \\ uncertainty (1$\sigma$)} &  \pbox{20cm}{Model \\ uncertainty (1$\sigma$)} &  \pbox{20cm}{Parameter \\ uncertainty (1$\sigma$)} \\
   \midrule
15.7  & 46.5  & 46.44  & 2.5 \\
16.0  & 52.9  & 52.84  & 2.9 \\
16.2  & 54.4  & 54.27  & 3.0 \\
16.8  & 62.5  & 62.35  & 3.7 \\
17.1  & 66.1  & 66.00  & 4.1 \\
17.3  & 66.9  & 66.81  & 4.3 \\
17.7  & 72.0  & 71.86  & 4.8 \\
17.9  & 76.0  & 75.87  & 5.1 \\
18.2  & 80.9  & 80.72  & 5.5 \\
18.4  & 83.9  & 83.73  & 5.9 \\
19.0  & 90.3  & 90.05  & 7.0 \\
19.1  & 87.9  & 87.57  & 7.2 \\
19.3  & 85.1  & 84.76  & 7.7 \\
19.5  & 85.4  & 85.01  & 8.3 \\
20.0  & 98.7  & 98.18  & 9.9 \\
20.3  & 99.6  & 99.05  & 10.6 \\
20.5  & 101.2  & 100.64 & 11.0 \\
20.9  & 104.1  & 103.44 & 11.9 \\ 
21.0  & 105.4  & 104.75 & 12.1 \\ 
21.2  & 108.8  & 108.06 & 12.5 \\
21.4  & 108.9  & 108.12 & 12.8 \\ 
21.5  & 107.2  & 106.40 & 12.9 \\
22.1  & 101.5  & 100.61 & 13.7 \\
22.6  & 97.9   & 96.93  & 13.9 \\
23.4  & 99.4   & 98.42  & 14.1 \\
24.0  & 107.6  & 106.67 & 14.3 \\
24.5  & 118.2  & 117.28 & 14.5 \\
25.3  & 120.1  & 119.27 & 14.4 \\
25.8  & 109.4  & 108.43 & 14.2 \\
26.3  & 103.2  & 102.27 & 14.0 \\
27.0  & 97.8   & 96.85  & 13.8 \\
27.5  & 96.6   & 95.59  & 13.7 \\
27.9  & 97.2   & 96.23  & 13.6 \\
28.7  & 102.6  & 101.74 & 13.5 \\
29.1  & 107.2  & 106.38 & 13.5 \\
\bottomrule 
\end{tabular}
\end{table}

A similar table showing the model and parameter uncertainties extracted for energies from 9 to 60 MeV in the case of the $^{58}$Ni(p,non-el) cross section, in Table~\ref{modelparam_uncert_MT003}. From the table, it can be observed that the parameter uncertainties are generally lower than the model uncertainties as expected, ranging from 23.6 to 59.8 representing respectively. These values represent 3.4\% to 7.4\% of the cross section at the considered energies. In contrast, as expected, the model uncertainties are generally large ranging from 19.56\% to 26.40\% of the cross section. An observed trend in the table is the increase in both model and parameter uncertainties with increasing energy. Additionally, both the model and parameter spreads are narrow in the lower energy regions, widening as the energy increases.

% These large uncertainties were however able to overlap all of the experimental data available as expected.  

\begin{table}[htb]
 \centering
  \caption{1$\sigma$ model and parameter uncertainties for selected incident energies for the $^{58}$Ni(p,non-el) cross section. The model uncertainty was computed using Eq.~\ref{modelUncert}.}
  \label{modelparam_uncert_MT003}
  \begin{tabular}{cccc}  % Use 'c' to center the column, 'l' to have it left adjusted and 'r' for right adjusted
  \toprule
   \pbox{20cm}{ Incident energy \\ (MeV)}  &  \pbox{20cm}{Total \\ uncertainty (1$\sigma$)} &  \pbox{20cm}{Model \\ uncertainty (1$\sigma$)} &  \pbox{20cm}{Parameter \\ uncertainty (1$\sigma$)} \\
   \midrule
9.14 & 137.5 & 135.42 & 23.6 \\
22.70 & 182.9 & 178.95 & 38.0 \\
25.10 & 188.0 & 183.28 & 41.9 \\
30.00 & 197.4 & 192.08 & 45.3 \\
30.10 & 197.5 & 192.24 & 45.4 \\
34.80 & 205.0 & 199.02 & 49.3 \\
39.70 & 211.2 & 204.47 & 53.0 \\
40.00 & 211.5 & 204.74 & 53.2 \\
45.20 & 216.0 & 208.66 & 55.9 \\
47.90 & 217.5 & 209.93 & 56.9 \\ 
49.50 & 218.5 & 210.77 & 57.5 \\
60.80 & 221.3 & 213.02 & 59.8 \\
\bottomrule
\end{tabular}
\end{table}

\begin{figure*}[htb] %tb]
  \centering
  \includegraphics[trim = 0mm 20mm 0mm 10mm, clip, width=0.48\textwidth]{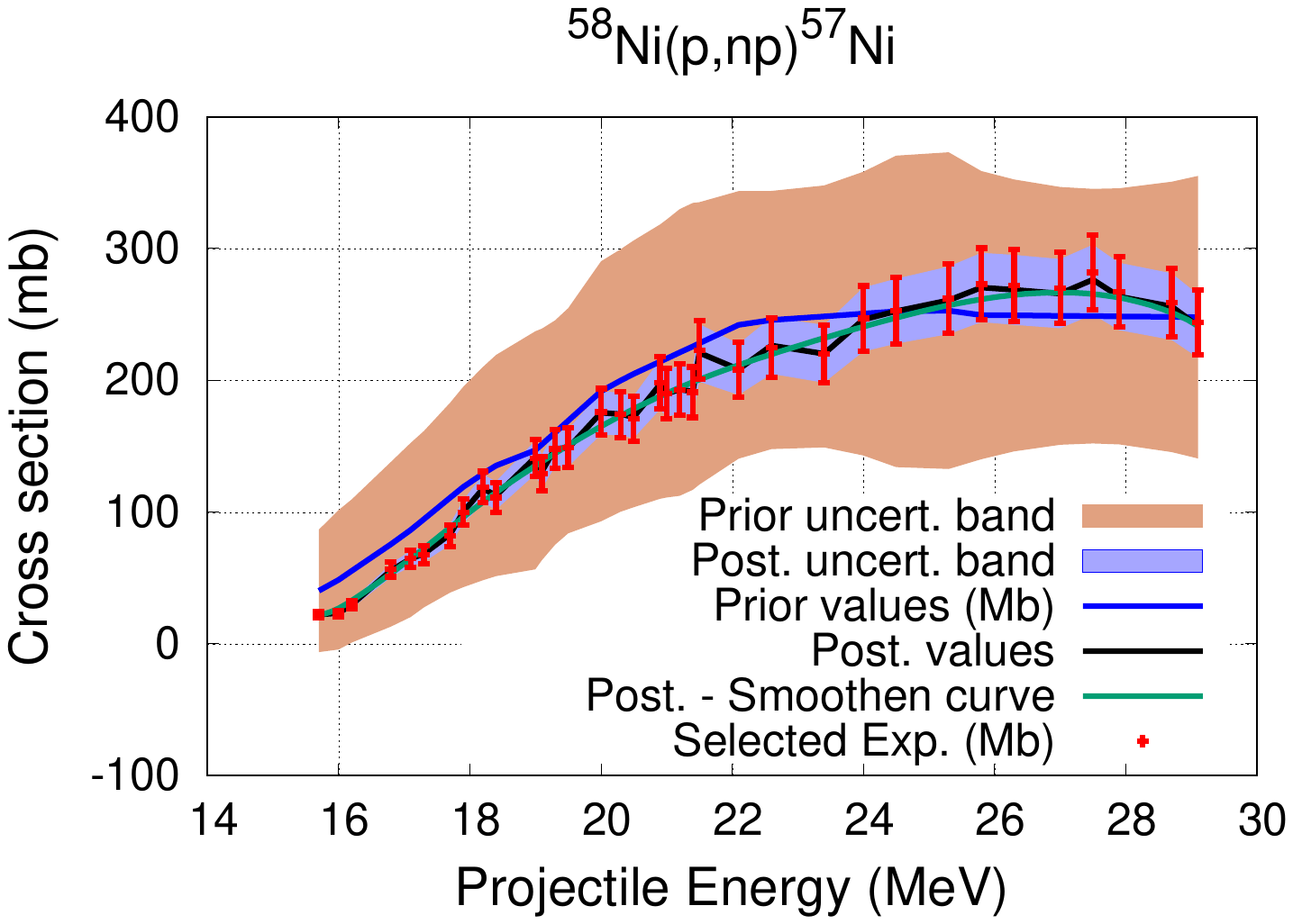}
  \includegraphics[trim = 0mm 20mm 0mm 10mm, clip, width=0.48\textwidth]{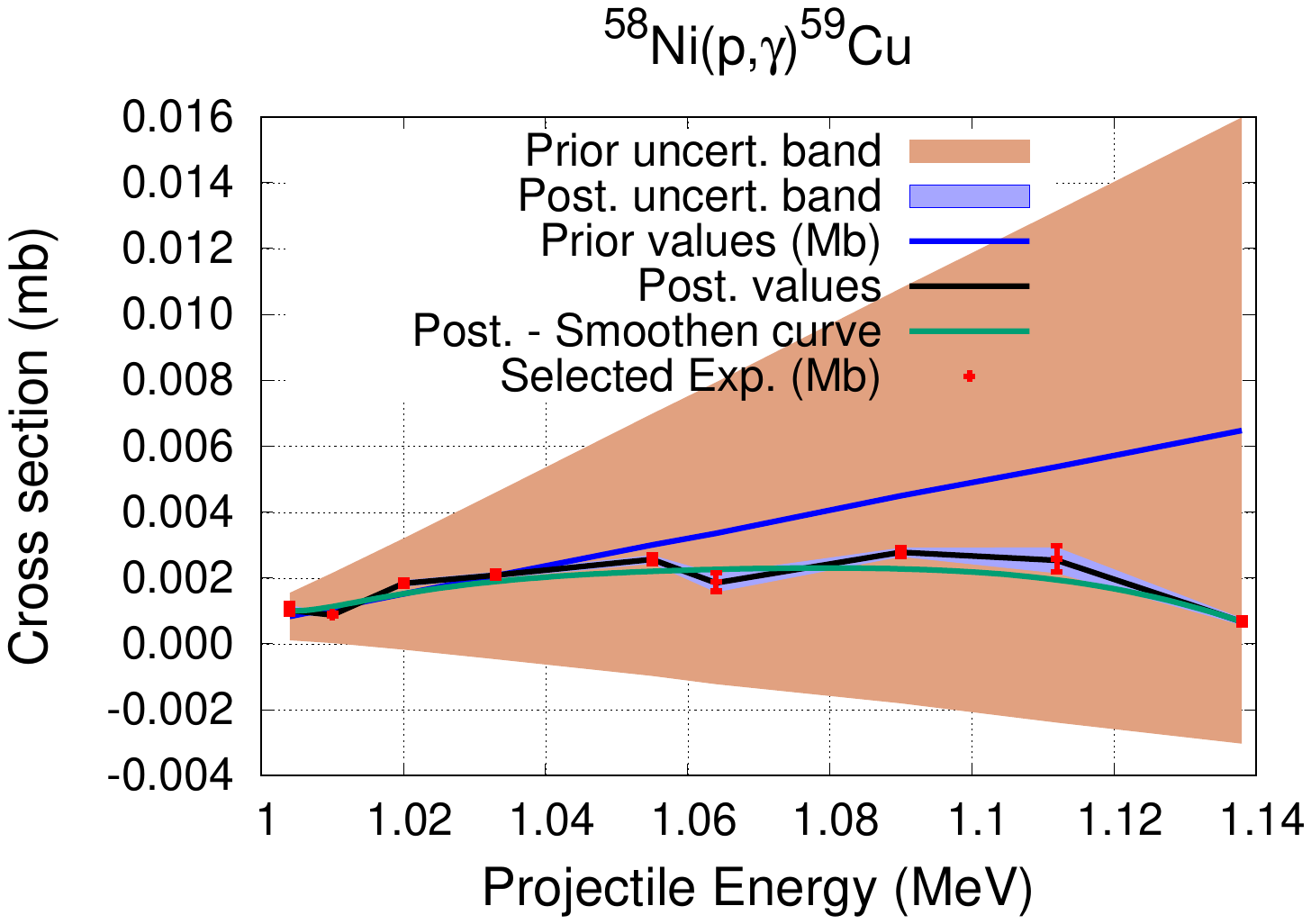}
  \includegraphics[trim = 0mm 20mm 0mm 10mm, clip, width=0.48\textwidth]{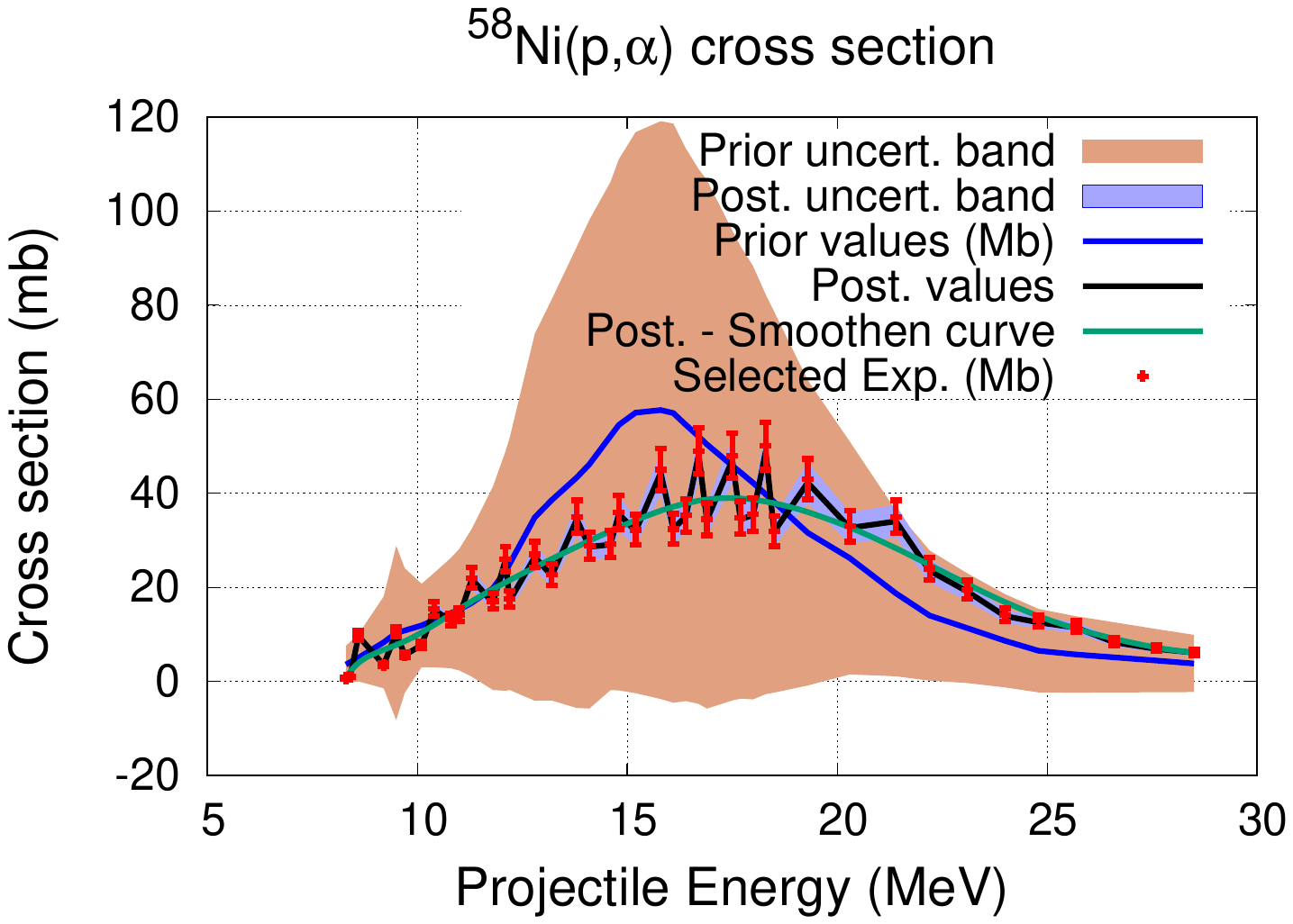}
  \includegraphics[trim = 0mm 20mm 0mm 10mm, clip, width=0.48\textwidth]{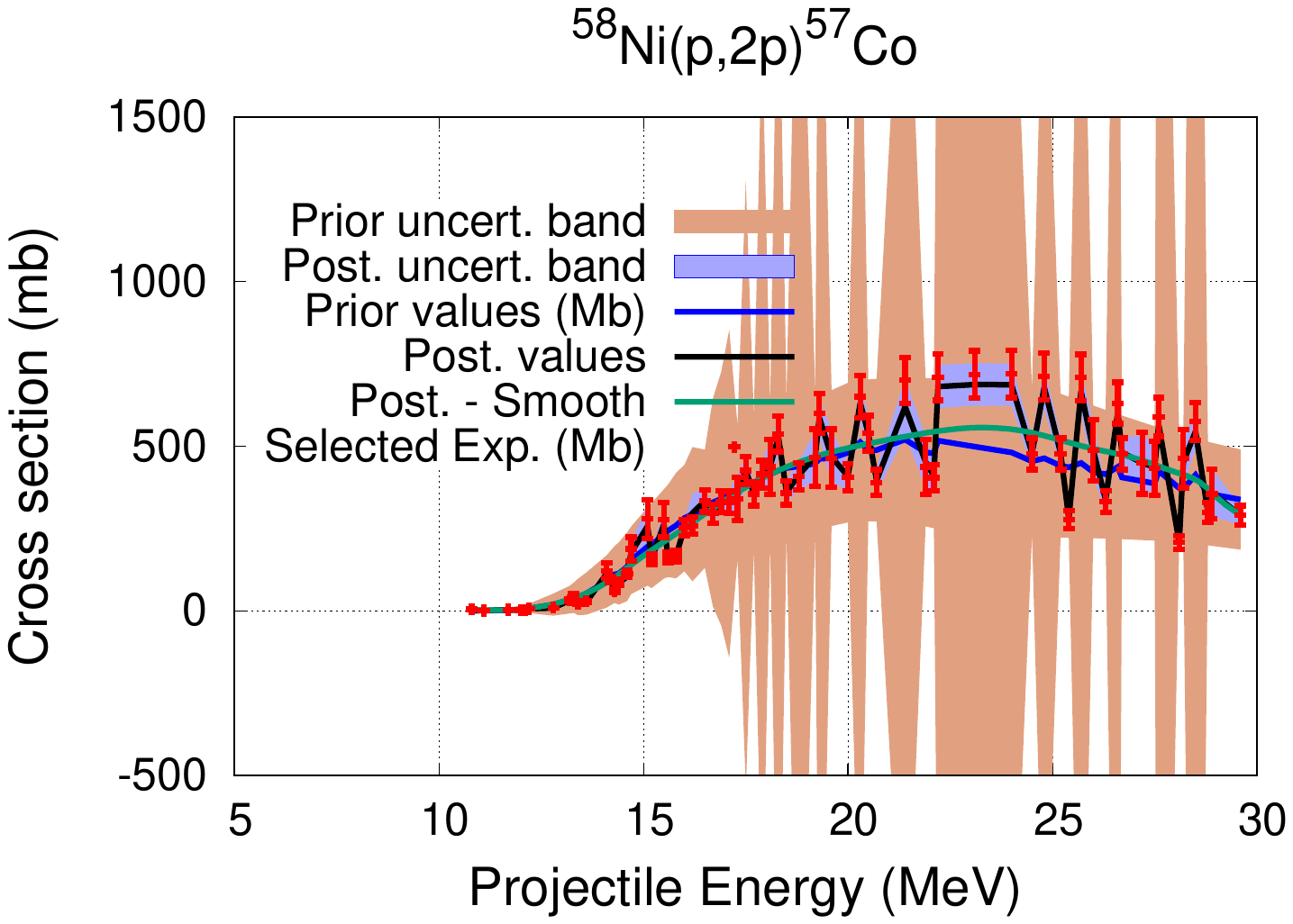}
  \caption{Prior and posterior means with their corresponding uncertainties compared with experimental data for the following cross sections: $^{58}$Ni(p,np)$^{57}$Ni, $^{58}$Ni(p, $\gamma$), (p,$\alpha$) $^{58}$Ni(p,2p). Note: the prior and posterior uncertainty bands are $\pm$1$\sigma$ uncertainty.}
  \label{xs_bma1}
 \end{figure*}

 \begin{figure}[t] %tb]
  \centering
  \includegraphics[trim = 0mm 020mm 0mm 10mm, clip, width=0.48\textwidth]{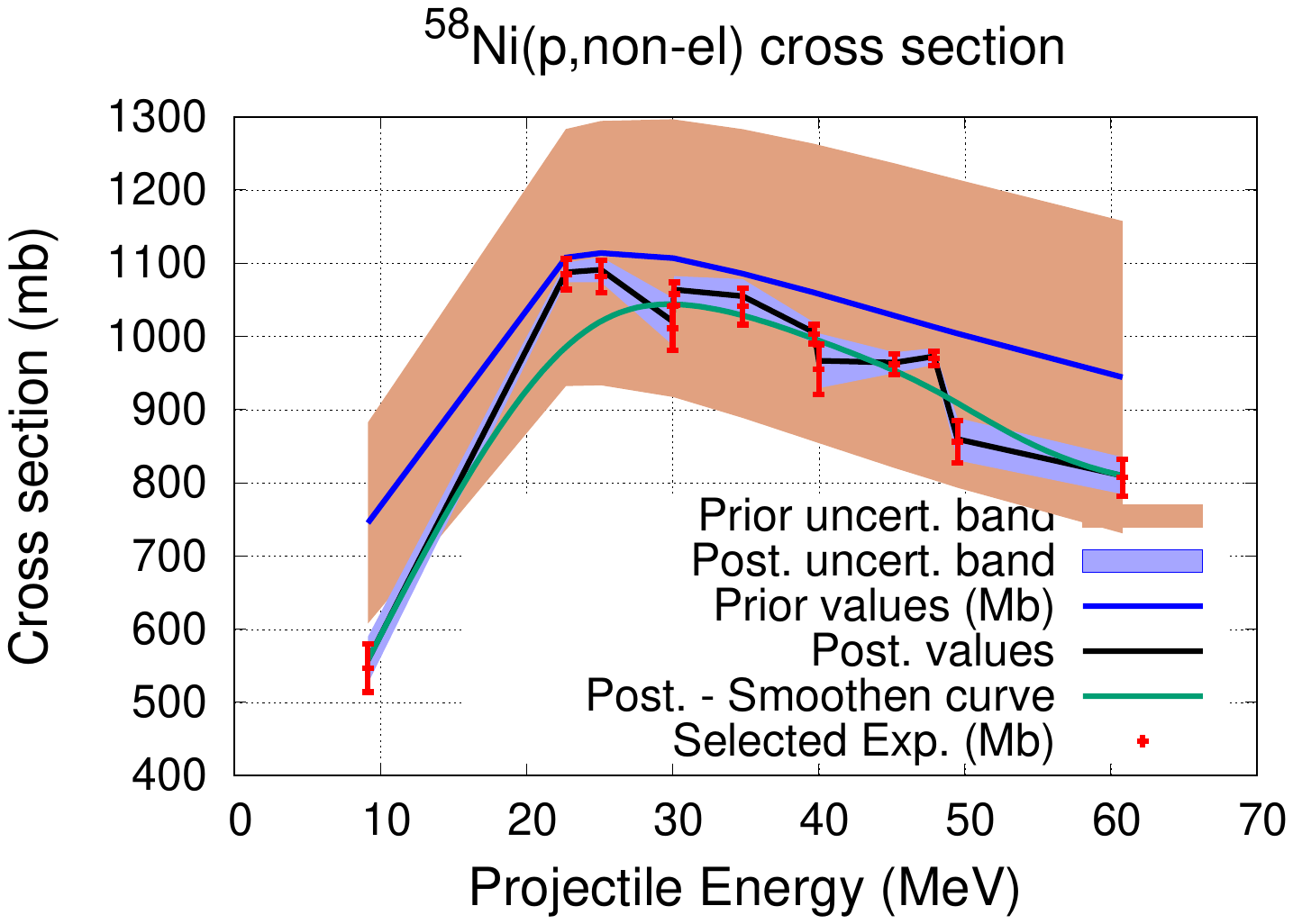}
  \caption{Prior and posterior means with their corresponding uncertainties compared with experimental data for the $^{58}$Ni(p,non-el) cross section. Note: the prior and posterior uncertainty bands are $\pm$1$\sigma$ uncertainty and hence only 68.27\% of the random cross section values lie within this estimate.}
  \label{bma_MT003_1}
 \end{figure}

 \begin{figure*}[t] %tb]
  \centering
  \includegraphics[trim = 0mm 20mm 0mm 10mm, clip, width=0.4\textwidth]{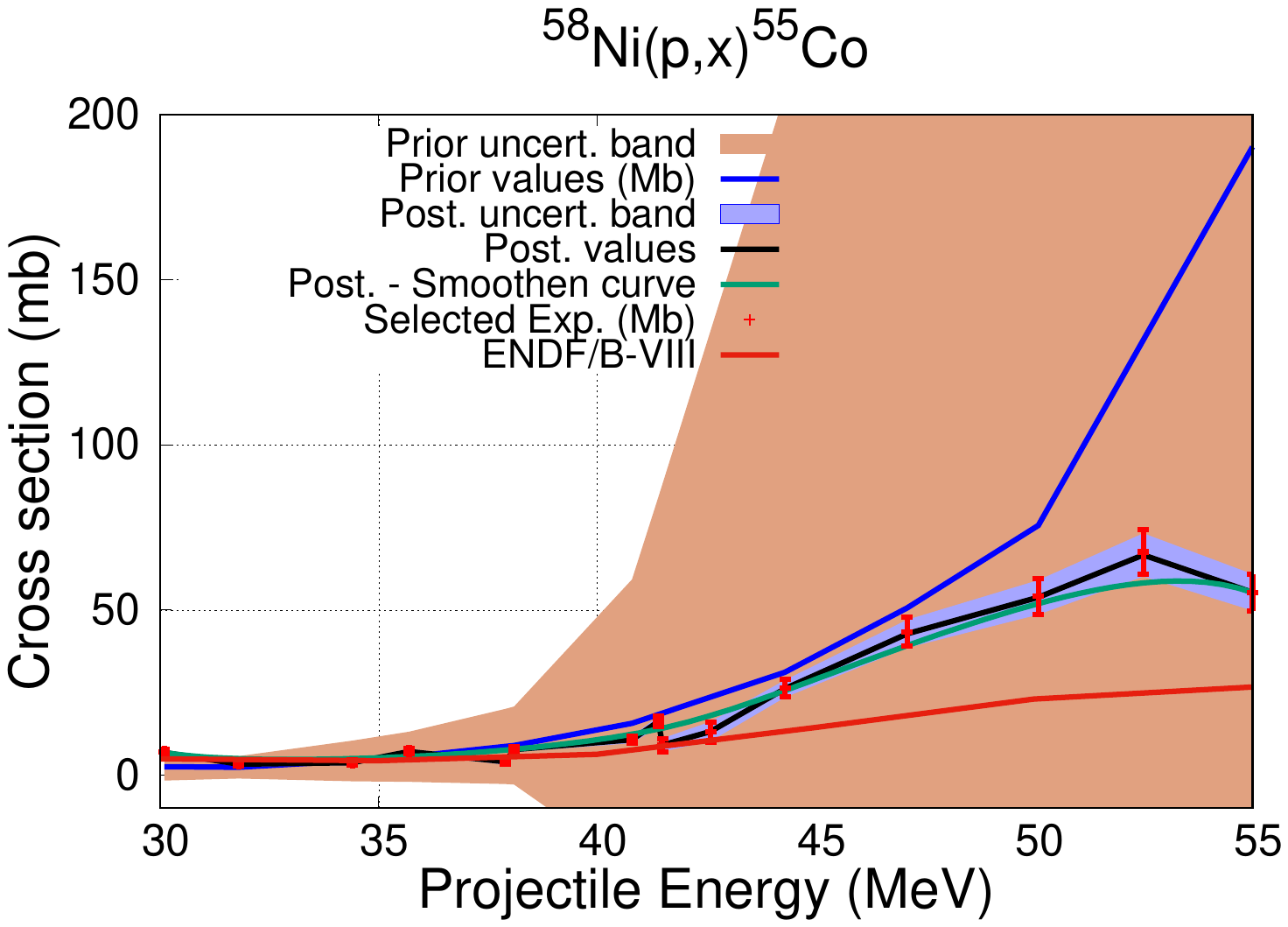}
  \includegraphics[trim = 0mm 20mm 0mm 10mm, clip, width=0.4\textwidth]{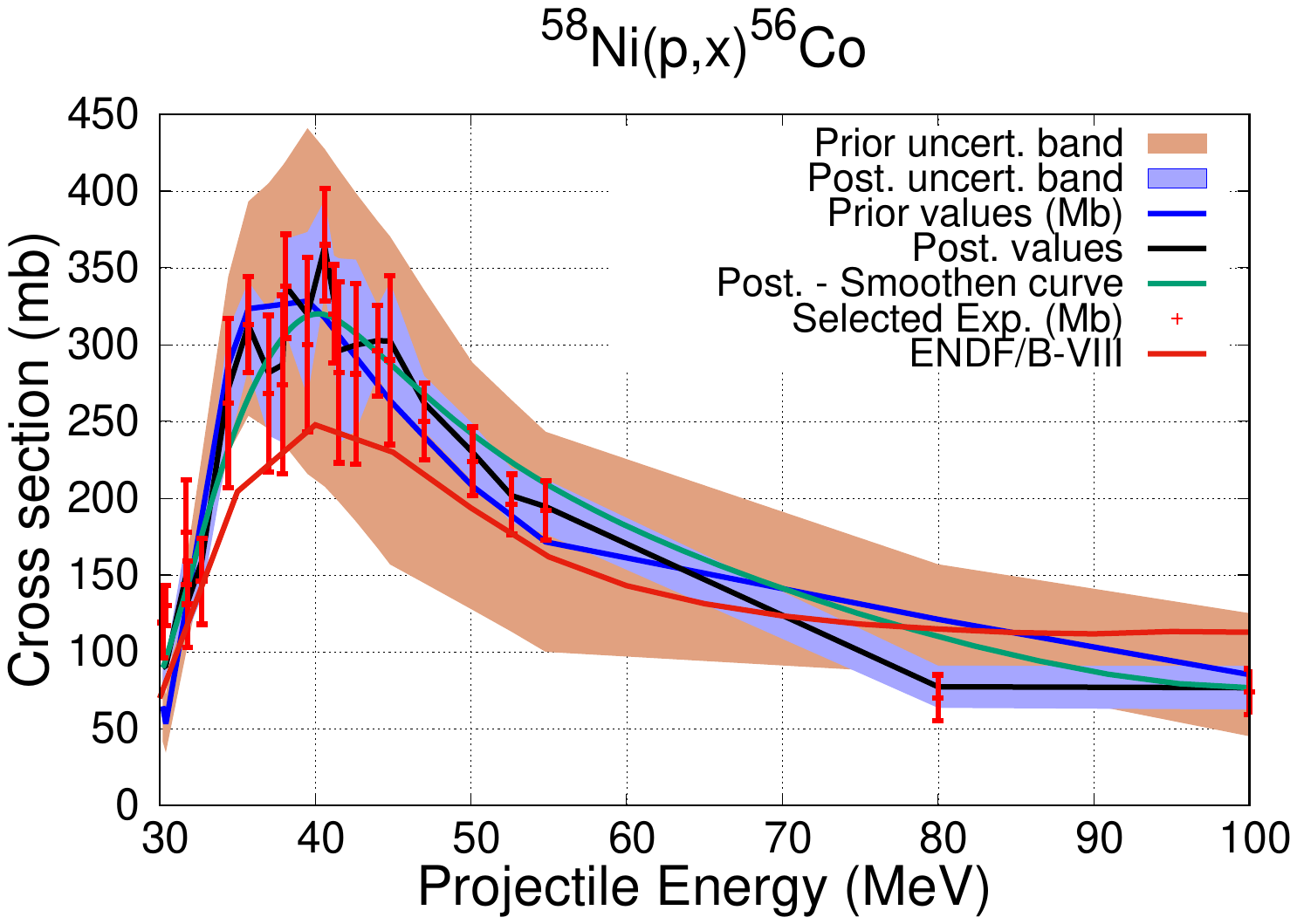}
  \includegraphics[trim = 0mm 20mm 0mm 10mm, clip, width=0.4\textwidth]{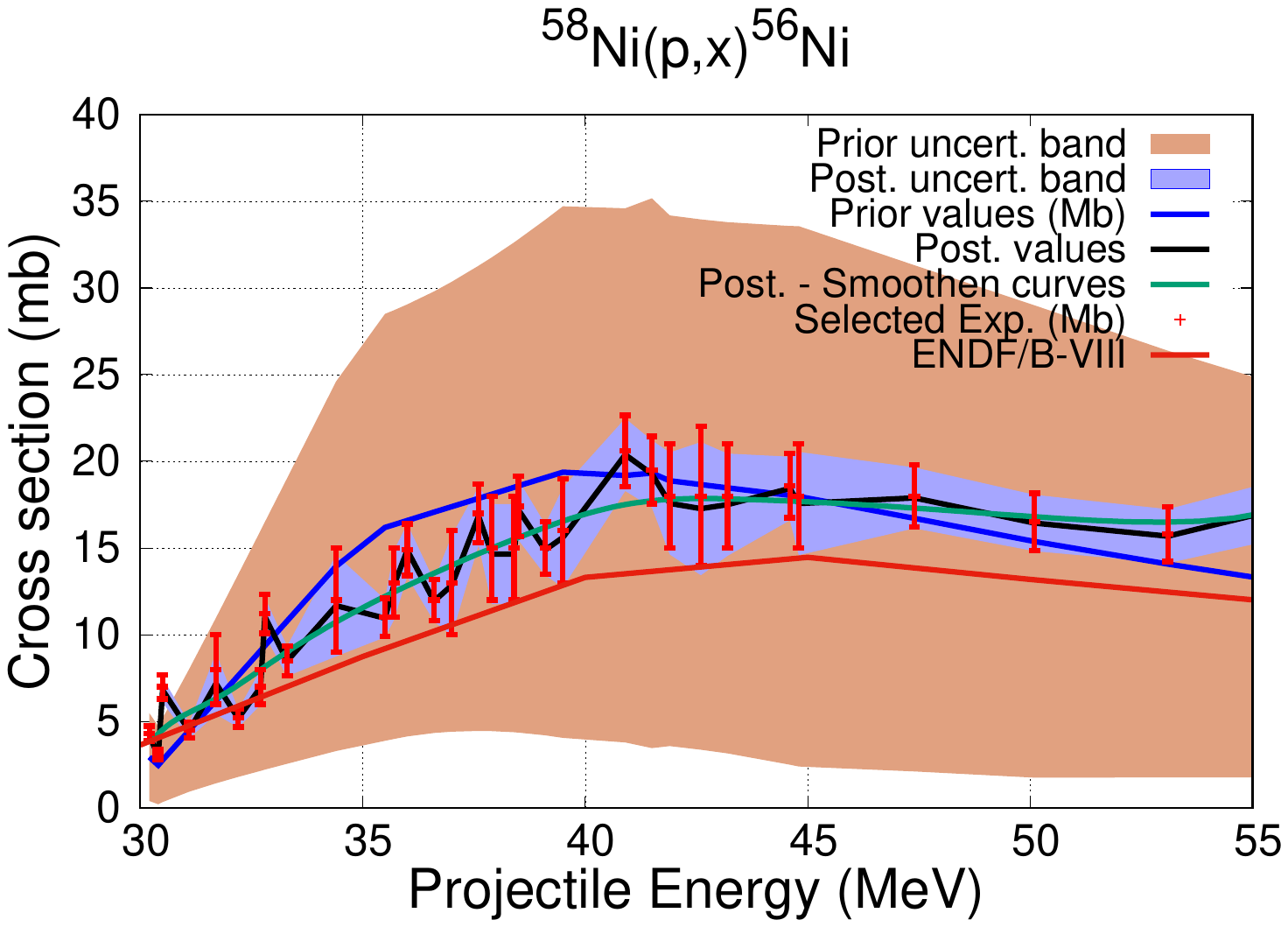}
  \includegraphics[trim = 0mm 20mm 0mm 10mm, clip, width=0.4\textwidth]{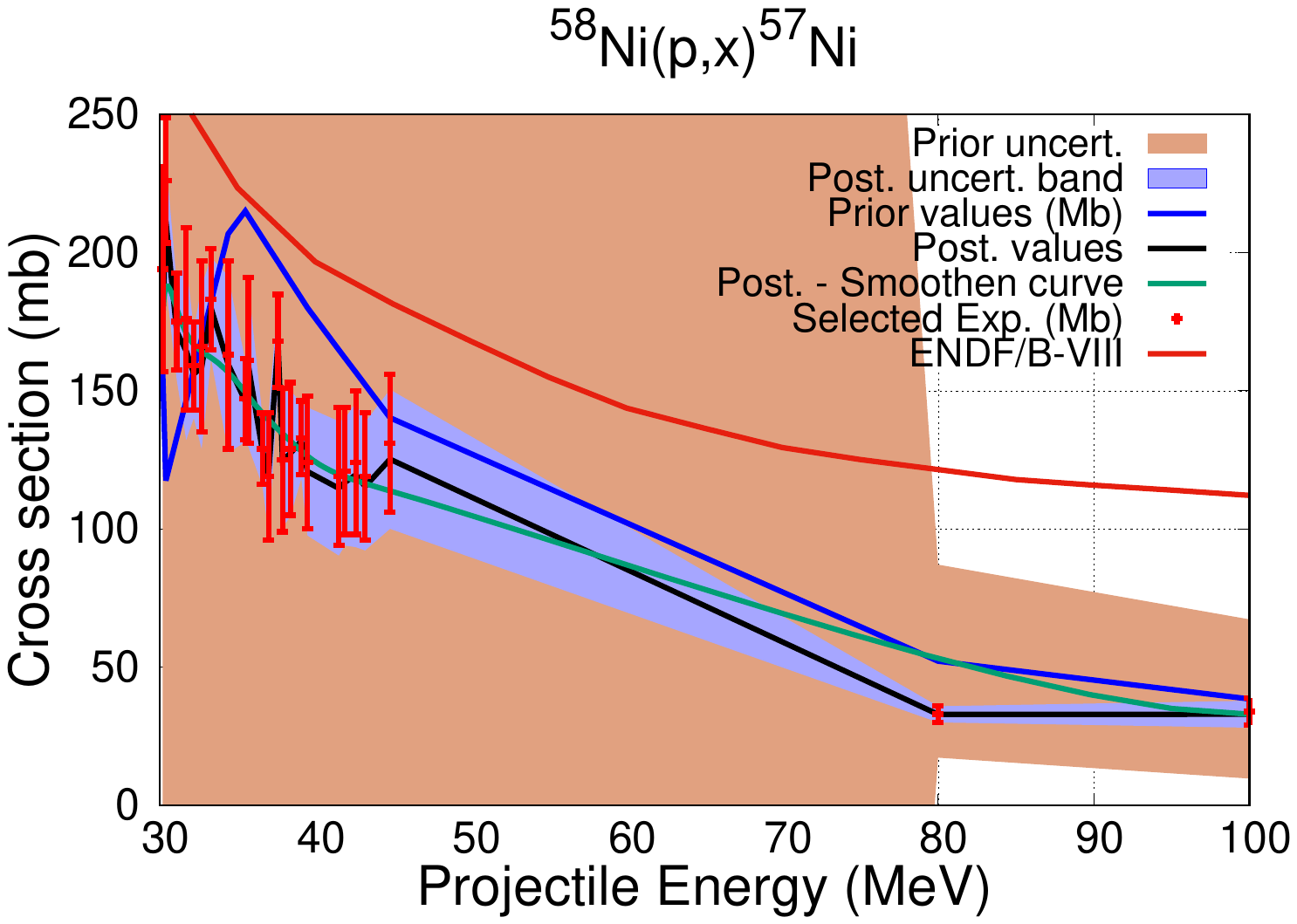}
  \caption{Prior and posterior means with their corresponding uncertainties compared with experimental data for the following residual production cross sections: $^{58}$Ni(p,x)$^{55}$Co, $^{58}$Ni(p,x)$^{56}$Co, $^{58}$Ni(p,x)$^{56}$Ni $^{58}$Ni(p,x)$^{57}$Ni. Note: the prior and posterior uncertainty bands are $\pm$1$\sigma$ uncertainty.}
  \label{bma_prodxs}
  \end{figure*}

  \begin{figure*}[htb!] %tb]
  \centering
  \includegraphics[trim = 10mm 20mm 10mm 20mm, clip, width=0.85\textwidth]{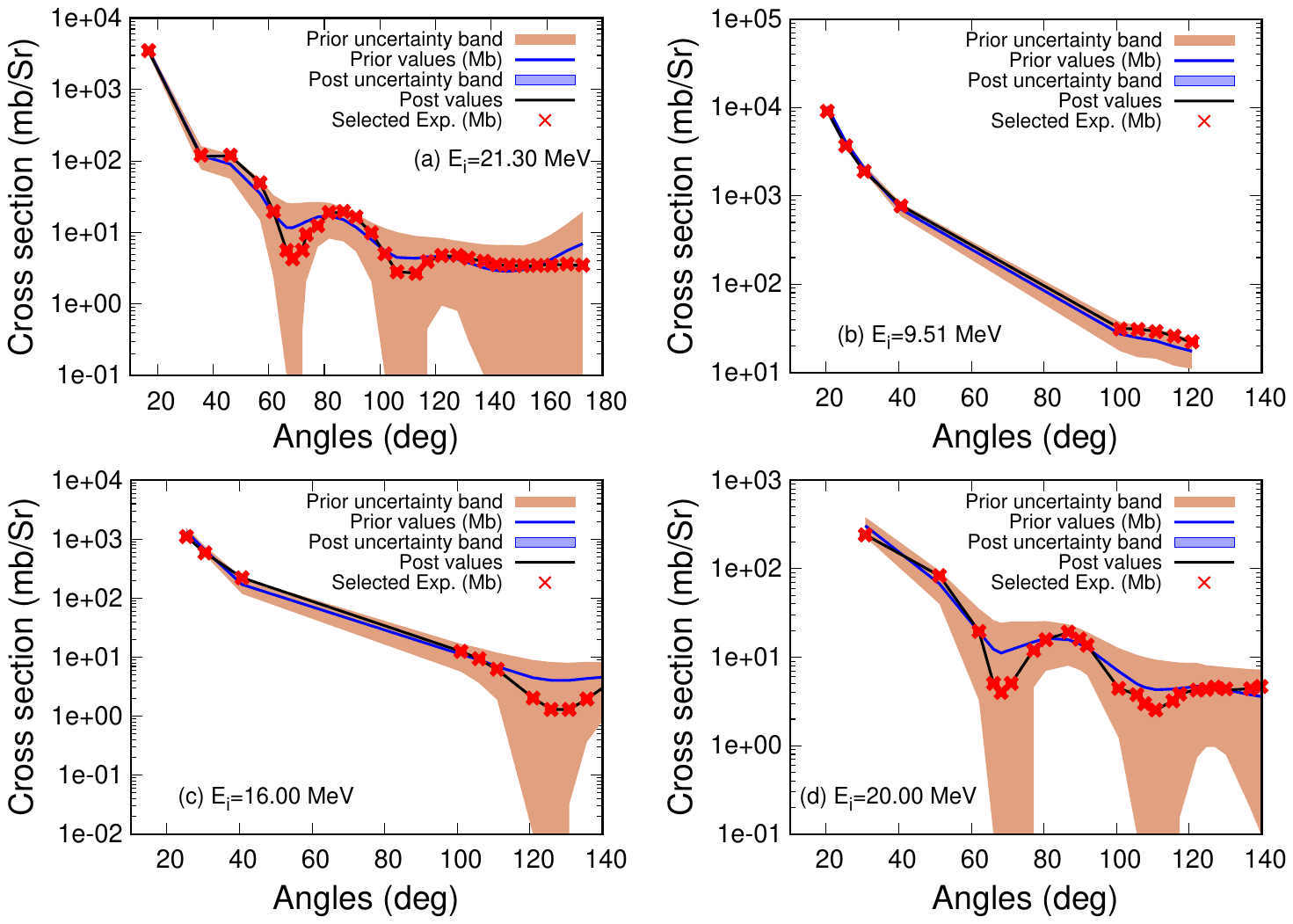} 
  \caption{Prior and posterior means with their corresponding uncertainties for the elastic angular distributions at selected incident energies: (a) 21.30 MeV, (b) 9.51 MeV, (c) 16.0 MeV and (d) 20.0 MeV for p+$^{58}$Ni.}
  \label{file_performance_da1}
   \end{figure*} 
 
  \begin{figure*}[htb] %tb]
  \centering
  \includegraphics[trim = 10mm 20mm 10mm 20mm, clip, width=0.85\textwidth]{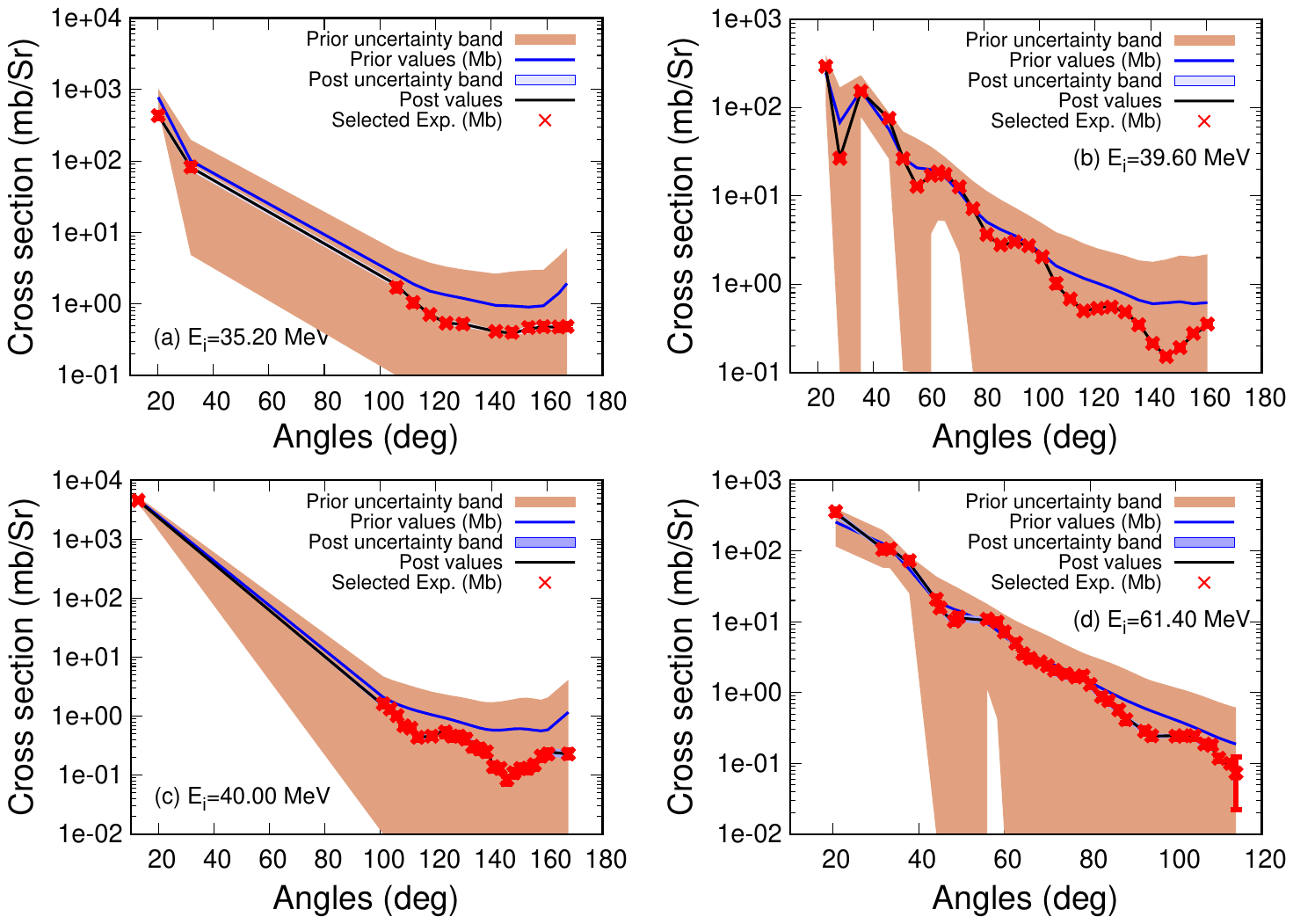} 
  \caption{Prior and posterior means with their corresponding uncertainties for the elastic angular distributions at selected incident energies: (a) 35.20 MeV, (b) 39.60 MeV, (c) 40.0 MeV and (d) 61.40 MeV for p+$^{58}$Ni.}
  \label{file_performance_da2}
  \end{figure*}

  \begin{figure*}%[htb] %tb]
  \centering
  \includegraphics[trim = 20mm 22mm 0mm 10mm, clip, width=0.45\textwidth]{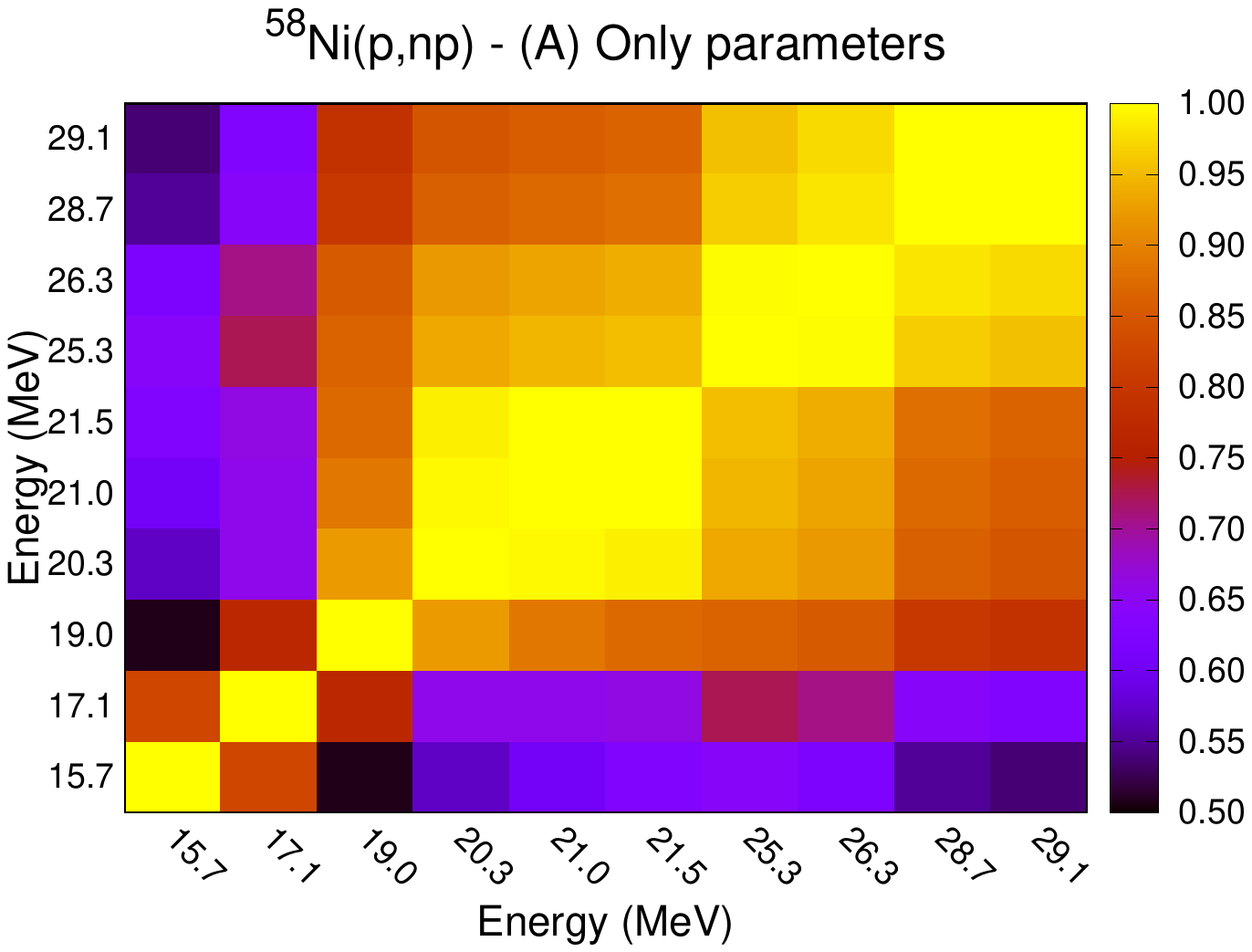}
  \includegraphics[trim = 20mm 22mm 0mm 10mm, clip, width=0.45\textwidth]{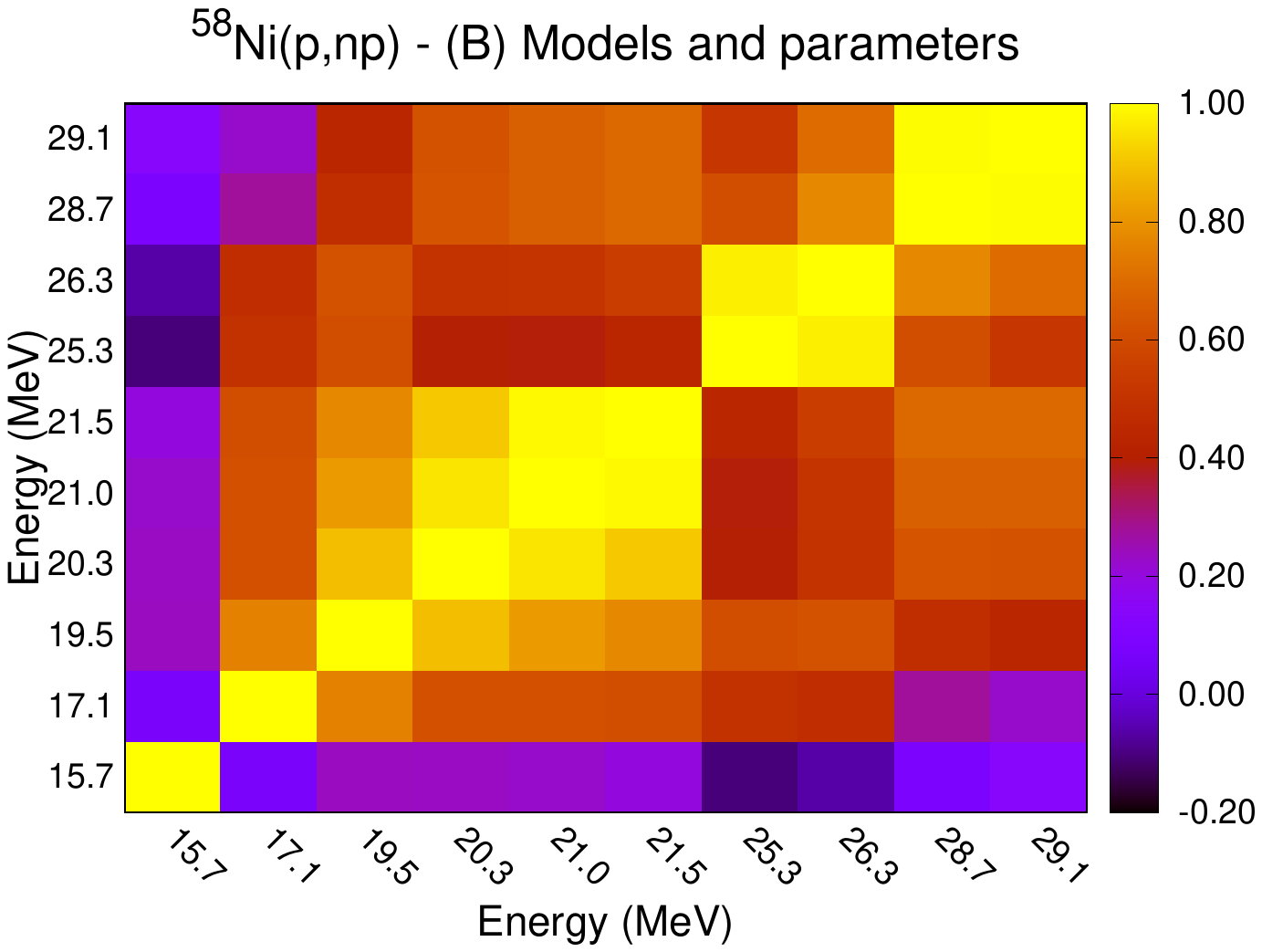}
  \caption{Example of $^{58}$Ni(p,np) correlations based on (A) the variation of only model parameters and (B) based on the variations of both models and their parameters.}
  \label{prior_corr_matrxMT028_Ni058_Gen}
  \end{figure*}

  \begin{figure*}%[htb] %tb]
  \centering
  \includegraphics[trim = 20mm 22mm 0mm 10mm, clip, width=0.45\textwidth]{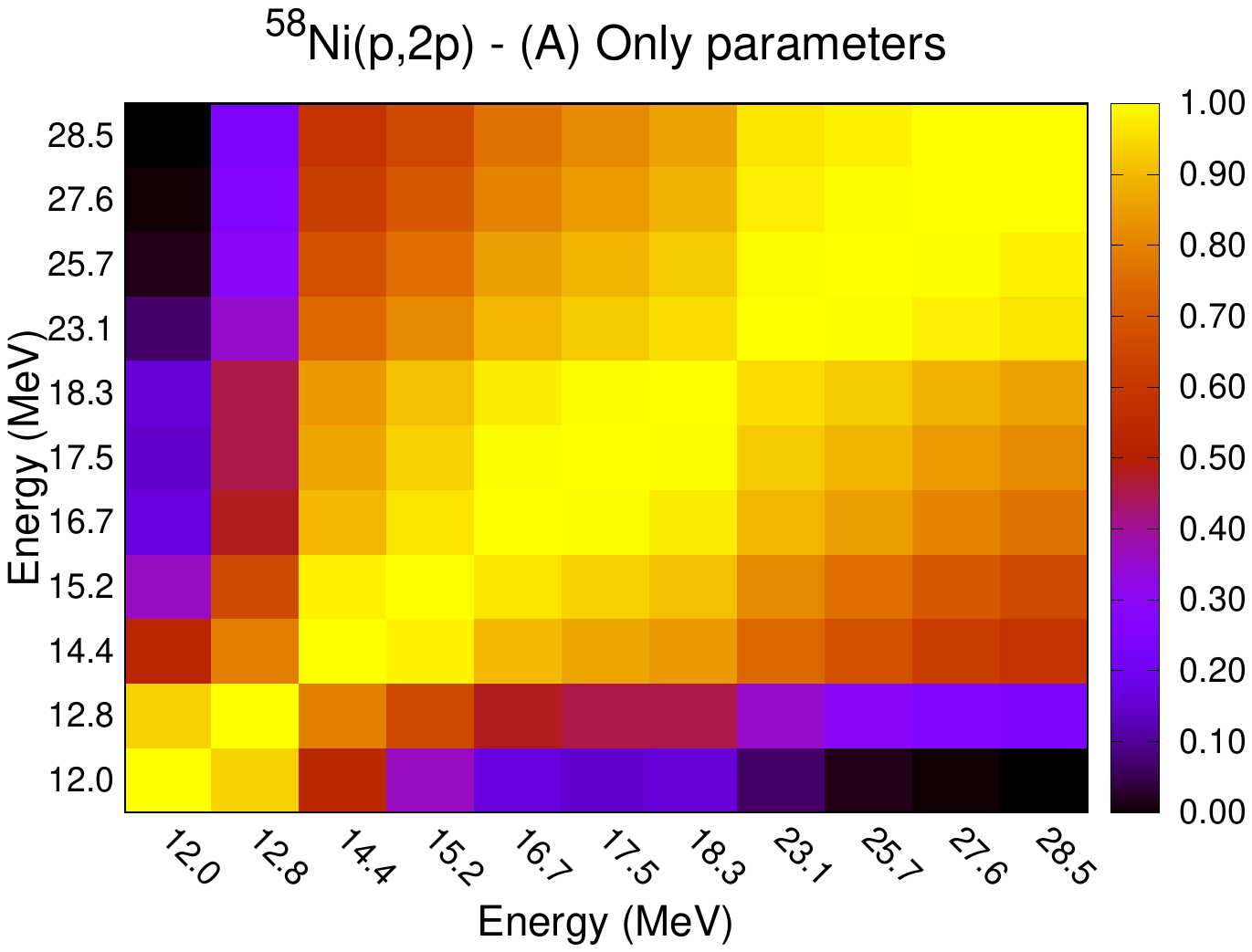}
  \includegraphics[trim = 20mm 22mm 0mm 10mm, clip, width=0.45\textwidth]{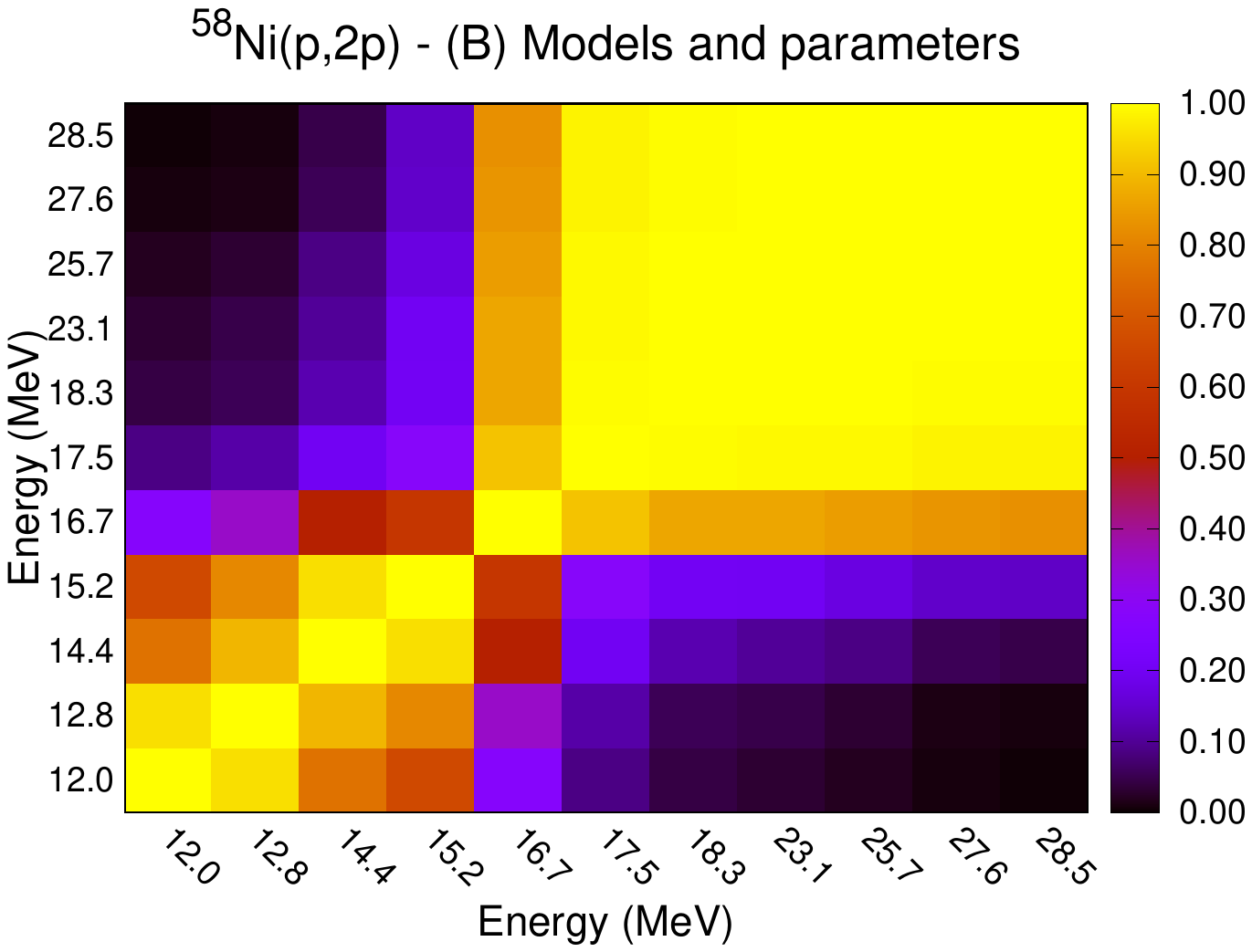}
  \caption{Example of $^{58}$Ni(p,2p) correlations based on (A) the variation of only model parameters and (B) based on the variations of both models and their parameters.}
  \label{prior_corr_matrxMT111_Ni058_Gen}
  \end{figure*} 
  
  \begin{figure*}%[htb] %tb]
  \centering
  \includegraphics[trim = 20mm 22mm 0mm 10mm, clip, width=0.45\textwidth]{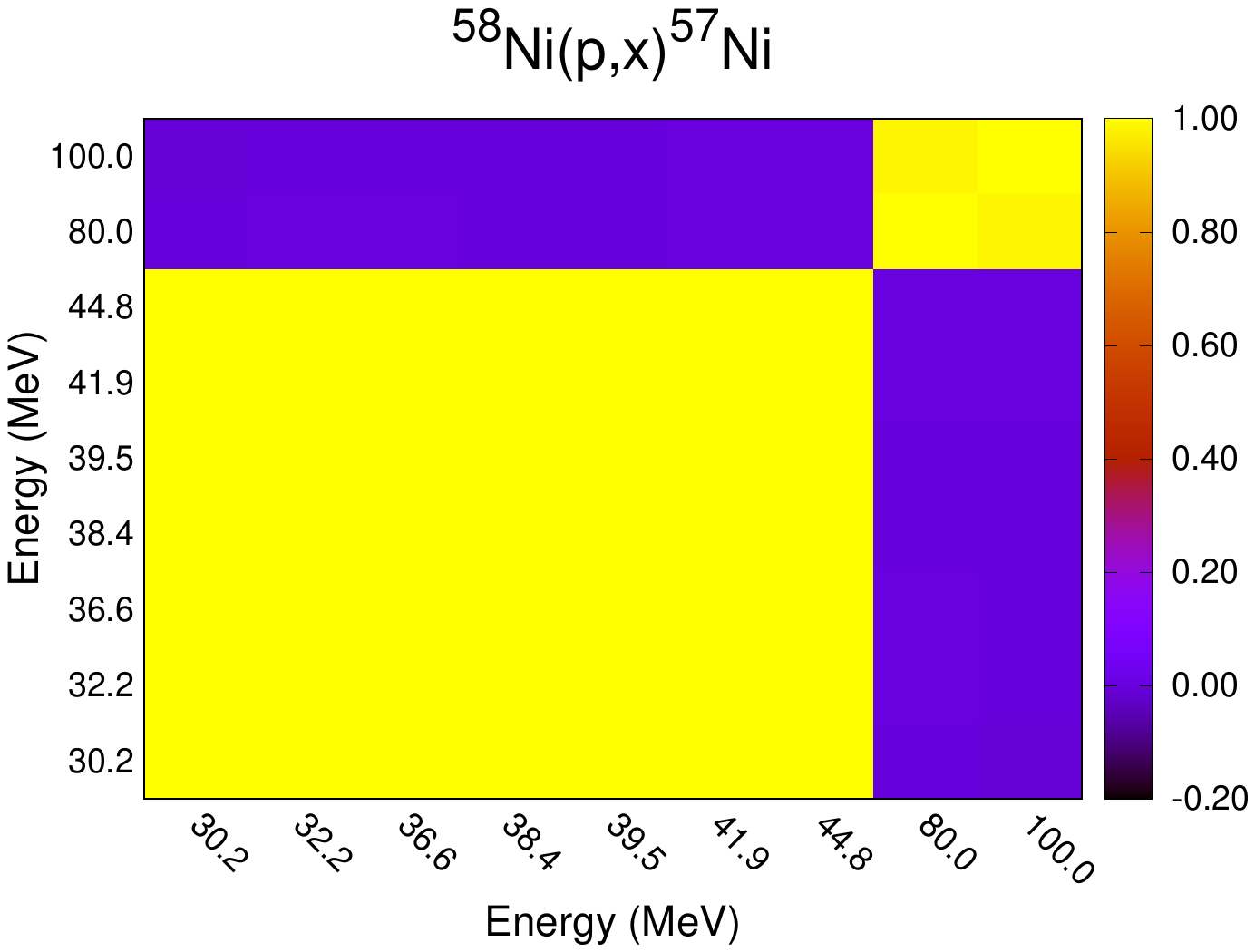}
  \includegraphics[trim = 20mm 22mm 0mm 10mm, clip, width=0.45\textwidth]{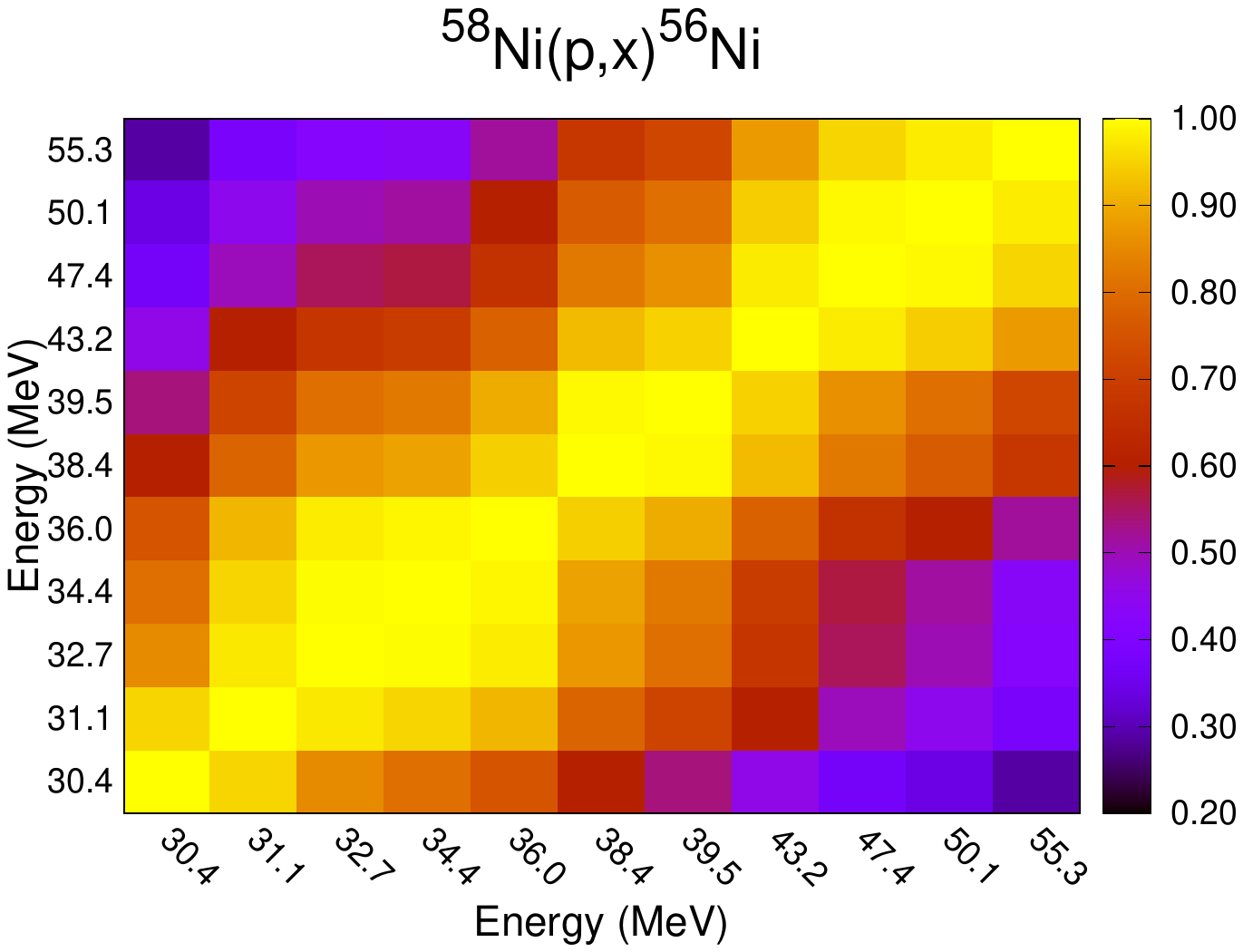}
  \caption{Example of correlations computed for the following residual production cross sections: $^{58}$Ni(p,x)$^{57}$Ni (left) and $^{58}$Ni(p,x)$^{56}$Ni (right) based on the simultaneous variations of many models and their parameters.}
  \label{prior_corr_matrxMT028057_Ni058_Gen}
  \end{figure*}

In Fig.~\ref{xs_bma1}, we present the Bayesian Model Averaging (BMA) results for a selected cross sections: $^{58}$Ni(p,np)$^{57}$Ni, $^{58}$Ni(p, $\gamma$), (p,$\alpha$) $^{58}$Ni(p,2p) illustrating the prior and posterior means along with their corresponding $\pm$1$\sigma$ uncertainties. For the (p,$\alpha$) and (p,$\gamma$) cross sections, the prior means appear off within certain energy ranges where the model uncertainties are large. This is however different for the (p,np) cross section where the uncertainties appear relatively constant over the considered energy region, with the prior mean comparing favorably with the experimental data. The posterior means for all channels presented consistently reproduce experimental data; however, localized adjustments at each energy point as carried out in this work introduces kinks, particularly noticeable in the $^{58}$Ni(p,non-el) cross section. These kinks can be attributed to the imperfections in our experimental data and the absence of experimental correlation considerations in the computation of the reduced chi-square. In this work, spline interpolation was employed to smooth out these kinks. It can be observed that generally smaller posterior uncertainties were obtained after experimental data were taken into account. This can be attributed to a number of reasons. First, this gives an indication that perhaps, the model vectors used have similar performance with respect to the experimental data and hence resulted in similar weights being assigned to each model vector. Additionally, if one of the model vectors (and parameters as well) is supported by strong evidence in the data, this could result in this model set being assigned with large weights compared to the other model vectors. Consequently, this would result in a reduction of the uncertainty associated with the choice of the model (and/or parameters) leading to a small posterior spread. Furthermore, since a large uninformed prior was used in this work, the posterior distribution was dominated by the experimental data through the likelihood function and hence, the impact of the prior was less insignificant. It should also be noted that since similar nuclear reaction models were used, the model predictions would be correlated and hence could result in similar weights. In such cases, the uncertainty in model selection diminishes, resulting in smaller posterior spreads. Although posterior uncertainties are generally small and, in some instances, smaller than experimental uncertainties, it important to note that only 1$\sigma$ uncertainties are reported here. For example, in Fig.\ref{bma_MT003_1}, the experimental data at 10 MeV appears outside the prior uncertainty band of the models. However, if the prior uncertainty were extended to 3$\sigma$, the experimental data would fall within the expected uncertainty band as expected (see $^{58}$Ni(p,non-el) cross section in Fig.~\ref{MT003_modelparam}). Visible kinks in the posterior curve, as seen in Fig.\ref{bma_MT003_1}, necessitate the use of a smoothing function. It's crucial to acknowledge that the smoothing process might skip some experimental data points, as observed in Fig.\ref{bma_MT003_1}. The posterior uncertainties, which took the differential data into account, are small, and the weighted average cross sections are in close agreement with experimental data. 

In Fig.~\ref{bma_prodxs}, we compare the prior and posterior means, along with their corresponding uncertainties, to experimental data and the ENDF/B-VIII evaluation for the residual production cross sections: $^{58}$Ni(p,x)$^{55}$Co, $^{58}$Ni(p,x)$^{56}$Co, $^{58}$Ni(p,x)$^{56}$Ni, and $^{58}$Ni(p,x)$^{57}$Ni. It is observed that the prior mean, representing the average values over all models without incorporating experimental data, outperformed the ENDF/B-VIII evaluation for all the considered cross sections. This is particularly surprising as the prior did not yet account for experimental data. Additionally, both the smoothed and non-smoothed versions of the posterior were observed to reasonably describe the experimental data. In Figs.~\ref{file_performance_da1} and \ref{file_performance_da2}, we present the prior and posterior means along with their corresponding uncertainties, for the elastic angular distributions at selected incident energies for p+$^{58}$Ni. As expected, it is observed that the posterior mean compares favorably with experimental data. Generally, small uncertainties are observed at smaller angles but they increase at certain angles, particularly at high angles. This is consistent with what was observed in Ref.~\cite{bib:1aa} where it was noted that TALYS faced difficulties in reproducing experiments in the high angle regions. In Fig.\ref{prior_corr_matrxMT028_Ni058_Gen}, we present an example of $^{58}$Ni(p,np) correlations based on the variation of only model parameters and on the variations of both models and their parameters. Similar correlation plots are presented in Fig.\ref{prior_corr_matrxMT111_Ni058_Gen} for the $^{58}$Ni(p,2p) cross section. Additionally, in Fig.~\ref{prior_corr_matrxMT028057_Ni058_Gen}, correlation matrices are presented for the following residual production cross sections: $^{58}$Ni(p,x)$^{57}$Ni (left) and $^{58}$Ni(p,x)$^{56}$Ni (right) based on the variations of both models and their parameters.

As expected, high correlations are observed in both cases, especially close to the diagonal. The correlations observed in the variation of only model parameters can be attributed to the use of the same models but with varying parameters. In the case of the variation of both models and their parameters, although different models were used, it is known that the models in each combination, made use of the same model parameters, inputs, and approaches in their solutions. These factors introduce correlations in the prior distribution. It must be noted that these prior correlations are taken into account through the simultaneous variations of both the models and the parameters. Additionally, as a consequence of the method, posterior correlations and covariances can be obtained. These prior and posterior correlations can be utilized for sampling and for the generation of random cross sections for the purpose of nuclear data uncertainty propagation to applications~\cite{bib:1m}.

Since we select models at each incident energy point rather than globally, the proposed BMA method naturally accounts for overfitting as well as underfitting issues as overfitted (and underfitted) models would have lower assigned posterior probabilities, and hence, their contributions to the final evaluation is tempered by the model averaging process. It is however important to note that the choice of models and prior model distributions is important as overly complex and poorly regularized models risks overfitting.

\section{Conclusion}
In traditional BMC approach, a single "best" model is often chosen to make predictions. However, this approach has been observed to be sensitive to the specific choice of the model. Additionally, the uncertainties related to model selection are not explicitly considered. In this work, we proposed a nuclear data evaluation method based on Bayesian Model Averaging (BMA) tailored to the fast energy region. Our proposed approach involves the use of a very large non-informative prior derived from sampling numerous models along with their parameters. In addition, instead of selecting a single "winning" model set for the entire energy range of interest, we select the models locally at each incident energy based on experimental data. The final evaluation is a weighted average over all considered models, with weights determined by the likelihood function values. Since the cross-sections and angular distributions were updated on a per-energy-point basis, the BMA approach typically results in discontinuities, or "kinks," in the cross sections or angular distributions curves. To address these kinks, a smoothing function was applied. In the future, we intend to explore other methods for the smoothing of the cross sections such as the Nadaraya Watson kernel regression using energy dependent weights. Furthermore, both prior and posterior covariances were obtained for each evaluation. The proposed method has been applied to the evaluation of p+$^{58}$Ni from 1 to 100 MeV energy range. The results demonstrate favorable comparisons with experimental data, as well as with the TENDL-2021 evaluation. \\

\section{Acknowledgment}
This work was done with funding from the Paul Scherrer Institute, Switzerland through the NES/GFA-ABE Cross Project.

% \section{Appendix}

\end{document}